\documentclass[moor]{informs1} 

\usepackage[utf8]{inputenc}
\usepackage[english]{babel}
\usepackage{textcomp}
\usepackage{mathtools}
\usepackage{enumerate}
\usepackage{accents}
\usepackage{enumitem}
\usepackage{natbib}
 \NatBibNumeric
 \def\bibfont{\small}%
 \def\newblock{\ }%
 \bibpunct[, ]{[}{]}{,}{n}{}{,}%

\usepackage[colorlinks=true,breaklinks=true,bookmarks=true,urlcolor=blue,
     citecolor=blue,linkcolor=blue,bookmarksopen=false,draft=false]{hyperref}

\def\EMAIL#1{\href{mailto:#1}{#1}}

\TheoremsNumberedThrough     

\EquationsNumberedThrough    

\newcommand{\w}{\omega}
\newtheorem{axiom}{Axiom}

\begin{document}

\TITLE{\begin{LARGE} Nonconcave Robust Utility Maximization under Projective Determinacy\end{LARGE}}

\ARTICLEAUTHORS{%
\AUTHOR{Laurence Carassus}
\AFF{ MICS, Centrale-Sup\'{e}lec, Universit\'{e} Paris-Saclay, France
 \EMAIL{laurence.carassus@centralesupelec.fr}}
\AUTHOR{Massinissa Ferhoune}
\AFF{LMR, UMR9008 CNRS and Universit\'{e} de Reims
Champagne-Ardenne, France, \EMAIL{fmassinissa@free.fr}}
} 
\ABSTRACT{We study a general robust utility maximization problem in a discrete-time frictionless market. The investor is assumed to have a possibly infinite, random, nonconcave, and nondecreasing utility function defined on the whole real line. She also faces model ambiguity on her beliefs about the market, which is modelled through a set of priors. 
We assume that the utility and the prices are projective functions of the path, while the graphs of the local priors are projective sets. Our other assumptions are stated on a prior-by-prior basis and correspond to generally accepted assumptions in the literature on markets without ambiguity. Under the set-theoretic axiom of Projective Determinacy (PD), our main result is the existence of an optimal investment strategy when the utility function is also upper-semicontinuous.  
 We further provide several counterexamples justifying our assumptions. }

\MSCCLASS{Primary: 91B16, 91G80, 54H05, 28B20, 60A10, 93E20}

\maketitle

\section{Introduction}
We are interested in the existence of an optimal investment strategy in a discrete-time, frictionless market, where there is uncertainty on the true probability of the events. 
Take the example of an urn with red and black balls from which a ball is drawn. There is only risk, and no uncertainty (this is called the ``known unknown"), if you know precisely the composition, e.g., 50 red and 50 black balls.  There is uncertainty if the composition is unknown, e.g., 100 red and black balls, but in a ratio entirely unknown to you. This is called the ``unknown unknown", and is named Knightian uncertainty after the work of Knight (see \cite{refkni_pj}).
Elsberg (see \cite{El61}) observes that people prefer the first urn to the second one, even this one may be more advantageous, showing uncertainty aversion. In line with this observation, 
Gilboa and Schmeidler \citep{refgs_pj} propose axioms so that the individuals' preferences have the following numerical representation: $X\mapsto \inf_{Q\in\mathcal{Q}} \mathbb{E}_Q U(X)$, where $U$ is a utility function, and $\mathcal{Q}$ is the set of probability measures (called priors) that model all the individual's beliefs about the future. Thus, the existence of an optimal investment strategy is equivalent to the existence of a solution to a maxmin expected utility problem. This is also called distributionally robust
optimization (DRO).  When $\mathcal{Q}$ is reduced to a singleton, there is no uncertainty, and the problem amounts to solving a classical utility maximization problem. We refer to \citep{refFS_pj} and the references therein for a detailed overview of these results in discrete time. 

The first attempts to solve the maxmin utility maximization problem, when $\mathcal{Q}$ is not a singleton, assumed that $\mathcal{Q}$ is dominated by a given probability measure. We refer to \citep{refdoml_pj} for a comprehensive survey of the dominated case. However, this setting excludes models with uncertainty about volatility, and Bouchard and Nutz propose the notion of quasi-sure uncertainty in \citep{refbn1_pj}. First, random sets of ``local" priors are defined, representing the investor's beliefs between times $t$ and $t+1$. The Fubini products of these ``local'' priors then form the set of inter-temporal priors $\mathcal{Q}$. The framework of \citep{refbn1_pj} assumes that the graphs of these random sets are analytic sets. Apart from this, $\mathcal{Q}$ is neither assumed to be compact, nor dominated by any particular measure.  Since beliefs are uncertain, utility functions are usually considered to be random. In the discrete time quasi-sure setting, \citep{ref9_pj} solves the maxmin utility maximization problem for concave and bounded from above utility functions, when only positive wealth is admissible, i.e., the utility functions are defined on the positive axis. The same result is later proved in \citep{ref4_pj} for concave and unbounded utility functions. We now focus on the case of potentially negative wealth. 
The existence of a solution to the maxmin utility maximization problem is proved in \cite{ref11_pj} for unbounded and concave utility functions, but in a one-period market. This result is then extended in \citep{refnotre_pj} to a general multiperiod market.  For nonconcave and bounded from above utility functions, \citep{refns_pj} obtains existence under the strong assumption that the investor's asset positions belong to a discrete set. All these results are proved using primal methods. 
Using a dual approach, \citep{refbar1_pj} solves the maxmin problem for an exponential utility function assuming, a strong local no-arbitrage condition. Some of these results are extended in \citep{refbar3_pj} to concave and bounded utility functions, under the assumption that medial limits exist. The existence of medial limits is a set-theoretic assumption, which cannot be proved in the usual Zermelo-Fraenkel set theory with the Axiom of Choice (ZFC).  
 Finally,  \citep{ref8_pj} derived existence results for concave and unbounded utility functions in a completely different framework, where uncertainty is represented by a set of stochastic processes.\\ 
When we finished the first version of this work, there were no general results on the existence of an optimal investment strategy in the discrete-time nondominated quasi-sure setting of \citep{refbn1_pj}, when the utility function is nonconcave, and the wealth of the investor can be negative. Since \citep{neufeldsester} have proved 
a dynamic programming principle, when the controls and the ``local'' priors are compact-valued (with a distance constraint on the priors), hemicontinuous, and the utility function is (separately) Lipschitz continuous.  Our result is of a different nature as we stay at a high level of generality. 
Our main contribution is to remove the concavity and the continuity assumptions on $U(\w,\cdot)$: we only assume that it is nonincreasing. 
Nonconcave utility functions are more realistic to represent an individual's beliefs. Indeed, in \citep{refkahn_pj}, it has been shown experimentally that there should be a wealth threshold under which an individual is risk seeking (her utility function is convex below that threshold), and above which she is risk averse (her utility function is concave above that threshold). This is the so-called S-shape utility functions. It is also natural that an individual may have jumps in her preferences, when certain levels of wealth are reached. 

One may wonder what kind of measurability is required when dealing with Knightian uncertainty. In the case of bounded utility functions, \citep{ref9_pj} and \citep{refns_pj} assume that $U$ is lower-semianalytic (lsa). In the unbounded case,  \citep{ref4_pj} and \citep{refnotre_pj} suppose that $U(\cdot,x)$ is Borel, which implies that $U$ is jointly Borel  (and thus lsa) as $U(\w,\cdot)$ is also assumed to be upper-semicontinuous and nondecreasing. Moreover, 
upper semianalyticity  (usa) is the measurability standard for derivatives and prices in this literature. However, the composition of two usa (or lsa) functions may fail to be usa (or lsa). As the composition of a Borel function with a usa (resp. lsa) one is usa (resp. lsa), the price process is assumed to be Borel. 
Summing up, in this literature, the utility function and the prices are Borel, the graphs of the set of ``local'' priors are analytic, while the optimal strategies, or the stochastic kernels,  obtained by measurable selection thanks to Jankov-von Neumann theorem, are only analytically, and thus, universally measurable. 
Here, we propose a new setup that generalises the one of \citep{refbn1_pj}, and which is homogeneous in terms of measurability requirements. Indeed, the basic inputs of the model, as well as the outputs, will have the same level of measurability. This setup is based on projective sets, and we call it the projective setup. 

The projective sets are a known generalization of Borel sets, whose construction is made recursively. Since Suslin in 1917, it is known that the projection of Borel sets may not be Borel. The projections of some Borel sets are the analytic sets.  The class of complements of analytic sets,  called coanalytic sets, does not coincide with the class of analytic sets and is not stable under projection. An analytic set of level 2 is then the projection of some coanalytic set, and coanalytic sets of level 2 are the complements of analytic sets of level 2. 
The classes of analytic and coanalytic sets of level $n$ are defined recursively, and their intersection is called $\Delta_n^1$. A set is called projective if it belongs to $\Delta_n^1$ for some $n$.  Contrary to the class of analytic sets, the class of projective sets is stable under complement; however, it is only stable under finite union or intersection. 
Borel and analytic sets are projective sets, and also universally measurable sets. However, ZFC is not enough to ensure that all projective sets are universally measurable. 
A well-accepted set-theoretic axiom that implies this result is the axiom of Projective Determinacy (PD). This axiom postulates that certain set-theoretic two-player games played on a projective set are determined in the sense that one of the two players always has a winning strategy. It amounts to admitting that Morgan's law applies formally to countable sequences of $\forall$ and $\exists$ in the case of projective sets. The seemingly only other use of the (PD) axiom in mathematical finance is in \citep{refprjnoarb_pj}, which, in a different uncertainty setting called model-independent, provides a pointwise fundamental theorem of asset pricing, as well as superhedging duality results assuming that the set of scenarios is analytic. 
Non-trivial results from set theory show that the (PD) axiom is implied by the existence of infinitely many Woodin cardinals (see \citep{reflargedet_pj}), or by the Proper Forcing axiom.
Quoting Woodin \lq\lq{}Projective Determinacy is the correct axiom for the projective sets; the ZFC axioms are obviously incomplete and, moreover, incomplete in a fundamental way.\rq\rq{}

To generalize lower and upper-semianalytic functions (lsa or usa), we will use projective functions. A function $f$ is projective, if there exists some $n$ such that $f$ is  $\Delta_n^1$-measurable. 
We will rely on the properties proved in the companion paper \cite{refprojnotre_pj} (the first--very long-- version of this work was including \cite{refprojnotre_pj} and the present paper). For example, the class of projective functions is stable under sums, differences, products, finite suprema and infima, and composition. So, contrary to lsa or usa functions, composition and difference of projective functions remain projective. 
We also have that Borel, as well as lsa and usa functions, are projective.  Moreover, under the (PD) axiom, the integrals of projective functions remain projective, and a projectively measurable $\epsilon$-optimal selector exists for projective functions. 

The projective setup assumes that the graph of ``local'' priors is a projective set (rather than an analytic one), and that the price process and the utility function are projective (instead of Borel). So, the projective setup generalizes the quasi-sure one of \cite{refbn1_pj}, without the (PD) axiom. 
We show that an optimal investment strategy exists under the (PD) axiom, and well-accepted conditions on the market and on the nonconcave utility function. 
Our two main results are Theorems \ref{one_step_strategy_pj} and \ref{optimality_M_typeA_pj}. Theorem \ref{one_step_strategy_pj} gives the existence of an investment strategy such that, if admissible, a bound on the optimality error can be derived that depends mainly on the jumps of the utility function. In particular, if the utility function is upper-semicontinuous (usc), then this investment strategy is optimal. In Theorem \ref{optimality_M_typeA_pj}, we show that for a given type of nonconcave and usc random utility, called of type (A) (introduced in \citep{refnotre_pj}), which includes $S$-shaped functions, the optimal solution of Theorem \ref{one_step_strategy_pj} is automatically admissible and, thus, optimal. This is done under additional integrability assumptions on the market, weaker than the ones of \citep{refnotre_pj}. We now comment on the assumptions of Theorem \ref{one_step_strategy_pj}, starting with three that can be tested directly, when the utility and the market are specified. The first is that $U(\w,\cdot)$ is nondecreasing. 
Note again that we do not assume that $U(\w,\cdot)$ is concave or continuous. The second one is the classical Asymptotic Elasticity constraints introduced in \citep{refAE1_pj} and \citep{refAE2_pj}.
The third one requires that $U$ is ``negative enough",  and is automatically satisfied for deterministic unbounded from below utility functions. 
We show that if one of these two last assumptions is not verified, then an optimal solution may not exist. 
We also assume the quasi-sure no-arbitrage condition. This condition is equivalent to $\mathcal{H} \neq \emptyset$, where $\mathcal{H}$ is the set of priors $P$, for which the $P$ no-arbitrage condition holds in a quasi-sure sense. The measures in $\mathcal{H}$ are very important ingredients of our work, and where introduced in \citep{ref4_pj}. 
Our last two conditions are not directly verifiable. 
The first asserts that the $P$ prior value function $U_0^{P}$ at time 0 is finite for each prior $P\in \mathcal{H}$. This is the well-accepted assumption, when there is no uncertainty, of the seminal work \citep{ref2_pj} (extended by \citep{ref7_pj} to nonconcave functions). This assumption provides a control from above for the value functions. The last assumption asserts that the expectations of the $P$-prior value functions $U_t^P$ are well-defined for all time $t$, and for each prior $P\in \mathcal{H}$. These assumptions are made on a prior-by-prior basis, and not for the supremum over all priors. Thus, they can be verified as in a setting without ambiguity. They are of course satisfied if $U$ is bounded from above.\\ 
Theorem \ref{one_step_strategy_pj} is proved by dynamic programming. First, we solve a one-period maximization problem (without the (PD) axiom). Here, the value function is not the ``obvious" maxmin, but is defined using an additional closure. This choice is because our utility function is not assumed to be regular, and is, in particular, nonconcave. Note that the ``obvious" maxmin one-period problem of \citep{refnotre_pj} may not have a solution here.  Most assumptions postulated in the one-period case are taken from \citep{ref7_pj}, and stated for each prior in $\mathcal{H}$. The method for finding a one-period optimal strategy uses arguments of regularity and coercivity obtained through the definition of the value function, as well as the assumptions of quasi-sure no-arbitrage and asymptotic elasticity. We then return to the general multiperiod problem and specify the (multiperiod) value functions. Our dynamic programming procedure is much simpler than those in \citep{ref3_pj} and \citep{refnotre_pj}. Indeed, our value functions do not involve a countable supremum or closure. The (PD) axiom makes such a choice possible, ensuring that these value functions exist and are projective. The (PD) axiom also provides a measurable selection theorem, which we apply to multiperiod versions of our one-period problem to find one-step optimal projective strategies. The resulting ``glued" strategy is the desired investment strategy of Theorem \ref{one_step_strategy_pj}, and its admissibility and the optimal one-step strategies' properties provide the desired bound.\\
One may wonder if our multi-period DRO problem could be solved in ZFC. Theorem \ref{not_provable} provides a negative answer to this question, if the utility function is analytically measurable, i.e., less than lsa. We show that in the constructible universe $L$ (a model of ZFC inconsistent with the (PD) axiom), the unique optimal strategy is not Lebesgue measurable. Manipulating integrals involving these functions is difficult, if not impossible, in a model without (PD). \\
The machinery developed here can be applied to other nonconcave and nonsmooth DRO problems, i.e., when the control does not appear as integrand in a (discrete) stochastic integral. The no-arbitrage condition may be replaced by some compactness assumption on the controls. 
The sharpest conditions for the dynamic programming principle to hold are left for further study. \\
The rest of this article is organized as follows. Section \ref{sett_main_pj} introduces the projective setup, the financial model, the assumptions, and the main results. In Section \ref{op_pj}, we solve a utility maximization problem in a one-period market, while in Section \ref{muti_per_pj}, we prepare the dynamic programming procedure. Section \ref{proof_th1_pj} provides the proof of Theorem \ref{one_step_strategy_pj}, while Section \ref{proof_optimality_M_pj} the one of Theorem \ref{optimality_M_typeA_pj}. Section \ref{contreex} gives several counterexamples. In the appendix, we recall results on projective sets and functions, and provide further proofs and results.

\section{Setting and main result}
\label{sett_main_pj}
\subsection{Projective setup}
\label{pd_exp_lab}
\label{subsec_def_proj_pj}
In this section, we introduce projective sets and functions, the (PD) axiom, and the two crucial consequences of the (PD) axiom.
\begin{definition}
Let $X$ be a Polish space. Let $\mathcal{N}:= \mathbb{N}^\mathbb{N}$ be the Baire space. First, $\Sigma_1^1(X)$ is the class of analytic sets of $X,$ while $\Pi_1^1(X)$ is the class of coanalytic sets of $X$:
\begin{eqnarray}
\Sigma_{1}^1(X) := \{\textup{proj}_X(C),\;  C\in {\cal B}(X\times \mathcal{N})\} \quad 
\Pi_{1}^1(X) := \{X\setminus C,\; C\in\Sigma_{1}^1(X)\}. 
\label{level1}
\end{eqnarray}
 For each $n\geq 2$, we define recursively the classes $\Sigma_n^1(X)$ and $\Pi_n^1(X)$ of $X$ as follows:
\begin{eqnarray*}
\Sigma_{n}^1(X):= \{\textup{proj}_X(C),\;  C\in\Pi_{n-1}^1(X\times \mathcal{N})\} \quad 
\Pi_{n}^1(X) := \{X\setminus C,\; C\in\Sigma_{n}^1(X)\}. 
\end{eqnarray*}
Then, for all $n\geq 1$, we set 
\begin{eqnarray*}
\Delta_{n}^1(X)&:=& \Sigma_n^1(X)\cap \Pi_n^1(X).
\label{Delta_pj}
\end{eqnarray*}
The class $\mathbf{P}(X)$ of projective sets of $X$ is defined by
\begin{eqnarray*}
\mathbf{P}(X) := \bigcup_{n\geq 1} \Delta_n^1(X).
\end{eqnarray*}
\label{def_proj_set_pj}
\end{definition}
The Borel sets are the only sets that are analytic and coanalytic (see \cite[Theorem 14.11, p88]{refproj_pj}).  They are at the first level of the projective hierarchy: 
\begin{equation}\label{equation_borel_set}
    \mathcal{B}(X)=\Sigma^1_{1}(X)\cap\Pi^1_{1}(X)=\Delta_1^1(X). 
\end{equation}
So, Borel sets, analytic and coanalytic sets are projective sets. We now define the projective functions.
\begin{definition}
Let $X$ and $Y$ be Polish spaces and $D\subset X$. A function $f : D \to Y$ is projective (or projectively measurable) if $D\in\textbf{P}(X)$ and if there exists some $n\geq 1$ such that $f$ is $\Delta_n^1(X)$-measurable in the sense that $f^{-1}(B):=\{x\in D, f(x)\in B\}\in\Delta_n^1(X),$ for all $B\in\mathcal{B}(Y)$. 
\label{proj_fct_def_pj}
\end{definition}
So, Borel, lower-semianalytic (lsa) and upper-semianalytic (usa) functions are projective, see \eqref{level1} and \eqref{equation_borel_set}.\\
In the appendix, we give a ``toolbox\rq\rq{} proposition with the key properties of the projective sets and functions, see Proposition \ref{base_hierarchy_pj}. In particular, projective functions are stable under classical operations, but in contrast with usa or lsa functions, they are stable under difference and composition. 

We now introduce the axiom of Projective Determinacy (PD), which will be assumed for our multi-step results, Theorems \ref{one_step_strategy_pj} and \ref{optimality_M_typeA_pj}.  Let $A \subset \mathcal{N}$ be a non-empty set. Consider a two-player infinite game played as follows. 
Player I plays $a_0\in \mathbb{N}$, then Player II plays $b_0\in \mathbb{N}$, then Player I plays $a_1\in \mathbb{N}$, etc. A play is thus a sequence $(a_0,b_0,a_1,b_1,\cdots) \in \mathcal{N}$.  
Player I wins the game if  $(a_0,b_0,a_1,b_1,\cdots) \in A$. Else, if $(a_0,b_0,a_1,b_1,\cdots) \in \mathcal{N} \setminus A,$ 
Player II wins. A winning strategy for Player I (resp. II) is a strategy under which Player I  (resp. II) always wins, i.e., whatever Player II  (resp. I) plays, the result of the game always belongs to  $A$ (resp. $\mathcal{N} \setminus A$). 
 We say that $A$ is \textit{determined} if a winning strategy exists for one of the two players. 
\begin{axiom}
If $A \in \mathbf{P}(\mathcal{N})$, $A$ is determined.
\label{projective_determinacy_pj}
\end{axiom}
\begin{remark}
We refer to \cite{refprojnotre_pj} and the references therein for a detailed discussion on the (PD) axiom. 
The (PD) axiom is fruitful as it allows us to answer old, fundamental questions on projective sets, see Theorem \ref{proj_is_univ_pj} below. 
What makes this axiom plausible is that it is implied by other axioms of descriptive set theory that, a priori, have no direct connection with projective sets. For example, it is implied by the Proper Forcing axiom. The (PD) axiom is also closely related to the existence of Woodin cardinals. 
The relative consistency of large cardinals with ZFC has been extensively studied via inner model theory and forcing techniques. No contradictions are known, assuming that ZFC itself is consistent (see \cite{refwood1_pj}). In Section \ref{ZFC_proof},  we show that the ZFC theory is insufficient to solve a dynamic programming problem.  A VOIR
 \end{remark}

We will not use the (PD) axiom directly, but the following two consequences. The first one asserts that projective sets are universally measurable, and the second one that measurable selections on projective sets are possible, like in the Jankov von Neumann theorem, see \citep[Proposition 7.49, p182]{ref1_pj}. This is less strong than the (PD) axiom, as it is implied, for example, by the existence of an infinite sequence $\kappa_0 <\kappa_1<\ldots$ of cardinals, with supremum $\lambda$, such that each $\kappa_n$ is $\lambda$-strong. 
\begin{theorem}
Assume the (PD) axiom.\\ 
(i) Let $X$ be a Polish space, then $\textbf{P}(X) \subset \mathcal{B}_c(X)$.\\
(ii) Let $X$ and $Y$ be Polish spaces and $A \in \mathbf{P}(X \times Y)$. Then, there exists a projective function $\phi : \textup{proj}_X(A) \to Y$ such that $\textup{Graph}(\phi) \subset A$.
\label{proj_is_univ_pj}
\end{theorem}
\proof{Proof.}
See \citep[Theorem 38.17, p326]{refproj_pj} and \citep[Proposition 7]{refprojnotre_pj}. \Halmos \\ \endproof
Theorem \ref{proj_is_univ_pj}, together with $\Delta_n^1(X) \subset \textbf{P}(X)$ for all $n\geq 1$, ensure that projective functions are universally measurable. \\
The (PD) axiom will be assumed for our multi-step financial results, but not in the one-step Section \ref{op_pj}.  We will always write ``under the (PD) axiom" to stress where this axiom is indeed used.

\subsection{Financial Setting}
\label{sett_pj}
We fix a time horizon $T$, and introduce a family of Polish spaces $(\Omega_t)_{1\leq t \leq T}$. For some $0\leq t\leq T$, let $\Omega^t:=\Omega_1 \times \cdot\cdot\cdot \times \Omega_t$, with the convention that $\Omega^0$ is a singleton. For a Polish space $X$, we denote by $\mathfrak P(X)$ the set of all probability measures defined on the measurable space $(X,\mathcal{B}(X))$, where $\mathcal{B}(X)$ is the Borel sigma-algebra on $X$. We denote by $\mathcal{B}_c(X)$ the completion of $(X,\mathcal{B}(X))$  with respect to all $P\in\mathfrak{P}(X)$. 
Let $S:=(S_t)_ {0\leq t\leq T}$ be a $\mathbb{R}^d$-valued process representing the discounted price of $d$ risky assets over time. 
\begin{assumption}
For all $0\leq t \leq T$, $\Omega^t\ni \w^t \mapsto S_t (\w^t)\in \mathbb{R}^d$ is a projective function.
\label{S_borel_pj}
\end{assumption}
In the setting of Bouchard and Nutz, $S_t$ is usually assumed to be Borel. As Borel functions are projective without the (PD) axiom (see \eqref{equation_borel_set}), our assumption is thus weaker.
We consider a random utility function defined on the whole real line, which models the investor's preference on the market in the case of possible negative wealth.
\begin{definition}
A random utility $U$: $\Omega^T \times \mathbb{R} \to \mathbb{R}\cup\{-\infty,+\infty\}$ is  such that $U$ is a projective function and $U(\w^T,\cdot)$ is nondecreasing, for all $\w^T \in \Omega^T$. \\ 
\label{U_hp_pj}
\end{definition}

The utility function is chosen random, because the law of nature is uncertain.  Usually $U(\cdot,x)$ is assumed to be Borel  and $U(\w^T,\cdot)$ to be nondecreasing and upper-semicontinuous (usc), see \citep{ref3_pj}, \citep{ref4_pj}, \citep{ref10_pj}, \citep{refnotre_pj} and \citep{ref9_pj}. This implies that $U$ is $\mathcal{B}(\Omega^T) \otimes \mathcal{B}(\mathbb{R})$-measurable (see \citep[Lemmata 5.10 and 5.13]{ref10_pj}).  So, $U$ is projective (see \eqref{equation_borel_set} without the (PD) axiom) and our assumption is again weaker. Moreover, we do not assume that $U(\w^T,\cdot)$ is continuous or usc. However, as $U(\w^T,\cdot)$ is nondecreasing, $U(\w^T,\cdot)$ is continuous except on a countable set of points. And of course, we do not assume that $U(\w^T,\cdot)$ is concave.\\ 

We now construct the set $\mathcal{Q}^T$ of all prevailing priors. The set $\mathcal{Q}^T$ captures all the investor's beliefs about the law of nature. For all $0\leq t \leq T-1$, let\footnote{The notation $\twoheadrightarrow$ stands for set-valued mapping.} $\mathcal{Q}_{t+1}: \Omega^t \twoheadrightarrow \mathfrak P(\Omega_{t+1})$, where $\mathcal{Q}_{t+1}(\w^t)$ can be seen as the set of all possible priors for the $t+1$-th period, given the state $\w^t$ at time $t$. The following assumption allows us to perform measurable selection, see Theorem \ref{proj_is_univ_pj} (ii).
\begin{assumption}
The set $\mathcal{Q}_1$ is nonempty and convex. 
For all $\;1\leq t \leq T-1$, $\mathcal{Q}_{t+1}$ is a nonempty  and convex-valued 
random set such that $\textup{Graph}(\mathcal{Q}_{t+1}):=\{(\w^t,p)\in \Omega^t \times \mathfrak P(\Omega_{t+1}), p\in\mathcal{Q}_{t+1}(\w^t)\}$ is a projective set.
\label{analytic_graph_pj}
\end{assumption}
Explicit examples of nondominated financial markets satisfying Assumption \ref{analytic_graph_pj} can be adapted from  \citep{ref4_pj} and \citep{refbar1_pj}. Among them is a robust discrete-time Black-Scholes model and a robust binomial model, where the uncertainty affects the probability of jumps and their size.\\
In the Bouchard and Nutz setting (see, among others \citep{refbn1_pj}, \citep{ref9_pj}, \citep{ref3_pj}, \citep{refbar1_pj}, \citep{ref4_pj}, \citep{refns_pj} and \citep{refnotre_pj}), $\textup{Graph}(\mathcal{Q}_{t+1})$ is assumed to be an analytic set. By definition, an analytic set is a projective set (again without the (PD) axiom, see \eqref{level1}). Thus, in the ZFC theory, our setting includes the classical quasi-sure one, without postulating continuity and concavity assumptions on $U(\w^T,\cdot)$. \\ 

For all $0\leq t\leq T-1$, $SK_{t+1}$ is the set of projectively measurable stochastic kernels on $\Omega_{t+1}$ given $\Omega^t$. This means that $q_{t+1}\in SK_{t+1}$ if $q_{t+1}[\cdot|\cdot] : \mathcal{B}(\Omega_{t+1})\times \Omega^t \to \mathbb{R}$ is such that for all $\w^t\in\Omega^t$, $q_{t+1}[\cdot|\w^t] \in \mathfrak{P}(\Omega_{t+1})$ and $\Omega^t \ni \w^t \mapsto q_{t+1}[\cdot|\w^t]\in \mathfrak{P}(\Omega_{t+1})$ is projectively measurable.  
Under the (PD) axiom, if $q_{t+1}\in SK_{t+1}$, $ \w^t \mapsto q_{t+1}[\cdot|\w^t]$ is universally measurable (see  Theorem \ref{proj_is_univ_pj} (i)) and $q_{t+1}$ is an universally measurable stochastic kernel, see \citep[Definition 7.12, p134]{ref1_pj}.
Moreover, still under the (PD) axiom, Assumption \ref{analytic_graph_pj} and Theorem \ref{proj_is_univ_pj} (ii) show that there exists $q_{t+1}\in SK_{t+1}$ such that for all $\w^t\in \Omega^t$ (recall that $\mathcal{Q}_{t+1}\neq \emptyset$), $q_{t+1}[\cdot|\w^t]\in \mathcal{Q}_{t+1}(\w^t).$ Now, for all $1\leq t\leq T$, we can use Fubini's theorem (see Remark \ref{rem_proj_int_pj}) and define the product measure $q_1 \otimes \cdots \otimes q_t$, which belongs to $\mathfrak{P}(\Omega^t)$. The set $\mathcal{Q}^t \subset \mathfrak{P}(\Omega^t)$ is defined by 
\begin{eqnarray}
\mathcal{Q}^t &:=& \{q_1 \otimes q_2 \otimes \cdot\cdot\cdot \otimes q_t, q_1\in \mathcal{Q}_1,\; q_{s+1}\in SK_{s+1}, q_{s+1}[\cdot|\w^s]\in \mathcal{Q}_{s+1}(\w^s),\; \forall \w_s\in \Omega^s,\forall 1\leq s \leq t-1\}. \label{def_set_Q_pj}
\end{eqnarray}
We also set $\mathcal{Q}^0:= \{ \delta_{\w_0}\}$, where $\delta_{\w_0}$ is the Dirac measure on the single element $\w_0$ of $\Omega^0$.
If $P:= q_1 \otimes q_2 \otimes \cdot\cdot\cdot \otimes q_T \in\mathcal{Q}^T$, we write for any $1\leq t\leq T$, $P^t:= q_1 \otimes q_2 \otimes \cdot\cdot\cdot \otimes q_t$ and $ P^t\in\mathcal{Q}^t$. In this paper, we mostly work directly on the disintegration of $P$ rather than on $P$. Thus, from now on, we will precise the fixed disintegration for which the required result holds.
\begin{remark}
Let $X$ be a Polish space. Let $f : X \to \mathbb{R}\cup\{-\infty,+\infty\}$ be a  universally measurable function, and let $p\in \mathfrak{P}(X)$. 
We define the $(-\infty)$ integral, denoted by $\int_{-} f dp$,  
as follows.  When $\int f^+ dp<+\infty$ or $\int f^- dp<+\infty$, $\int_{-} f dp$ equals the extended integral of $f$, i.e.
\begin{eqnarray*}
\int_{-} f dp 
:=\int f^+ dp - \int f^- dp. 
\end{eqnarray*}
Otherwise, $\int_{-} f dp := -\infty$. 
This means that $\int_{-}$ is computed using the convention :
\begin{eqnarray}
\label{cvt_inf_pj}
-\infty+\infty=+\infty-\infty=-\infty.
\end{eqnarray}
This is the usual convention for maximization problems. Note, however, that \citep{ref1_pj} adopts the opposite convention. 
We have seen in Theorem \ref{proj_is_univ_pj} that, under the (PD) axiom, any projective set $A$ is universally measurable. This allows us to define $p[A]$ for any probability measure $p$ and, more generally, to use classical measure theory results in the projective context. Under the (PD) axiom, any projective function $f$ is universally measurable, so that $\int_{-} f dp$ is well-defined, and the integral of a projective function is projective, see Proposition \ref{univ_cvt_pj} . 
Moreover, it is possible to construct a unique probability measure on the product space from projectively measurable stochastic kernels and also, to use Fubini\rq{}s theorem when $f$ is projective (see \cite[Proposition~7.45~p.175]{ref1_pj}). An important consequence is that the sets $(\mathcal{Q}^t)_{0\leq t\leq T}$ (see \eqref{def_set_Q_pj}) are indeed well-defined. From now on, we will write $\int$ (or $\mathbb{E}$) instead of $\int_{-}$ except when we need to clarify the difference between both integrals.\\ 
\label{rem_proj_int_pj}
\end{remark}
 
Trading strategies are represented by $d$-dimensional processes $\phi:=\{\phi_t,\; 1\leq t\leq T\}$, representing the investor's holdings in each of the $d$ risky assets over time. We assume that the functions $\phi_t : \Omega^{t-1}\to \mathbb{R}^d$ are projective, for all $1\leq t\leq T$. The set of all such trading strategies is denoted by $\Phi$. Under the (PD) axiom, if $\phi\in\Phi$, then $\phi$ is universally measurable (see  Theorem \ref{proj_is_univ_pj} (i)), which is the usual assumption in the quasi-sure literature. So,  our assumption on strategies is weaker, but we assume the (PD) axiom this time. Trading is assumed to be self-financing, and the value at time $t$ of a portfolio $\phi\in\Phi$, starting from initial capital $x\in \mathbb{R}$, is thus given by $$V_t^{x,\phi}=x+\sum_{s=1}^{t} \phi_s \Delta S_s.$$
Note that if $x,y\in \mathbb{R}^d$ then the concatenation $xy$ stands for their scalar product. The symbol $|\,.\,|$ refers to the Euclidean norm on $\mathbb{R}^d$ (or on $\mathbb{R}$) and $|\,.\,|_1$ is the norm on $\mathbb{R}^d$ defined by $|\,x\,|_1 :=\sum_{i=1}^d |x_i|$ for all $x\in \mathbb{R}^d$.\\ 

Recall that a set $A\subset \Omega^T$ is a $\mathcal{Q}^T$-polar set if there exists $N\in \mathcal{B}_c(\Omega^T)$ such that $A \subset N$ and $P[N]=0$ for all $P\in \mathcal{Q}^T$. A property holds $\mathcal{Q}^T$-quasi-surely (q.s.) if it holds outside of a $\mathcal{Q}^T$-polar set. The complement of a $\mathcal{Q}^T$-polar set is called a $\mathcal{Q}^T$-full-measure set. Of course, any $\mathcal{Q}^T$-full-measure set is a $P$-full-measure set for all $P\in\mathcal{Q}^T$. Under the (PD) axiom, if $A\subset \Omega^T$ is a projective set, then $A\in \mathcal{B}_c(\Omega^T)$ (see Theorem \ref{proj_is_univ_pj} (i)), and $A$ is a $\mathcal{Q}^T$-full-measure set if $P[A] = 1$ for all $P\in\mathcal{Q}^T$.
We recall the $NA(\mathcal{Q}^T)$-condition of Bouchard and Nutz \citep[Definition 1.1]{refbn1_pj} as well as the classical uni-prior no-arbitrage condition.
\begin{definition}
\label{defNA_pj}
The $NA(\mathcal{Q}^T)$ condition holds true if $V_T^{0,\phi}\geq 0$ $\mathcal{Q}^T$-$\mbox{q.s.}$ for some $\phi\in \Phi$ implies that $V_T^{0,\phi}= 0$ $\mathcal{Q}^T\mbox{-q.s.}$\\
Let $P \in \mathfrak P(\Omega^{T})$. The $NA(P)$ condition holds true if $V_T^{0,\phi}\geq 0$ $P$-$\mbox{a.s.}$ for some $\phi\in \Phi$ implies that $V_T^{0,\phi}= 0$ $P\mbox{-a.s.}$\\
\end{definition}
 \begin{assumption}
The $NA(\mathcal{Q}^T)$ condition holds true. 
\label{H_nonempty_pj}
\end{assumption}
We will use the alternative form of $NA(\mathcal{Q}^T)$ given in Theorem~\ref{main_result_NA}, which was proved by Blanchard and Carassus in the setup of Bouchard and Nutz (see \cite[Theorem 3.29]{ref4_pj}). For that, we need the conditional support of the price increments. Let $0\leq t\leq T-1$ and $P\in\mathfrak{P}(\Omega^T)$ with the fixed disintegration $P:=q_1^P \otimes \cdot\cdot\cdot \otimes q_T^P$. The multiple-priors conditional support $D^{t+1} : \Omega^t \twoheadrightarrow \mathbb{R}^d$  and the conditional support relatively to $P$, $D^{t+1}_P : \Omega^t \twoheadrightarrow \mathbb{R}^d$  are defined by
\begin{eqnarray}
D^{t+1}(\w^t)&:=&\bigcap \{A\subset \mathbb{R}^d,\; \mbox{closed},\; p[\Delta S_{t+1}(\w^t,\cdot)\in A]=1,\;\forall p \in \mathcal{Q}_{t+1}(\w^t)\} \label{support_def_pj}\\
D^{t+1}_P(\w^t)&:=&\bigcap \{A\subset \mathbb{R}^d,\; \mbox{closed},\; q_{t+1}^P[\Delta S_{t+1}(\w^t,\cdot)\in A|\w^t]=1\}.\label{support_defP_pj} 
\end{eqnarray}
Additionally, for some $R\subset \mathbb{R}^d$, let
\begin{eqnarray*}
\mbox{Aff}(R)&:=&\bigcap \{A\subset \mathbb{R}^d,\; \mbox{affine},\; R\subset A\}\quad
\mbox{Conv}(R):=\bigcap \{C\subset \mathbb{R}^d,\; \mbox{convex},\; R\subset C\}
\end{eqnarray*}
and if $R$ is convex, $\mbox{ri}(R)$ is the interior of $R$ relatively to $\mbox{Aff}(R)$.
\begin{theorem}
\label{main_result_NA}
    Assume the (PD) axiom. Assume that Assumptions~\ref{S_borel_pj} and \ref{analytic_graph_pj} holds true. 
    Then, $NA(\mathcal{Q}^T)$ is equivalent to $\mathcal{H}_T \neq \emptyset$, where
    \begin{eqnarray}
\mathcal{H}^T :=\big\{P\in \mathcal{Q}^T,\; 0\in \textup{ri}\big(\textup{conv}(D_P^{s+1})\big)(\cdot)\; \mathcal{Q}^s \textup{-q.s.},\;\textup{Aff}(D_P^{s+1})(\cdot)=\textup{Aff}(D^{s+1})(\cdot)\; \mathcal{Q}^s \textup{-q.s.},  \forall 0\leq s \leq T-1\big\}. 
\label{H_def_pj}
\end{eqnarray}
\end{theorem}
\proof{Proof.}
The fact that $\mathcal{H}_T \neq \emptyset$ implies $NA(\mathcal{Q}^T)$ is proved in Lemma \ref{lemmaH_nonempty_pj}, see Appendix \ref{miss_proof0_pj}. The other direction is proved in \cite[Theorem 2]{bci25}. 
\Halmos \\ \endproof
We can introduce the set of admissible strategies for the utility maximization problem. 
\begin{definition}
Let $U$ be a random utility function as in Definition \ref{U_hp_pj}. Let $P\in\mathfrak{P}(\Omega^T)$ and $x\in\mathbb{R}$. 
\begin{eqnarray*}
\Phi(x,U,P) &:=& \big\{\phi\in\Phi,\; \mathbb{E}_P U^-\big(\cdot,V_T^{x,\phi}(\cdot)\big)<+\infty\big\}\\
\Phi(x,U,\mathcal{Q}^T)&:=&\bigcap_{P\in \mathcal{Q}^T}\Phi(x,U,P).
\end{eqnarray*}
\label{admissibility_def_pj}
\end{definition} 
We are now in a position to state our utility maximization problem, when the uncertainty about the true probability of future events is modelled by $\mathcal{Q}^T$ :
\begin{eqnarray}
u(x):=\sup_{\phi\in \Phi(x,U,\mathcal{Q}^T)}\inf_{P\in \mathcal{Q}^T}\mathbb{E}_P U\big(\cdot,V_T^{x,\phi}(\cdot)\big).
\label{RUMP_pj}
\end{eqnarray} 
Note that $u(x)=\sup_{\phi\in \Phi}\inf_{P\in \mathcal{Q}^T}\mathbb{E}_P U(\cdot,V_T^{x,\phi}(\cdot)).$ Indeed, if $\phi\in \Phi\setminus \Phi(x,U,\mathcal{Q}^T)$, then $\inf_{P\in\mathcal{Q}^T}\mathbb{E}_P U(\cdot,V_T^{x,\phi}(\cdot))=-\infty$ thanks to convention \eqref{cvt_inf_pj}.\\ 

We now introduce the different assumptions needed for the existence of an optimal strategy in \eqref{RUMP_pj}.

\subsection{Direct Assumptions on $U$} 

We start with the assumptions that can be directly (and easily) checked. The first one is on the asymptotic behaviour of the random utility $U$. This kind of condition already appeared in \citep[Proposition 4]{ref7_pj} for a finite nondecreasing continuous and non necessarily concave function in the uni-prior setting. See also \citep[Proposition 5.1]{ref2_pj},  \citep[Proposition 3.24]{ref3_pj} and \citep[Assumption 3]{refnotre_pj}.
\begin{assumption}
There exist $0<\underline{\gamma}<\overline{\gamma}$ and a 
projective function $C : \Omega^T \to \mathbb{R}^+ \cup\{+\infty\}$ such that $\sup_{P\in\mathcal{Q}^T} \mathbb{E}_P C <+\infty$, and such that for all $\w^T\in\Omega^T$ satisfying $C(\w^T)<+\infty$, for all $\lambda\geq 1$ and $x\in\mathbb{R}$,
\begin{eqnarray}
U(\w^T,\lambda x) \leq  \lambda^{\overline{\gamma}}\big(U(\w^T,x)+C(\w^T)\big) & \mbox{and} &
U(\w^T,\lambda x) \leq  \lambda^{\underline{\gamma}}\big(U(\w^T,x)+C(\w^T)\big). 
\label{elas_gammaf_pj} \label{elas_gammaf_pj22}
\end{eqnarray}
\label{AE_pj}
\end{assumption}
Recall from Remark \ref{rem_proj_int_pj} that $\mathbb{E}_P C $ exists for all $P\in\mathcal{Q}^T$. 
A consequence of Assumption \ref{AE_pj} is that a one-step strategy must be bounded to be optimal for problem \eqref{cl_pb_one_pj} (see Proposition \ref{sub_optimal_pj}), and this compactness result will allow us to prove the existence of an optimal strategy for \eqref{RUMP_pj}. \\
Note that \citep[Assumption 3]{refnotre_pj} is a weaker version of Assumption \ref{AE_pj}. Indeed, when $U(\w^T,\cdot)$ is assumed to be concave, only one inequality in Assumption \ref{AE_pj} suffices (see \citep[Lemma 7]{refnotre_pj}). Moreover,  Assumption \ref{AE_pj} is related to the notion of Reasonable Asymptotic Elasticity (RAE) introduced in  \cite{refAE1_pj} and \cite{refAE2_pj}.  Assumption \ref{AE_pj} asserts that $\overline{\gamma}$, the Asymptotic Elasticity (AE) in $-\infty$ must be greater that $\underline{\gamma}$, the AE in $+\infty$. We refer to \citep{refnotre_pj} and \citep{ref7_pj} for a detailed discussion on the link between RAE and Assumption \ref{AE_pj}.

The next assumption ensures that $U$ takes negative values and is similar to \citep[Assumption 4]{refnotre_pj}.
\begin{assumption}
There exists a projective function $\underline{X}: \Omega^T\to \mathbb{R}$ such that $\underline{X}(\cdot)<0$ and $U(\cdot,\underline{X}(\cdot))\leq -C(\cdot) -1$ $\mathcal{Q}^T \mbox{q.s.}$, where $C(\cdot)\geq 0$ has been introduced in Assumption \ref{AE_pj}.
\label{nncst_pj}
\end{assumption}
Assumption \ref{nncst_pj} is of course satisfied if $U$ is deterministic, nondecreasing and such that $\lim_{x\to -\infty} U(x) =-\infty$. 
Moreover, if Assumption \ref{nncst_pj} is not satisfied, then \eqref{RUMP_pj} may have no solution, see \citep[Remark 5]{refnotre_pj} in a one-period concave and usc setting. 
\subsection{Value functions and Assumption on $U_0^P$}

We first introduce the dynamic programming procedure and the associated value functions. We do it for the multiple-priors utility maximization problem \eqref{RUMP_pj} with the value functions $U_t$, and also for the utility problem related to a given prior $P$ with the value functions $U_t^P$. Fix $P\in\mathcal{Q}^T$ with fixed disintegration $P:=q_1^P\otimes \cdots \otimes q_T^P$. For all $0\leq t\leq T-1$, we define
\begin{eqnarray}
&&\Bigg\{
    \begin{array}{ll}
   	
U_T^P(\w^T,x)&:=U(\w^T,x)\\
u_t^P(\w^t,x,h)&:= \mathbb{E}_{q_{t+1}^P[\cdot|\w^t]} U_{t+1}^P\big(\w^t,\cdot,x+h\Delta S_{t+1}(\w^t,\cdot)\big) \label{u_t_p_pj}\\
U_t^P(\w^t,x)&:=\sup_{h\in\mathbb{R}^d }u_t^P(\w^t,x,h)\label{state_val_t_p_pj}\\
    \end{array}
\\ 
\nonumber \\
&&\Bigg\{
	\begin{array}{ll}
U_T(\w^T,x)&:=U(\w^T,x)\\
u_t(\w^t,x,h)&:= \inf_{p\in\mathcal{Q}_{t+1}(\w^t)} \mathbb{E}_p U_{t+1}\big(\w^t,\cdot,x+h\Delta S_{t+1}(\w^t,\cdot)\big)\label{u_t_rob_pj}\\
U_t(\w^t,x)&:=\sup_{h\in\mathbb{R}^d} u_t(\w^t,x,h).\label{state_val_t_rob_pj}
    \end{array}
\end{eqnarray}
The existence and the measurability of $U_t$ and $U_t^P$ are not trivial, especially as we take uncountable suprema. These results will be proved in Section \ref{muti_per_pj} under the (PD) axiom. Here, no regularity is assumed for the random utility function, and the (PD) axiom provides a measurability framework, where the uncountable supremum or infimum of measurable functions remains measurable, and measurable selection can be performed. Our definition of the value functions differs from the usual one as we take suprema over all $h$ in $\mathbb{R}^d$ and not in $\mathbb{Q}^d$, and as we do not take the closure of these suprema, see \citep{ref3_pj}, \citep{refnotre_pj}, \citep{refns_pj} and \citep{ref9_pj}. Taking a countable supremum and the closure ensures that $U_t$ is lower-semianalytic, i.e., the measurability of $U_t$ stays at the lowest level of the projective hierarchy. Low enough to use measurable selection methods in the ZFC theory only. One may wonder if a better result can be obtained in Theorem \ref{one_step_strategy_pj}, taking a countable infimum in \eqref{u_t_p_pj} and \eqref{u_t_rob_pj}, and the closure of this $U_t$ as is usually done. The answer is no, see Remarks \ref{rem_star} and \ref{rem_link_value_art_pj}.\\

We now introduce an assumption that provides a control from above on each $U_t^P$ for $P\in\mathcal{H}^T$.
\begin{assumption}
For all $P\in \mathcal{H}^T$, $U_0^P(1)<+\infty$.
\label{U0_pj}
\end{assumption}

Assumption \ref{U0_pj} is similar to \citep[Assumption 2.3 (1)]{ref2_pj}, and thus well-accepted in the uni-prior setting. What is nice here is that the assumption is postulated prior by prior, and not uniformly on all of them. This assumption is obviously satisfied when $U$ is bounded from above. Otherwise, Assumption \ref{U0_pj} is not easy to verify. We will propose a general context, where it holds automatically true (see Theorem \ref{optimality_M_typeA_pj}). Assumption \ref{U0_pj} was also postulated in \cite[Assumption 6]{refnotre_pj} or \citep[Assumption 2 (11)]{ref7_pj}. We refer to \citep[Lemma 2]{refnotre_pj}, and the discussion after, for a detailed comparison of Assumption \ref{U0_pj} to the similar assumptions in the literature. Note that $U_0(1)<+\infty$ is not enough to get existence of an optimal strategy, see the counter-example in \citep[Remark 2]{refnotre_pj}.\\ 

The last assumption states that the $P$-prior  problems are well-defined in the following sense.
\begin{assumption}
For all $1\leq t \leq T$, $P\in\mathcal{H}^T$ and $\theta\in \{-1,1\}^d \cup \{0\}$, $\mathbb{E}_{P} U_t^P(\cdot,1+\theta \Delta S_{t}(\cdot))$ is well defined in the generalized sense i.e.
We have that
\begin{eqnarray*}
\mathbb{E}_{P} (U_t^P)^+\big(\cdot,1+\theta \Delta S_{t}(\cdot)\big)<+\infty \quad \mbox{or} \quad\mathbb{E}_{P} (U_t^P)^-\big(\cdot,1+\theta \Delta S_{t}(\cdot)\big)<+\infty.
\end{eqnarray*}
\label{well_def_hp_pj}
\end{assumption}
This assumption is satisfied if either $U$ is bounded from above or below. 
We compare Assumption \ref{well_def_hp_pj} to \citep[Assumption 5]{refnotre_pj}, i.e., that  
$\sup_{P\in\mathcal{Q}^T}\mathbb{E}_P U^-(\cdot,x+h\Delta S_t(\cdot))<+\infty$ for all $1\leq t\leq T$, $(x,h)\in \mathbb{Q} \times \mathbb{Q}^d.$  
 One trivial observation is that contrary to \citep[Assumption 5]{refnotre_pj}, Assumption \ref{well_def_hp_pj} allows that $U=-\infty$ on a non-polar set. Moreover, using \citep[(52) and Proposition 6 (iv)]{refnotre_pj}, we can show that Assumption \ref{well_def_hp_pj} holds true under Assumptions \ref{S_borel_pj}, \ref{analytic_graph_pj}, \ref{AE_pj}, and \citep[Assumption 5]{refnotre_pj}, when $U$ is as in \citep[Definition 1]{refnotre_pj}.

\subsection{Main result}
As $U(\w^T,\cdot)$ is not continuous, our main result is not the existence of an optimal solution for \eqref{RUMP_pj}. Such a result will be obtained if $U(\w^T,\cdot)$ is usc, or if the candidate for the optimal wealth does not belong to the set of discontinuity of $U(\w^T,\cdot)$, see \eqref{eg_fdmt_ok_pj}. What we obtain in general is a strategy $\phi^{*,x}$ for which a bound on the optimality error, i.e., on $u(x) -\inf_{P\in \mathcal{Q}^T}\mathbb{E}_P U\big(\cdot,V_T^{x,\phi^{*,x}}(\cdot)\big)$
can be derived. This bound will depend on the jump size of $U(\w^T,\cdot)$, see \eqref{bound_for_opt_pj}. However, the main technical obstacle for solving $u(x)$ is not the absence of continuity (which could be solved by taking the closure of $U_t$ as usual), but rather the absence of concavity. The concavity permits to prove that $u_t(\w^t,\cdot,h)$ is finite, concave, and thus usc. The absence of concavity leads us to solve a one-step optimisation problem in $h$, which implies the closure of $u_t$, see \eqref{utcl-pb_pj}. 
So, we first need to introduce the closure operator for a function $F : \Omega^t\times \mathbb{R}^p \to \mathbb{R}\cup\{-\infty,+\infty\}$ with $p\geq 1$. Fix $\w^t\in\Omega^t$. Then, $\mathbb
{R}^p \owns x \mapsto F_{\w^t}(x):=F(\w^t,x)$ is an extended real-valued function, and its closure, denoted by $\textup{Cl}(F_{\w^t})$, is the smallest usc function $w$ : $\mathbb{R}^p\to \mathbb{R}\cup\{-\infty,+\infty\}$ such that $F_{\w^t}\leq w$. Now $\textup{Cl}(F)$ : $\Omega^t\times \mathbb{R}^p \to \mathbb{R}\cup\{-\infty,+\infty\}$ is defined by $\textup{Cl}(F)(\w^t,x):=\textup{Cl}(F_{\w^t})(x)$. We will use the closure operator for functions defined on $\Omega^t\times \mathbb{R}$, or on $\Omega^t\times \mathbb{R}\times\mathbb{R}^d$. 
When $F$ is defined on  $\Omega^t\times \mathbb{R}$, and $F(\omega^t,\cdot)$ is nondecreasing,  using \citep[1(7)]{ref5_pj} applied to $-F(\w^t,\cdot)$, we get that 
\begin{eqnarray}
\textup{Cl}(F)(\w^t,x) = \lim_{\epsilon\to 0^+} F(\w^t,x+\epsilon)=F(\w^t,x)+ \Delta_{+}F(\w^t,x),
\label{UeqClU}
\end{eqnarray}
where $ \Delta_{+}F(\w^t,x):=\lim_{\epsilon\to 0^+} F(\w^t,x+\epsilon)-F(\w^t,x)$ is the jump of $F(\w^t,\cdot)$ at $x$. 

As explained above, we solve the following one-step optimization problem associated to $\textup{Cl}(u_t)$ for all $0\leq t \leq T.$ 
\begin{eqnarray}
u_t^{cl}(\w^t,x):= \sup_{h\in\mathbb{R}^d} \textup{Cl}(u_t)(\w^t,x,h). \label{utcl-pb_pj} 
\end{eqnarray}
As $u_t$ has no reason to be usc, we do not have that $U_t = u_t^{cl}$, and we do not solve \eqref{u_t_rob_pj} directly. This is discussed again in Section \ref{op_pj} (see Problem \eqref{cl_pb_one_pj}).\\

We are now in position to state our first main theorem.
\begin{theorem}
Assume the (PD) axiom. Let $U$ be a random utility function (see Definition \ref{U_hp_pj}).  Assume that Assumptions \ref{S_borel_pj}, \ref{analytic_graph_pj}, \ref{H_nonempty_pj}, \ref{AE_pj}, \ref{nncst_pj}, \ref{U0_pj} and \ref{well_def_hp_pj} hold. \\
(i) There exist some $(\phi^{*,x})_{x\in\mathbb{R}}\subset \Phi$, and for all $0\leq t \leq T-1$, a $\mathcal{Q}^t$-full measure set $\widehat{\Omega}^t \in \mathcal{B}(\Omega^t)$ such that for all $x\in\mathbb{R}$ and $\w^t\in\widehat{\Omega}^t$, we have that $\phi^{*,x}_{t+1}(\w^t)\in \textup{Aff}(D_{t+1})(\w^t)$, and that

\begin{small}
\begin{equation}
\begin{aligned}
U_t\big(\w^t,V_t^{x,\phi^{*,x}}(\w^t)\big) \leq \textup{Cl}(U_t)\big(\w^t,V_t^{x,\phi^{*,x}}(\w^t)\big)\leq  u_t^{cl}\big(\w^t,V_t^{x,\phi^{*,x}}(\w^t)\big) &= \sup_{h\in\mathbb{R}^d}\textup{Cl}(u_t)\big(\w^t,V_t^{x,\phi^{*,x}}(\w^t), h\big)\label{one_step_strategy_temp2_pj}\\ 
&= \textup{Cl}(u_t)\big(\w^t,V_t^{x,\phi^{*,x}}(\w^t), \phi_{t+1}^{*,x}(\w^t)\big).
\end{aligned}
\end{equation}
\end{small}
(ii) If $\phi^{*,x}\in \Phi(x,U,\mathcal{Q}^T)$, then
\begin{eqnarray}
u(x)&\leq &U_0(x) \leq \textup{Cl}(U_0)(x) \leq u_0^{cl}(x) =\textup{Cl}(u_0)(x)\leq \inf_{P\in \mathcal{Q}^T}\mathbb{E}_P \textup{Cl}(U)\big(\cdot,V_T^{x,\phi^{*,x}}(\cdot)\big).
\label{eg_fdmt_pj}
\end{eqnarray}
We have the following bounds for the optimality error 
\begin{eqnarray}
0 \leq u(x) - \inf_{P\in \mathcal{Q}^T}\mathbb{E}_P U\big(\cdot,V_T^{x,\phi^{*,x}}(\cdot)\big) \leq  \sup_{P\in \mathcal{Q}^T}\mathbb{E}_P \Delta_+ U\big(\cdot,V_T^{x,\phi^{*,x}}(\cdot)\big),
\label{bound_for_opt_pj}
\end{eqnarray}
(iii) If $\phi^{*,x}\in \Phi(x,U,\mathcal{Q}^T)$ and if  $U(\w^T,\cdot)$ is usc  for all $\w^T$ in a $\mathcal{Q}^T$-full measure set, or $V_T^{x,\phi^{*,x}}(\cdot) \notin \mathcal{D}(\cdot)$ $\mathcal{Q}^T$-q.s., where $\mathcal{D}(\w^T)$ is the (countable) set of discontinuity of $U(\w^T,\cdot)$, then $U_0$ is usc in $x$ and 
\begin{eqnarray}
u(x)=U_0(x)=\textup{Cl}(U_0)(x) =u_0^{cl}(x) =\textup{Cl}(u_0)(x)= \inf_{P\in \mathcal{Q}^T}\mathbb{E}_P U\big(\cdot,V_T^{x,\phi^{*,x}}(\cdot)\big). \label{eg_fdmt_ok_pj}
\end{eqnarray}
\label{optimality_pp_pj}
\label{one_step_strategy_pj}
\end{theorem}
\proof{Proof.}
Part (iii) of Theorem \ref{one_step_strategy_pj} is an easy consequence of part (ii). Assume first that $U(\w^T,\cdot)$ is usc  for all $\w^T$ in a $\mathcal{Q}^T$-full measure set. We have that 
\begin{eqnarray}
U\big(\cdot,V_T^{x,\phi^{*,x}}(\cdot)\big) = \textup{Cl}(U)\big(\cdot,V_T^{x,\phi^{*,x}}(\cdot)\big)\;\; \mathcal{Q}^T\mbox{-q.s}. \label{UeqClUbis}
\end{eqnarray}
So, \eqref{eg_fdmt_pj} shows the desired result as  $\phi^{*,x}\in \Phi(x,U,\mathcal{Q}^T)$, see \eqref{RUMP_pj}. Now, if $V_T^{x,\phi^{*,x}}(\cdot) \notin \mathcal{D}(\cdot)$ $\mathcal{Q}^T$-q.s., \eqref{UeqClUbis} still holds true using 
\eqref{UeqClU}. \\
The rest of the proof of Theorem \ref{optimality_pp_pj} is quite involved and is delayed to Section \ref{proof_th1_pj}. We will use the dynamic programming, i.e., glue together all the one-step optimal strategies constructed in Section \ref{op_pj} for Problem \eqref{utcl-pb_pj}.
\Halmos \\ \endproof
\begin{remark}
\label{rem_star}
One may wonder if we can prove that $\phi^{*,x}$ is an optimal strategy for Problem \eqref{RUMP_pj}. The answer is no without further assumptions. To prove that $\phi^{*,x}$ is optimal for \eqref{RUMP_pj}, we should first prove that $\phi_{t+1}^{*,x}$ is an optimal one-step strategy between $t$ and $t+1$, starting from an initial wealth equal to $x+\sum_{s=1}^t \phi^{*,x}_s \Delta S_s$, i.e., if one have already followed the strategies $(\phi^{*,x}_{1},\cdot\cdot\cdot,\phi^{*,x}_{t})$ until time $t$. This is to say that 
\begin{equation}
\begin{aligned}
U_t\big(\w^t,V_t^{x,\phi^{*,x}}(\w^t)\big)& = \sup_{h\in\mathbb{R}^d} \inf_{p\in\mathcal{Q}_{t+1}(\w^t)} \mathbb{E}_p U_{t+1}\big(\w^t,\cdot, V_t^{x,\phi^{*,x}}(\w^t) + h \Delta S_{t+1}(\w^t,\cdot)\big)\\
&=\inf_{p\in\mathcal{Q}_{t+1}(\w^t)} \mathbb{E}_p U_{t+1}\big(\w^t,\cdot, V_t^{x,\phi^{*,x}}(\w^t) + \phi^{*,x}_{t+1}(\w^t) \Delta S_{t+1}(\w^t,\cdot)\big).\label{eq_fdmt_t_usc_pj}
\end{aligned}
\end{equation}
We show in the Example\rq{}s section that the last equality in  \eqref{eq_fdmt_t_usc_pj} can be false, see Example  \ref{CE_no_cl_pj}. We also prove that we can have that $u(x) > \inf_{P\in\mathcal{Q}^T} \mathbb{E}_{P} U\big(\cdot,V_T^{x,\phi^{*,x}}(\cdot)\big).$\\
Now, if $u_t(\w^t,\cdot,\cdot)$ is  usc (recall \eqref{u_t_rob_pj}), then \eqref{one_step_strategy_temp2_pj} implies that \eqref{eq_fdmt_t_usc_pj} holds true. However, $u_t(\w^t,\cdot,\cdot)$ has no reason to be usc. One might believe that if $U_{t+1}(\w^{t+1},\cdot)$ is usc, then $u_t(\w^t,\cdot,\cdot)$ would also be usc. This is not necessarily true with our one-period assumptions, even if $U_{t+1}(\w^{t+1},\cdot)$ is assumed to be concave and finite (and in particular usc), see Example \ref{CE_no_cl_2} in the Example\rq{}s section. So, even if $U_{t+1}$ was defined in a closure ``way", i.e., as the closure of $x \mapsto \sup_{h\in\mathbb{R}^d} u_{t+1}(\w^{t+1},x,h)$ for all $\w^{t+1}\in\Omega^{t+1}$ as in \citep{ref3_pj} and \citep{refnotre_pj}, it would not guarantee 
$u_t(\w^t,\cdot,\cdot) $ to be usc for $\w^t\in\Omega^t$ in some full-measure set. In any case, the closures in \eqref{one_step_strategy_temp2_pj} should still be needed.\\
\end{remark}

\subsection{Application}
\label{sec ilus}
The strategy $\phi^{*,x}$ obtained in Theorem \ref{optimality_pp_pj} (i) belongs to $\Phi$, but may fail to be admissible. This is not specific to our quasi-sure setting. Already in \citep[Theorem 2.7]{ref2_pj} or \citep[Theorem 1]{ref7_pj}, one had to assume that $\phi^{*,x}$ is  admissible  in order to be optimal (see \citep[Remark 15]{ref7_pj}). 
 If $\phi^{*,x}$ is admissible and if $V_T^{x,\phi^{*,x}}(\w^T)$ does not belong to the set of discontinuity of $U(\w^T,\cdot)$, then $\phi^{*,x}$ is a solution of \eqref{RUMP_pj}. The condition that $\phi^{*,x}\in \Phi(x,U,\mathcal{Q}^T)$ is obviously not easy to verify. However, we prove below that Theorem \ref{one_step_strategy_pj} applies to a broad class of market models and random utility functions. For that, we first define some sets of random variables that are integrable enough.
 \begin{definition}
Fix $0\leq t\leq T$ and $P\in\mathfrak{P}(\Omega^t)$.
\begin{eqnarray*}
\mathcal{M}^t(P) &:=& \big\{X : \Omega^t \to \mathbb{R}\cup\{-\infty, +\infty\}\; \mbox{projective such that } \mathbb{E}_P |X|^r <+\infty,\; \forall r\geq 1\big\}\\
\mathcal{M}^t &:=& \bigcap_{P\in\mathcal{Q}^t}\mathcal{M}^t(P).
\end{eqnarray*}
\label{def_W_pj}
\label{M_tP_pj}
\end{definition}
Note that $\mathcal{M}^0=\mathcal{M}^{0}(P)=\mathbb{R}$. It is also clear that $\mathcal{M}^t\subset \mathcal{M}^t(P)$ for all $P\in\mathcal{Q}^t$. The set $\mathcal{M}^T$ is a robust extension in the projective setting of the set $\mathcal{M}$ defined in \citep{ref2_pj} (see also \citep{ref7_pj}). 
\begin{definition}
A random utility, as in Definition \ref{U_hp_pj}, is of type (A), if $U(\w^T,\cdot)$ is usc for all $\w^t\in\Omega^T$, $U^+(\cdot,1)\in \mathcal{M}^T$, Assumption \ref{AE_pj} holds for some $C\in\mathcal{M}^T$,  Assumption \ref{nncst_pj} holds for some $\underline{X}\in \mathcal{M}^T$ and if there exist $p\geq 1$ and a non-negative and projective function $D_1\in \mathcal{M}^T$, such that for all $\w^T\in\Omega^T$ and $x\in\mathbb{R}$
\begin{eqnarray}
U(\w^T,x)\geq -D_1(\w^T)(1+|x|^p).
\label{ineq_b_inf_det_A_pj}
\end{eqnarray}
\label{typeA_pj}
\end{definition}
Random utility function of type (A) was already defined in \citep[Definition 5]{refnotre_pj}, but for concave utility function, and assuming that $U(\cdot,x)$ is Borel and $D_1\in\mathcal{W}^T,$ where $\mathcal{W}^T$ is the set of $X$ such that $\sup_{P\in\mathcal{Q}^T} \mathbb{E}_P |X|^r <+\infty$ for all $r\geq 1$ (see \citep[Definition 4]{refnotre_pj}). It is better to work with $\mathcal{M}^T$, as the integrability conditions of $\mathcal{M}^T$ can be checked on a prior-by-prior basis. \\
An example of utility functions of type (A) is utility functions with random benchmark, see \citep[Definition 6 and Proposition 13]{refnotre_pj}, where the utility function is no longer concave, and the benchmark is only assumed to be a projective function in $\mathcal{M}^T$.\\

We propose the following theorem that shows the existence of an optimal strategy for random utility of type $(A)$, under some integrability conditions on the market, especially on the process $\alpha^P$ introduced below and related to the ``quantitative" no-arbitrage condition, see \citep[Definition 3.19]{ref4_pj}. This lemma generalizes \citep[Proposition 3.35]{ref4_pj} to our projective setup
\begin{lemma}
\label{simi_qt_na_pj}
Assume the (PD) axiom. Assume that Assumptions \ref{S_borel_pj}, \ref{analytic_graph_pj} and \ref{H_nonempty_pj} hold. Fix $P:=q_1^P\otimes \cdots \otimes q_T^P\in \mathcal{H}^T$. For all $0\leq t \leq T-1$, there exists some projective function $\alpha_t^P(\cdot)\in (0,1]$ such that $\Omega^{t,P}_{qNA}$ is a $\mathcal{Q}^t$-full-measure set, where
\begin{eqnarray}
\Omega^{t,P}_{qNA}:=\big\{\w^t\in\Omega^t,\forall h\in\mbox{Aff}(D^{t+1})(\w^t), h\neq 0, q_{t+1}^P\big[h\Delta S_{t+1}(\w^t,\cdot)<-\alpha_t^P(\w^t)|h| | \w^t \big]\geq \alpha_t^P(\w^t)\big\}.
\label{set_qt_na_H_pj}
\end{eqnarray}
\end{lemma}
\proof{Proof.}
See Appendix \ref{miss_proof0_pj}. 
\Halmos \endproof

\begin{theorem}
Assume the (PD) axiom. Let $U$ be a random type $(A)$ utility. Assume that  Assumptions \ref{S_borel_pj}, \ref{analytic_graph_pj} and \ref{H_nonempty_pj} hold. Moreover, suppose that for all $P\in \mathcal{H}^T$, and for all $0\leq t \leq T-1$, $1/\alpha_{t}^P\in \mathcal{M}^{t}$ and $|\Delta S_{t+1}|\in\mathcal{M}^{t+1}$.
Then, for all $x\in \mathbb{R}$, there exists $\phi^{*,x}\in \Phi(x,U,\mathcal{Q}^T)$ such that 
\begin{eqnarray*}
u(x)=\sup_{\phi\in \Phi(x,U,\mathcal{Q}^T)}\inf_{P\in \mathcal{Q}^T}\mathbb{E}_P U\big(\cdot,V_T^{x,\phi}(\cdot)\big)
&=& \inf_{P\in \mathcal{Q}^T}\mathbb{E}_P U\big(\cdot,V_T^{x,\phi^{*,x}}(\cdot)\big).
\end{eqnarray*}
\label{optimality_M_typeA_pj} 
\end{theorem}
The integrability conditions of Theorem \ref{optimality_M_typeA_pj} are quite classical in the uni-prior literature on unbounded utility functions, see for example \citep[Proposition 7.1]{ref2_pj}, \citep[Theorem 4.16]{ref10_pj} and \citep[Proposition 7]{ref7_pj}, and also close to the ones of the concave, multiple-priors literature, see for example \citep[Theorem 3.6]{ref3_pj}, \citep[Corollary 3.16]{ref4_pj}, \citep[Theorem 2]{refnotre_pj} and \citep[Theorem 3.11]{ref8_pj}. Theorem \ref{optimality_M_typeA_pj} extends \citep[Theorem 2]{refnotre_pj} to utility functions that are no longer necessarily concave and to integrability conditions stated through $\mathcal{M}^t$ and not $\mathcal{W}^t$.

\section{One period case}
\label{op_pj}

As mentioned after Theorem \ref{one_step_strategy_pj}, the candidate for optimal solutions of \eqref{RUMP_pj} will be constructed by gluing together one-step optimal strategies for \eqref{utcl-pb_pj}. So, we start with a one-period model. 
Let $\overline{\Omega}$ be a Polish space, $\mathfrak{P}(\overline{\Omega})$ be the set of all probability measures on $(\overline{\Omega},\mathcal{B}(\overline{\Omega}))$ and, $\mathcal{Q}$ be a nonempty convex 
subset of $\mathfrak{P}(\overline{\Omega})$. We will not assume the (PD) axiom in the one-period model, because we don't do measurable selection. 
We will also not use projective sets or functions.  Thus, here the $\mathbb{R}^d$-vector $Y(\cdot):=(Y_1(\cdot),...,Y_d(\cdot))$, which could represent the change of values of the price process during the period, is assumed to be universally measurable, and not projective. 
Like previously, $D\subset \mathbb{R}^d$ is the support of the distribution of $Y(\cdot)$ under $\mathcal{Q}$, and $D_p\subset \mathbb{R}^d$ is the one of $Y(\cdot)$ under $p\in\mathcal{Q}$. We first state the one-period counterpart of Assumption \ref{H_nonempty_pj}.
\begin{assumption}
There exists $p^*\in \mathcal{Q}$ such that $0\in \textup{ri}(\textup{conv}(D_{p^*}))$ and $\textup{Aff}(D)= \textup{Aff}(D_{p^*})$.
\label{P^*_pj}
\end{assumption}
In the one-period case, Assumption \ref{P^*_pj} is equivalent to the $\mbox{NA}(\mathcal{Q})$ condition, see \citep[Lemma 2.7]{refBaZh_pj} and also \citep[Theorem 3.29]{ref4_pj} for $T=1$.
In the rest of this Section, we fix some $p^*$ as in Assumption \ref{P^*_pj}. 
As $NA(p^*)$ holds true, the ``quantitative" no-arbitrage condition is satisfied, and \citep[Proposition 3.3]{ref2_pj} shows that there exists some $0<\alpha^*\leq 1,$ such that for all $h\in \textup{Aff}(D_{p^*})=\textup{Aff}(D)$ (see Assumption \ref{P^*_pj}), $h\neq 0$, and 
\begin{eqnarray}
p^*[h Y<-\alpha^* |h|]\geq \alpha^*.
\label{qNA_pj}
\end{eqnarray} 

\begin{assumption}
A random utility $V$ : $\overline{\Omega}\times \mathbb{R} \to \mathbb{R}\cup\{-\infty,+\infty\}$ is a $\mathcal{B}_c(\overline{\Omega}\times \mathbb{R})$-measurable function, such that $V(\w,\cdot)$ is nondecreasing for all $\w\in\overline{\Omega}$.
\label{V_pj}
\end{assumption}
\begin{remark}
The value functions $U_{t+1}^P(\w^t,\cdot,\cdot)$ and $U_{t+1}(\w^t,\cdot,\cdot)$ (see \eqref{u_t_p_pj} and \eqref{u_t_rob_pj}) are the multiperiod counterparts of $V(\cdot,\cdot).$ They are projective functions (see Proposition \ref{U_t_well_pj} (i)), and thus under the (PD) axiom, $\mathcal{B}_c(\Omega^t\times \mathbb{R})$-measurable (see  Theorem \ref{proj_is_univ_pj} (i)).\\
One may wonder if the one-period analysis can be done with an arbitrary measurable space $(\overline{\Omega},\mathcal{G})$, and under the assumption that $V(\cdot,x)$ is $\mathcal{G}$-measurable and $V(\w,\cdot)$ is non-decreasing as in \citep[Section 3]{ref3_pj} and \citep[Section 3]{refnotre_pj}. The answer is no. As we do not assume here that $V(\w,\cdot)$ is usc, $V$ has no reason to be jointly measurable, and \eqref{v_pj} and \eqref{Phi_pj} may not exist.
  \end{remark}

The pendant of \eqref{RUMP_pj} is : 
\begin{eqnarray}
v(x) &:=& \sup_{h\in\mathbb{R}^d} \inf_{p\in \mathcal{Q}} \mathbb{E}_p V\big(\cdot, x+h Y(\cdot)\big).
\label{v_pj}
\end{eqnarray}
Let $p\in\mathcal{Q}$. Then, $\Psi_p$, $\Psi : \mathbb{R} \times \mathbb{R}^d \to \mathbb{R}\cup \{-\infty,+\infty\}$ 
are defined, for all $(x,h)\in \mathbb{R}\times\mathbb{R}^d$ by
\begin{eqnarray}
\Psi_p(x,h):=\mathbb{E}_p V\big(\cdot, x+h Y(\cdot)\big) & \quad& \Psi(x,h) := \inf_{p\in \mathcal{Q}} \Psi_p(x,h)=\inf_{p\in\mathcal{Q}} \mathbb{E}_p V\big(\cdot, x+h Y(\cdot)\big). 
\label{Phi_pj} \label{robust_phi_pj}
\end{eqnarray} 
Then, $v(x)= \sup_{h\in\mathbb{R}^d} \Psi(x,h)$. The function $\Psi$ (resp. $\Psi_p$)  is the pendant of $u_t$ (resp. $u^P_t$), while $v$ is the pendant of $U_t$, see \eqref{u_t_rob_pj}.  
The measurability of $V$ implies that the expectations in \eqref{v_pj} or \eqref{Phi_pj} are well-defined, in the generalized sense, using the convention $+\infty-\infty=-\infty+\infty=-\infty$. Indeed, let $(x,h)\in\mathbb{R}\times \mathbb{R}^d$. Noting that $\w \mapsto (\w, x+hY(\w))$ is $\mathcal{B}_c(\overline{\Omega})$-measurable (recall that $Y$ is $\mathcal{B}_c(\overline{\Omega})$-measurable), and using \citep[Proposition 7.44, p172]{ref1_pj}, we get that $V(\cdot, x+h Y(\cdot))$ is also $\mathcal{B}_c(\overline{\Omega})$-measurable. Thus, $\Psi_p$ and $\Psi$ are well-defined although potentially infinite. \\
Problem \eqref{v_pj} has already been solved in \citep{ref11_pj} and \citep{refnotre_pj}, in the case of concave and usc utility functions. The fact that $\Psi(x,\cdot)$ was usc played a significant role in proving an optimiser\rq{}s existence. In \citep[Lemma 3.5.12]{ref11_pj}, a strong integrability assumption on $V^+$ and the continuity and the concavity of $V(\w,\cdot)$ allow us to show that $\Psi(x,\cdot)$ is usc using Fatou's lemma. In \citep[Proposition 1]{refnotre_pj}, it is a strong integrability assumption on $V^-$ (and a weaker integrability assumption on $V^+$), as well as the concavity of $V(\w,\cdot)$, which imply that $\Psi(x,\cdot)$ is finite and concave, and thus usc.
Here, even if $V(\w,\cdot)$ is assumed to be usc and concave, there is no reason for $\Psi(x,\cdot)$ to be usc and to have a maximizer. 
We propose in the Example\rq{}s section, an illustration of this phenomenon, see Example \ref{CE_no_cl_2}.  Without the continuity and concavity assumption, we do not aim to solve \eqref{v_pj}, but instead, we focus on the related problem
\begin{eqnarray}
v^{cl}(x) := \sup_{h\in\mathbb{R}^d} \textup{Cl}(\Psi)(x,h).
\label{cl_pb_one_pj}
\end{eqnarray}
Except from Assumption \ref{V_pj}, our other assumptions are very similar to the ones of \citep[Section 3]{refnotre_pj}. We will comment as we go along on the differences.

\begin{assumption}
For all $\theta\in \{-1,1\}^d$, $\mathbb{E}_{p^*} V^+(\cdot,1+\theta Y(\cdot))<+\infty$. 
\label{integ_V+_pj}
\end{assumption}
Assumption \ref{integ_V+_pj} is similar to \citep[Assumption 10]{refnotre_pj}, \citep[Assumption 3.5.6 (3.19)]{ref11_pj} and \citep[Assumption 3.16]{ref3_pj} and provides some upper bound for the value function. As \citep[Assumption 10]{refnotre_pj}, Assumption \ref{integ_V+_pj} is stated only for $p^*$ in contrast to \citep[Assumption 3.5.6 (3.19)]{ref11_pj}, which is postulated for all $p\in\mathcal{Q}$. However, Assumption \ref{integ_V+_pj} is stronger than \citep[Assumption 10]{refnotre_pj}, which only requires that $\mathbb{E}_{p^*} V^+(\cdot,1)<+\infty$. This counterbalances \citep[Assumption 9]{refnotre_pj}, which is replaced by Assumption \ref{integ_V-_bis_pj}, as here we don't need a control from below on $v$ to transfer concavity results to the value function. We now make the assumption related to RAE in discrete time.

\begin{assumption}
There exist some constants $0<\underline{\gamma} < \overline{\gamma}$, and a $\mathcal{B}_c(\overline{\Omega})$-measurable  random variable $C : \overline{\Omega}\to \mathbb{R}^+ \cup \{+\infty\}$ such that $c^*:=\mathbb{E}_{p^*}(C)<+\infty$, and such that for all $\w\in \overline{\Omega}$ satisfying $C(\w)<+\infty$, for all $\lambda \geq 1$, $x\in \mathbb{R}$,
\begin{eqnarray}
V(\w,\lambda x)\leq  \lambda^{\underline{\gamma}}\big(V(\w,x)+C(\w)\big) \mbox{ and } 
V(\w,\lambda x)\leq  \lambda^{\overline{\gamma}}\big(V(\w,x)+C(\w)\big). \label{V_pos_pj}\label{V_minus_pj}
\end{eqnarray}
\label{AE_one_pj}
From now, we choose some $0<\eta<1$ such that $\underline{\gamma}<\eta \overline{\gamma}$.
\end{assumption}
The right-hand side of  \eqref{V_pos_pj} controls $V^-$, while the left-hand side controls $V^+$. Indeed, we see easily that for all $\w\in \overline{\Omega}$ such that $C(\w)<+\infty$, $x\in\mathbb{R}$ and $\lambda \geq 1$,
\begin{eqnarray}
V^-(\w,\lambda x) \geq \lambda^{\overline{\gamma}} (V^-(\w,x) -C(\w)) \mbox{ and } 
V^+(\w,\lambda x) \leq \lambda^{\underline{\gamma}} (V^+(\w,x)+C(\w)). \label{ineq_pos_part_pj}
\end{eqnarray}

The coefficient $\eta$ will play an important role in establishing bounds for the value function in \eqref{cl_pb_one_pj}, and the optimal strategy: it is crucial that the control $\overline{\gamma}$ on $V^-$ is strictly larger than the control $\underline{\gamma}$ on $V^+$. As $U$ is not concave, we don\rq{}t have the condition that $\underline{\gamma}\leq 1\leq  \overline{\gamma}$, which is required in \citep[Assumption 11]{refnotre_pj}. 
\begin{assumption}
There exists some $n_0^*\in \mathbb{N}\setminus \{0\}$ such that, 
\begin{eqnarray*}
p^*\Big[V(\cdot,-n_0^*)\leq -\Big(1+2\frac{c^*}{\alpha^*}\Big)\Big]\geq 1-\frac{\alpha^*}{2},
\end{eqnarray*}
where $\alpha^*$ is defined in (\ref{qNA_pj}) and $c^*$ in Assumption \ref{AE_one_pj}.
\label{pb_inequality_pj}
\end{assumption}
Assumption \ref{pb_inequality_pj} (see also \citep[Assumption 8]{ref7_pj} and  \citep[Assumption 12]{refnotre_pj}) is the one-period counterpart of Assumption \ref{nncst_pj},  and ensures that the functions $v$ and $v^{cl}$ (see \eqref{v_pj} and \eqref{cl_pb_one_pj}) can take arbitrary negative values.
Note that if $\lim_{x\to -\infty}V(\cdot,x)=-\infty$ $p^*-$almost surely, then Assumption \ref{pb_inequality_pj} is verified. If Assumption \ref{pb_inequality_pj} fails, then there may be no solution to the one-step utility maximization problem, even in the concave uni-prior case, see \citep[Remark 5]{refnotre_pj}. \\

\begin{proposition}

Assume that Assumptions \ref{P^*_pj}, \ref{V_pj},  \ref{integ_V+_pj} and \ref{AE_one_pj} hold.  Then, $\Psi_{p^*}<+\infty$ and for all $(x,h)\in \mathbb{R}\times\mathbb{R}^d$,
\begin{eqnarray}
\Psi_{p^*}(x,h)\leq \mathbb{E}_{p^*} V^+\big(\cdot, x+h Y(\cdot)\big) \leq (|h|\vee x^+ \vee 1)^{\underline{\gamma}} (l^*+c^*),
\label{inequ_imp_positive_pj}
\end{eqnarray}
where $c^*= \mathbb{E}_{p^*} C<+\infty$, $l^*:=\sum_{\theta\in\{-1,1\}^d} \mathbb{E}_{p^*}V^+(\cdot,1 + \theta Y(\cdot))<+\infty$, and $a\vee b= \max(a,b)$.\\ Moreover, the function $\Psi_{p^*}^{cl} : \mathbb{R} \times \mathbb{R}^d \to \mathbb{R}\cup\{-\infty, +\infty\}$ defined by 
\begin{eqnarray}
\Psi_{p^*}^{cl}(x,h) := \mathbb{E}_{p^*} \textup{Cl}(V)\big(\cdot, x+h Y(\cdot)\big) \label{psi_p*}
\end{eqnarray}
is usc.
\label{well_def_one_pj}
\end{proposition}

\proof{Proof.}
For $i\in\{1,\cdot\cdot\cdot,d\}$, let $\widehat{\theta}_i(\cdot):= \mbox{sgn}(Y_i(\cdot))$, where for all $y\in\mathbb{R}$, $\mbox{sgn}(y)=1$ if $y\geq 0$, and $\mbox{sgn}(y)=-1$ otherwise. Then, $\widehat{\theta}$ is a $\{-1,1\}^d$-valued process, and
\begin{eqnarray}
|Y(\cdot)|\leq |Y(\cdot)|_1=\sum_{i=1}^d \textup{sgn}\big(Y_i(\cdot)\big)Y_i(\cdot)= \widehat{\theta}(\cdot)Y(\cdot).
\label{Y_ineq_pj}
\end{eqnarray}
Let $\w\in \overline{\Omega}$ such that $C(\w)<+\infty$, and $(x,h)\in \mathbb{R}\times \mathbb{R}^d$. As $V^+(\w,\cdot)$ is nondecreasing, using Cauchy-Schwarz inequality, (\ref{ineq_pos_part_pj}) and (\ref{Y_ineq_pj}), we obtain that
\begin{eqnarray}
V^+(\w,x+h Y(\w))&\leq&  V^+\big(\w,(|h|\vee x^+ \vee 1) (1 + |Y(\w)|)\big)\nonumber\\
& \leq & (|h|\vee x^+ \vee 1)^{\underline{\gamma}} \big(V^+(\w,1 + |Y(\w)|)+C(\w)\big)\nonumber\\
& \leq & (|h|\vee x^+ \vee 1)^{\underline{\gamma}} \Big(\sum_{\theta\in\{-1,1\}^d}V^+\big(\w,1 + \theta Y(\w)\big)+C(\w)\Big).\label{temp_well_def_psi_pj}
\end{eqnarray}
As $c^*<+\infty$ by Assumption \ref{AE_one_pj}, $p^*[C<+\infty]=1$, and taking the expectation under $p^*$ in \eqref{temp_well_def_psi_pj} shows \eqref{inequ_imp_positive_pj}. As $l^*<+\infty$ by Assumption \ref{integ_V+_pj}, we conclude that $\Psi_{p^*}<+\infty$.\\ 
Let $(x,h)\in\mathbb{R}\times \mathbb{R}^d$ and $(x_n,h_n)\in\mathbb{R}\times \mathbb{R}^d$, such that $\lim_{n\to +\infty} (x_n,h_n) = (x,h)$. Let $M := \sup_{n\geq 0} |x_n| + \sup_{n\geq 0} |h_n|$. Then, $M<+\infty$.
Using \eqref{UeqClU} ($V(\w,\cdot)$ is nondecreasing), we get that for all $\w \in \overline{\Omega}$, and $x\in\mathbb{R}$ 
\begin{eqnarray*}
\textup{Cl}(V)(\w,x) = \inf_{n\geq 1}V\Big(\w,x+\frac{1}{n}\Big) \leq V(\w,x+1).
\end{eqnarray*}
Thus, as $V$ is $\mathcal{B}_c(\overline{\Omega}\times \mathbb{R})$-measurable,  $\textup{Cl}(V)$ is also $\mathcal{B}_c(\overline{\Omega}\times \mathbb{R})$-measurable. Using now \eqref{temp_well_def_psi_pj}, we find that for all $\w\in\overline{\Omega}$ such that $C(\w)<+\infty$,
\begin{eqnarray*}
(\textup{Cl}(V))^+\big(\w, x_n + h_n Y(\w)\big)\leq  (M + 1)^{\underline{\gamma}} \Big(\sum_{\theta\in\{-1,1\}^d}V^+\big(\w,1 + \theta Y(\w)\big)+C(\w)\Big).\nonumber
\end{eqnarray*}
As $c^*<+\infty$ and $l^*<+\infty$, we deduce that $(\textup{Cl}(V))^+(\cdot,x_n+h_n Y(\cdot))$ is dominated by a $p^*$-integrable random variable. So, (lim sup) Fatou's Lemma, and the fact that $\textup{Cl}(V)(\w,\cdot)$ is usc for all $\w\in\overline{\Omega}$, show that $$\underset{n\to +\infty}{\lim \sup} \; \Psi_{p^*}^{cl}(x_n,h_n) \leq \mathbb{E}_{p^*} \textup{Cl}(V)\big(\cdot, x+h Y(\cdot)\big)= \Psi_{p^*}^{cl}(x,h).$$ Thus, $\Psi_{p^*}^{cl}$ is usc.
\Halmos \\ \endproof

\noindent We give one additional assumption under which a polynomial control on $\textup{Cl}(\Psi)$, as well as a bound on the optimal strategies for $v^{cl}(x)$ in \eqref{cl_pb_one_pj}, can be derived.
\begin{assumption}
For all $k\in \mathbb{N}\setminus \{0\}$, $\inf_{p\in\mathcal{Q}} \mathbb{E}_p V(\cdot,-k) >-\infty$.
\label{integ_V-_bis_pj}
\end{assumption}
Assumption \ref{integ_V-_bis_pj} is trivially weaker than \citep[Assumption 9]{refnotre_pj}, which requires that 
 $\sup_{p\in\mathcal{Q}} \mathbb{E}_p V^-(\cdot,x+hY(\cdot))<+\infty$, for all  $(x,h) \in \mathbb{Q} \times \mathbb{Q}^d$. 
Note that Assumption \ref{integ_V-_bis_pj} is not required to get the existence of an optimal strategy in \eqref{cl_pb_one_pj}.

\begin{proposition}
Suppose that Assumptions \ref{P^*_pj}, \ref{V_pj}, \ref{integ_V+_pj}, \ref{AE_one_pj} and \ref{pb_inequality_pj} hold. Then, for all $x\in\mathbb{R}$, 
\begin{eqnarray}
v(x)&\leq & v^{cl}(x) = \sup_{h\in\mathbb{R}^d} \textup{Cl}(\Psi)(x,h)=\sup_{h\in  \textup{Aff}(D)}\textup{Cl}(\Psi)(x,h).\label{temp_aff_phi_pj}
\end{eqnarray}
For all $x\in\mathbb{R}$, we define $K_0$ and $K_1$ as follows
\begin{eqnarray*}
K_0(x)&:=& \max\Bigg(1,x^+,\frac{x^++n_0^*}{\alpha^*},\Big(\frac{x^++n_0^*}{\alpha^*}\Big)^{\frac{1}{1-\eta}}\Bigg)\\
K_1(x)&:=&\max\Bigg(K_0(x),\Big(\frac{6 l^*}{\alpha^*}\Big)^{\frac{1}{\eta \overline{\gamma}-\underline{\gamma}}},\Big(\frac{6c^*}{\alpha^*}\Big)^{\frac{1}{\eta \overline{\gamma}-\underline{\gamma}}},\Big(\frac{6}{\alpha^*} \Psi^-(x,0)\Big)^{\frac{1}{\eta\overline{\gamma}}}\Bigg),
\end{eqnarray*}
where $\alpha^*$ is defined in \eqref{qNA_pj}, $c^*$, $\eta$, $\overline{\gamma}$ and $\underline{\gamma}$ in Assumption \ref{AE_one_pj}, $l^*$ in Proposition \ref{well_def_one_pj} and $n_0^*$ in Assumption \ref{pb_inequality_pj}. 
Then, $K_0<+\infty$, and for all $h\in \textup{Aff}(D)$ and $x\in\mathbb{R}$, we get that
\begin{eqnarray}
|h|> K_0(x) &\implies& \textup{Cl}(\Psi)(x,h)\leq |h|^{\underline{\gamma}} (l^*+c^*)-|h|^{\eta\overline{\gamma}}\frac{\alpha^*}{2}.\label{coercive_ineq_pj}
\end{eqnarray}
Assume now that Assumption \ref{integ_V-_bis_pj} also holds. 
Then, $K_1<+\infty$, and  for all $h\in\textup{Aff}(D)$ and $x\in\mathbb{R}$,
\begin{eqnarray}
|h|>K_1(x) &\implies &\textup{Cl}(\Psi)(x,h)< \textup{Cl}(\Psi)(x,0).\label{strict_sub_opt_pj}
\end{eqnarray}
In this case, we also obtain that for all $x\in\mathbb{R}$, 
\begin{eqnarray}
v(x)&\leq &\sup_{h\in\mathbb{R}^d}  \textup{Cl}(\Psi)(x,h)=\sup_{h\in  \textup{Aff}(D)}\textup{Cl}(\Psi)(x,h) = \sup_{\substack{|h|\leq K_1(x) \\ h\in \textup{Aff}(D)}}\textup{Cl}(\Psi)(x,h).\label{eq_value_valboun1_pj}
\end{eqnarray}
\label{sub_optimal_pj}
\label{inequ_borne_pj}
\end{proposition}
\proof{Proof.}
We first show \eqref{temp_aff_phi_pj}.
As $v(x) = \sup_{h\in\mathbb{R}^d} \Psi(x,h)$ and $\Psi \leq \textup{Cl}(\Psi)$, the first inequality is immediate. Let $(x,h)\in\mathbb{R}\times \mathbb{R}^d$. Let $h^\perp$ be the orthogonal projection of $h$ on $\textup{Aff}(D)$. Using \citep[Remark 3.10]{ref3_pj}, we get that $hY=h^{\perp} Y$ $\mathcal{Q}-\mbox{q.s.}$ We remark now that 
\begin{eqnarray*}
\textup{Cl}(\Psi)(x,h) = \textup{Cl}(\Psi)(x,h^{\perp}).
\end{eqnarray*}
Indeed, using \citep[1(7)]{ref5_pj} (with $-\Psi$), we get that
\begin{eqnarray}
\textup{Cl}(\Psi)(x,h) = \inf_{\substack{\epsilon_1>0 \\ \epsilon_2>0}}\sup_{\substack{|x'-x|<\epsilon_1 \\ |h'-h|<\epsilon_2}}\Psi(x',h') \label{eq_cl1} &=& \inf_{\substack{\epsilon_1>0 \\ \epsilon_2>0}}\sup_{\substack{|x'-x|<\epsilon_1 \\ |h'-h|<\epsilon_2}}\Psi\big(x',h'+ (h^{\perp}-h)\big) \\&=& 
\inf_{\substack{\epsilon_1>0 \\ \epsilon_2>0}}\sup_{\substack{|x'-x|<\epsilon_1 \\ |h'-h^{\perp}|<\epsilon_2}}\Psi(x',h') = \textup{Cl}(\Psi)(x,h^{\perp}).\nonumber
\end{eqnarray}
Thus, the second equality in \eqref{temp_aff_phi_pj} follows. The proof that $K_0<+\infty$, and that for all $x\in\mathbb{R}$ and $h\in\textup{Aff}(D)$,
\begin{eqnarray}
|h|\geq K_0(x) \implies \Psi(x,h)\leq |h|^{\underline{\gamma}} (l^*+c^*)-|h|^{\eta\overline{\gamma}}\frac{\alpha^*}{2}
\label{temp_k0_psi_pj}
\end{eqnarray}
is exactly the same as the proof of \citep[Proposition 2 (29)]{refnotre_pj} and is thus omitted. We now prove \eqref{coercive_ineq_pj}. Let $(x,h)\in \mathbb{R}\times \mathbb{R}^d$. We set 
$$w(x,h):= |h^\perp|^{\underline{\gamma}} (l^*+c^*)-|h^\perp|^{\eta\overline{\gamma}}\frac{\alpha^*}{2}\mbox{ if }|h^\perp|>K_0(x),\mbox{ and }w(x,h):=+\infty \mbox{ otherwise.}$$ 
Then, as $hY=h^{\perp} Y$ $\mathcal{Q}-\mbox{q.s.}$, \eqref{temp_k0_psi_pj} shows that $\Psi(x,h)=\Psi(x,h^{\perp})\leq w(x,h)$. 
Assume for a moment that $w$ is usc. Then, by the closure definition, $\textup{Cl}(\Psi) \leq w$,
and \eqref{coercive_ineq_pj} is proved.\\ 
We show now that $w$ is usc. Let $(x,h)\in\mathbb{R}\times \mathbb{R}^d$ and $(x_n,h_n)\in\mathbb{R}\times \mathbb{R}^d$ such that 
$\lim_{n\to +\infty} (x_n,h_n) = (x,h)$. If $|h^\perp|\leq K_0(x)$, we have that $\lim \sup_{n\to +\infty} \; w(x_n,h_n) \leq +\infty = w(x,h).$\\ 
Assume now that $|h^\perp|>K_0(x)$. As $h \mapsto h^\perp$, and $x\mapsto K_0(x)$ are continuous, $\{(x,h)\in\mathbb{R}\times \mathbb{R}^d,\; |h^\perp|>K_0(x)\}$ is an open set, and there exists $k\geq 0$ such that for all $n\geq k$, $|h_n^\perp|>K_0(x_n)$. Thus, 
$$\underset{n\to +\infty}{\lim \sup} \; w(x_n,h_n) = \lim_{n\to +\infty} \; |h_n^\perp|^{\underline{\gamma}} (l^*+c^*)-|h_n^\perp|^{\eta\overline{\gamma}}\frac{\alpha^*}{2} = w(x,h).$$
So, $w$ is indeed usc.\\
From now on, Assumption \ref{integ_V-_bis_pj} is also postulated. 
Let $x\in\mathbb{R}$. As $V(\w,\cdot)$ is nondecreasing for all $\w\in \overline{\Omega}$, and as Assumption \ref{integ_V-_bis_pj} holds true, $\Psi(x,0)>\Psi(-\lceil |x| \rceil, 0)>-\infty$, where $\lceil x \rceil$ is the smallest natural number greater than $x$. Thus, $\Psi^-(x,0)<+\infty$. 
Recall that $K_0(x)<+\infty$. Moreover, $\alpha^*>0$ (see \eqref{qNA_pj}), $l^*<+\infty$ (see Assumption \ref{integ_V+_pj}),  $\eta\overline{\gamma}-\underline{\gamma}>0$ and $c^*<+\infty$ (see Assumption \ref{AE_one_pj}). Thus, $K_1(x)<+\infty$. Assume now that $h\in\textup{Aff}(D)$ and  $|h|> K_1(x).$ 
As $\underline{\gamma}<\eta \overline{\gamma}$, we get that $|h|> K_0(x),$
$$|h|^{\underline{\gamma}} l^* < |h|^{\eta\overline{\gamma}}\frac{\alpha^*}{6}\;\; \quad \;\;
        |h|^{\underline{\gamma}} c^* < |h|^{\eta\overline{\gamma}}\frac{\alpha^*}{6} \;\; \mbox{  and } \;\;
        \Psi^-(x,0)< |h|^{\eta\overline{\gamma}}\frac{\alpha^*}{6}.$$ 
Then, using \eqref{coercive_ineq_pj}, and the definition of the closure, we obtain that
\begin{eqnarray}
\textup{Cl}(\Psi)(x,h)< -|h|^{\eta\overline{\gamma}}\frac{\alpha^*}{6} < -\Psi^-(x,0)  \leq \Psi(x,0) \leq \textup{Cl}(\Psi)(x,0),
\label{temp_subH1_pj}
\end{eqnarray}
and \eqref{strict_sub_opt_pj} is proved. Assumption \ref{P^*_pj} shows that $0\in \textup{ri}(\textup{conv}(D_{p^*}))\subset\textup{Aff}(D_{p^*})=\textup{Aff}(D)$ and $\textup{Aff}(D)$ is a vector space. Thus, \eqref{temp_aff_phi_pj}, \eqref{temp_subH1_pj}, and the fact that $0\in\textup{Aff}(D)$, show \eqref{eq_value_valboun1_pj}. 
\Halmos \\ \endproof

\noindent We now show that under Assumptions \ref{P^*_pj} to \ref{pb_inequality_pj}, an optimal strategy exists for \eqref{cl_pb_one_pj}.
\begin{proposition}
Assume that Assumptions \ref{P^*_pj}, \ref{V_pj}, \ref{integ_V+_pj}, \ref{AE_one_pj} and \ref{pb_inequality_pj} hold true. Let $x\in\mathbb{R}$. Then, there exists $\hat{h}_x\in\textup{Aff}(D)$ such that 
\begin{eqnarray}
v(x)\leq \textup{Cl}(v)(x)\leq v^{cl}(x) = \sup_{h\in  \textup{Aff}(D)}\textup{Cl}(\Psi)(x,h) = \sup_{h\in\mathbb{R}^d}\textup{Cl}(\Psi)(x,h)= \textup{Cl}(\Psi)(x,\hat{h}_x).
\label{maximizer_pj}
\end{eqnarray}
If Assumption \ref{integ_V-_bis_pj} also holds true, then $|\hat{h}_x| \leq K_1(x)$, where $K_1$ is defined in Proposition \ref{sub_optimal_pj}.
\label{existence_uni_pj}
\end{proposition}
\proof{Proof.}
We start with the easy assertions in \eqref{maximizer_pj}.
The first inequality holds by definition of the closure, and the first and second equalities are proved in \eqref{temp_aff_phi_pj}.\\ 
Assume for a moment that $v^{cl}$ is usc. Recalling again \eqref{temp_aff_phi_pj}, we know that $v\leq v^{cl}$, and by definition of the closure, $\textup{Cl}(v) \leq v^{cl}$. The second inequality in \eqref{maximizer_pj} is proved. We now show that $v^{cl}$ is usc. Remark first that as $V(\w,\cdot)$ is nondecreasing, we have successively that $\Psi(\cdot,h)$ and $\textup{Cl}(\Psi)(\cdot,h)$ are nondecreasing for all $h\in\mbox{Aff}(D)$ (see \eqref{eq_cl1}), and thus $v^{cl}$ is nondecreasing. We show first that $v^{cl}$ is right-continuous, implying that $v^{cl}$ is usc.

\textit{$v^{cl}$ is right-continuous.}\\
Let $x\in\mathbb{R}$ and $(x_n)_{n\geq 0}\subset\mathbb{R}$, such that $x<x_n$ for all $n\geq 0$, and $\lim_{n\to +\infty}x_n = x$. We distinguish two cases.\\
\textit{Case 1 : There exists some $p\geq 0$ such that $v^{cl}(x_n)<+\infty$, for all $n\geq p$.}\\
\noindent We can extract a subsequence that we still denote by $(x_n)_{n\geq 0}$, such that $v^{cl}(x_n)<+\infty$ for all $n\geq 0$. 
As $v^{cl}$ is nondecreasing, and by definition of the sup in $v^{cl}(x_n)$, there exists $(h_n)_{n\geq 0} \subset \mbox{Aff}(D)$ such that for all $n\geq 0$,
\begin{eqnarray}
v^{cl}(x)\leq v^{cl}(x_n) \leq \textup{Cl}(\Psi)(x_n,h_n) + \frac{1}{n+1}. \label{temp_usc_w1_pj}
\end{eqnarray}
Assume that $(|h_n|)_{n\geq 0}$ is not bounded. Then, as $\lim_{n\to +\infty}K_0(x_n) = K_0(x)$, we can extract a subsequence, that we also denote by $(x_n,h_n)_{n\geq 0}$, such that $|h_n|$ goes to infinity, and $|h_n|\geq K_0(x) + 1 > K_0(x_n)$. 
Thus, \eqref{coercive_ineq_pj} in Proposition \ref{sub_optimal_pj} implies that for all $n\geq 0$,
$$v^{cl}(x)\leq v^{cl}(x_{n}) \leq \textup{Cl}(\Psi)(x_{n},h_{n}) + \frac{1}{{n}+1} \leq |h_{n}|^{\underline{\gamma}} (l^*+c^*)-|h_{n}|^{\eta\overline{\gamma}}\frac{\alpha^*}{2} + \frac{1}{{n}+1}.$$ 
As $\underline{\gamma}<\eta \overline{\gamma},$ taking the limit, we get that 
$v^{cl}(x)\leq \lim_{n\to +\infty}v^{cl}(x_n)\leq -\infty$,  and $v^{cl}$ is right-continuous in $x$.\\ 
Assume now that $(|h_n|)_{n\geq 0}$ is bounded. Then, one can extract a subsequence, that we still denote by $(x_n,h_n)_{n\geq 0}$, such that $\lim_{n\to +\infty} h_{n} = \overline{h}$ for some $\overline{h}\in\mathbb{R}^d$. Moreover, $\overline{h}\in \mbox{Aff}(D)$. As $\textup{Cl}(\Psi)$ is usc by definition of the closure, taking the limit in \eqref{temp_usc_w1_pj}, we get that 
$$v^{cl}(x)\leq \lim_{n\to +\infty}v^{cl}(x_n)\leq \underset{n\to +\infty}{\lim \sup} \; \textup{Cl}(\Psi)(x_{n},h_{n}) \leq \textup{Cl}(\Psi)(x,\overline{h})\leq v^{cl}(x).$$ 
Thus, $v^{cl}$ is again right-continuous in $x$.\\
\textit{Case 2 : One can extract a subsequence (still denoted by $(x_n)_{n\geq 0}$) such that $v^{cl}(x_n)=+\infty$, for all $n\geq 0$}.\\
\noindent Then, by definition of the sup in $v^{cl}(x_n)$, there exists $(h_n)_{n\geq 0} \subset \mbox{Aff}(D)$ such that for all $n\geq 0$,
\begin{eqnarray}
\textup{Cl}(\Psi)(x_n,h_n) \geq n. \label{temp_usc_w12_pj}
\end{eqnarray}
We show by contradiction that  $(h_n)_{n\geq 0}$ is bounded. Assume that $(h_n)_{n\geq 0}$ is unbounded. As done earlier in the proof, one can extract a subsequence, that we also denote by $(x_n,h_n)_{n\geq 0}$, such that $|h_n|$ goes to infinity, and $|h_n| > K_0(x_n)$. Using \eqref{temp_usc_w12_pj} and  \eqref{coercive_ineq_pj}, we obtain that for all $n\geq 0$,
$$n\leq \textup{Cl}(\Psi)(x_{n},h_{n}) \leq |h_{n}|^{\underline{\gamma}} (l^*+c^*)-|h_{n}|^{\eta\overline{\gamma}}\frac{\alpha^*}{2}.$$ 
As $\underline{\gamma}<\eta \overline{\gamma},$ taking the limit, we get that $+\infty \leq-\infty,$  a contradiction. Thus, $(h_n)_{n\geq 0}$ is bounded, and one can extract a subsequence, that we still denote by $(x_n,h_n)_{n\geq 0}$, such that $\lim_{n\to +\infty} h_{n} = \overline{h}$ for some $\overline{h}\in\textup{Aff}(D)$. 
As $\textup{Cl}(\Psi)$ is usc, taking the limit in \eqref{temp_usc_w12_pj}, we get that 
$$+\infty \leq \lim_{n\to +\infty}\textup{Cl}(\Psi)(x_{n},h_{n}) = \underset{n\to +\infty}{\lim \sup} \; \textup{Cl}(\Psi)(x_{n},h_{n}) \leq \textup{Cl}(\Psi)(x,\overline{h})\leq v^{cl}(x).$$ 
Thus, as $\textup{Cl}(\Psi)(x_{n},h_{n}) \leq v^{cl}(x_n)$, we obtain that $\lim_{n\to +\infty} v^{cl}(x_n) = +\infty = v^{cl}(x)$, and $v^{cl}$ is right-continuous in $x$.

\textit{Existence of a maximizer for $\textup{Cl}(\Psi)(x,\cdot).$}\\
Fix $x\in\mathbb{R}$. Let $w$ be defined by $w(h):=\textup{Cl}(\Psi)(x,h)$, when $h\in\textup{Aff}(D)$ and $w(h):=-\infty$ otherwise. Assume first that $-w$ is not proper in the sense of \citep[p5]{ref5_pj}, i.e., that for all $h\in\mathbb{R}^d$, $w(h)=-\infty$, or that there exists some $h^*\in\mathbb{R}$ such that $w(h^*)=+\infty$. 
In the first case, $\textup{Cl}(\Psi)(x,h) = -\infty$ for all $h\in\mathbb{R}^d$ using the second equality in \eqref{temp_aff_phi_pj}, and any $h\in\mathbb{R}^d$ is a maximizer. In the second case, we have that $w(h^*) = \textup{Cl}(\Psi)(x,h^*) =+\infty$, and $h^*$ is then a maximizer. 
Assume now that $-w$ is proper.  Remark that $w$ is usc as $\textup{Aff}(D)$ is a closed set, and $\textup{Cl}(\Psi)(x,\cdot)$ is usc. Now, using \eqref{coercive_ineq_pj}  and recalling that $\underline{\gamma}< \eta \overline{\gamma}$, we have that $\lim_{|h|\to +\infty} -w(h) = +\infty$. Thus, $-w$ is level bounded (see \citep[Definition 1.8, p11]{ref5_pj} and the short text below this definition), and  \citep[Theorem 1.9, p11]{ref5_pj} shows that there exists $\hat{h}_x\in\mathbb{R}^d$, such that $v^{cl}(x)=\sup_{h\in\mathbb{R}^d}w(h) = w(\hat{h}_x)>-\infty$. As a result, $\hat{h}_x\in\textup{Aff}(D)$, $w(\hat{h}_x)= \textup{Cl}(\Psi)(x,\hat{h}_x)$, and the last equality in \eqref{maximizer_pj} is proved.

If Assumption \ref{integ_V-_bis_pj} also holds, \eqref{strict_sub_opt_pj} immediately shows that $|\hat{h}_x|\leq K_1(x)$. \Halmos  \endproof

\section{Multiple-period case} 
\label{muti_per_pj} 

In this section, we prepare the proofs of Theorems \ref{one_step_strategy_pj} and \ref{optimality_M_typeA_pj}. For that, we will apply the one-period results in two contexts. The first one, called the robust context, assumes that $\mathcal{Q}:=\mathcal{Q}_{t+1}(\w^t)$, $V:= U_{t+1}(\w^t,\cdot,\cdot)$, and is used to prove Theorem \ref{one_step_strategy_pj}. For $P:=q_1^P\otimes \cdots \otimes q_T^P \in \mathcal{H}^T$, the second one, called the $P$-prior context, suppose that $\mathcal{Q}:=\{q_{t+1}^P[\cdot|\w^t]\}$, $V:= U_{t+1}^P(\w^t,\cdot,\cdot)$, and is used to prove Theorem \ref{optimality_M_typeA_pj}. 
Note that in the $P$-prior context, $\textup{Graph}(\mathcal{Q})=\textup{Graph}(q_{t+1}^P)$ may not be a projective set. This is not an issue, as we did not assume that $\textup{Graph}(\mathcal{Q})$ is a projective set in the one-period case.   
We will construct a $\mathcal{Q}^t$-full-measure set $\widetilde{\Omega}^t$ (resp. a $P^t$-full-measure set $\widetilde{\Omega}^{t,P}$), where Assumptions  \ref{P^*_pj} to \ref{pb_inequality_pj} hold true in the robust (resp. $P$-prior) context (see Lemma \ref{lemma_Assumption_true_pj} and Proposition \ref{U_t^-&C_pj}). To do that, we first introduce and prove properties for the dynamic version of $C$ that appears in Assumption \ref{AE_pj} (see Proposition \ref{CJ_t_pj}). Then, Proposition \ref{U_t_well_pj} gives fundamental properties of the value functions $U_t$ and $U_t^P$. 
We will use results on projective sets and functions, which are given in the Appendix (see Propositions \ref{base_hierarchy_pj} and \ref{univ_cvt_pj}, and  Lemma \ref{lemma_mes_u_pj}).  
 Recall that our approach differs from the ones of \citep{refnotre_pj} or \citep{ref3_pj}. It is simplest and more direct, given the results on projective sets and functions. For example, our definitions of $U_t$ and $U_t^P$ do not use closure, or sup on $\mathbb{Q}^d$, and the proof of Proposition \ref{U_t_well_pj} is much easy. 
Moreover, we don't need to introduce functions to control the value functions from below. We will comment on the other differences as we go along.\\ 
For the rest of this part, we fix some random utility function $U$ in the sense of Definition \ref{U_hp_pj}, and some random variable $C$ as in Assumption \ref{AE_pj}.\\ 
For $0\leq t\leq T$, we define by induction the function $C_t : \Omega^t \to \mathbb{R}\cup\{-\infty,+\infty\}$ as follows : 
\begin{eqnarray}
&&\bigg\{
    \begin{array}{ll}
    C_T(\w^T)&:=C(\w^T)\\
	C_t(\w^t)&:=
	\sup_{p\in\mathcal{Q}_{t+1}(\w^t)}\mathbb{E}_p C_{t+1}(\w^t,\cdot).
	\label{C_t_eq_pj}
    \end{array}
\end{eqnarray}
The function $C_t$ will appear in \eqref{elas_gammaf_pj}  in Assumption \ref{AE_pj} stated for $U_t$. It has already been introduced in \citep[(49)]{refnotre_pj}. 
\begin{proposition}
\label{CJ_t_pj}
Assume the (PD) axiom. Assume that Assumptions \ref{analytic_graph_pj} and \ref{AE_pj} hold. Then, for all $0\leq t \leq T$, $C_t$ is non-negative, projective and satisfies that
\begin{eqnarray}
C_t(\cdot)<+\infty \;\; \mathcal{Q}^t -q.s.
\label{finite_CJ_pj}
\end{eqnarray}

\end{proposition}
\proof{Proof.}
We show by backward induction that $C_t$ is non-negative, projective function satisfying  \begin{eqnarray}
\sup_{P\in\mathcal{Q}^t}\mathbb{E}_P C_t<+\infty.
\label{property_temp_pj}
\end{eqnarray}
The claim at time $T$ follows from Assumption \ref{AE_pj} as $C_T:=C$. Let $0\leq t\leq T-1$ and assume that the claim holds at time $t+1$. 
The non-negativity of $C_t$ follows from that of $C_{t+1}$. Considering the Borel, and thus projectively measurable (see \eqref{equation_borel_set}) stochastic kernel $q$ defined by $q[\cdot|\w^t,p]:=p[\cdot]$, and as $C_{t+1}$ is projective, Proposition  \ref{univ_cvt_pj}  (ii) shows that $\Omega^t\times \mathfrak{P}(\Omega_{t+1}) \ni (\w^t,p) \mapsto \mathbb{E}_p C_{t+1}(\w^t,\cdot)$ is projective. 
Then, Assumption \ref{analytic_graph_pj} and Proposition \ref{base_hierarchy_pj} (vii) show that $C_t$ is projective, and that given any $\epsilon>0$, there exists $q^\epsilon : \Omega^t \to \mathfrak{P}(\Omega_{t+1})$, which is projectively measurable, such that for all $\w^t\in\Omega^t$ (recall that $\mathcal{Q}_{t+1}\neq \emptyset$), $q^\epsilon[\cdot|\w^t]\in \mathcal{Q}_{t+1}(\w^t)$ and 
\begin{eqnarray}
\mathbb{E}_{q^{\epsilon}[\cdot|\w^t]} C_{t+1}(\w^t,\cdot)\geq \bigg\{
    \begin{array}{ll}
    \frac{1}{\epsilon} \;\; \mbox{if} \; C_t(\w^t)=+\infty,\\
	 C_t(\w^t)-\epsilon \;\; \mbox{otherwise.}
    \end{array}
\label{temp_J_qeps_pj}
\end{eqnarray} 
For all $\overline{P}\in\mathcal{Q}^t$,  $\overline{P}\otimes q^{\epsilon}\in\mathcal{Q}^{t+1}$. Taking the expectation under $\overline{P}$ in \eqref{temp_J_qeps_pj}, and using Fubini's theorem as $C_{t+1}\geq 0,$ we get 
\begin{eqnarray}
\sup_{P\in\mathcal{Q}^{t+1}}\mathbb{E}_{P} C_{t+1} \geq \mathbb{E}_{\overline{P}\otimes q^{\epsilon}} C_{t+1}\geq \frac{1}{\epsilon} \overline{P}[C_t =+\infty] + \mathbb{E}_{\overline{P}} [( C_t-\epsilon)1_{\{C_t<+\infty\}}] \geq \frac{1}{\epsilon} \overline{P}[C_t=+\infty]-\epsilon, \label{temp_CJt_gen_pj}
\label{ineq_temp_J_t_2_pj}
\end{eqnarray}
as  $C_t\geq 0$.  
If $\overline{P}[C_t =+\infty]>0$, taking the limit when $\epsilon$ goes to $0$ in \eqref{ineq_temp_J_t_2_pj}, we find that $\sup_{P\in\mathcal{Q}^{t+1}}\mathbb{E}_{P} C_{t+1} =+\infty$, which contradicts \eqref{property_temp_pj} at time $t+1$. Thus, $\overline{P}[C_t=+\infty]=0$ and  \eqref{temp_CJt_gen_pj} again implies that
\begin{eqnarray*}
\sup_{P\in\mathcal{Q}^{t+1}}\mathbb{E}_{P} C_{t+1} \geq  \mathbb{E}_{\overline{P}} C_t -\epsilon.
\end{eqnarray*}
So, letting $\epsilon$ go to $0$, taking the supremum over all $\overline{P} \in\mathcal{Q}^t$, and using \eqref{property_temp_pj} for $t+1$, we get that $\sup_{P\in\mathcal{Q}^{t}}\mathbb{E}_{P} C_t <+\infty$. This concludes the induction.\\ 
We now prove \eqref{finite_CJ_pj} by contradiction. If there exists $0\leq t\leq T-1$ and $P\in\mathcal{Q}^t$ such that $P[C_t=+\infty]>0$, then, as $C_t\geq 0$, $\mathbb{E}_P C_t = +\infty$, and also $\sup_{P\in\mathcal{Q}^t}\mathbb{E}_P C_t =+\infty$, contradicting \eqref{property_temp_pj}. 
\Halmos \\ \endproof

\noindent We now provide some fundamental properties of the value functions $U_t$ and $U_t^P$. 

\begin{proposition}
\label{U_t_well_pj}
Assume the (PD) axiom. Assume that Assumptions \ref{S_borel_pj} and  \ref{analytic_graph_pj} hold. Let $P\in\mathcal{Q}^T$, and $0\leq t \leq T$. We have that (i) $U_t$, $U_t^P$,  $\textup{Cl}(U_t)$ and $\textup{Cl}(U_t^P)$ are projective functions, (ii) for all $\w^t\in\Omega^t$, $U_t(\w^t,\cdot)$, $U_t^P(\w^t,\cdot)~:$ $\mathbb{R}\to \mathbb{R}\cup\{-\infty,+\infty\}$ are nondecreasing,  (iii) $U_t\leq U_t^P$, (iv)  
$$\textup{Cl}(U_t)(\w^t,x) \leq U_t(\w^t,x+1) \;\; \mbox{and}\;\;\textup{Cl}(U_t^P)(\w^t,x) \leq U_t^P(\w^t,x+1), \;\forall (\w^t\,x)\in\Omega^t \times \mathbb{R}.$$ 
Assume furthermore that Assumption \ref{AE_pj} holds. Then, for all $\w^t\in\Omega^t$ such that $C_t(\w^t)<+\infty$, $\lambda\geq 1$, $x\in\mathbb{R}$, we get that  
\begin{eqnarray}
U_t(\w^t,\lambda x)\leq \lambda^{\overline{\gamma}}\big(U_t(\w^t,x)+C_t(\w^t)\big) \;\; &\mbox{and}&\;\; U_t^P(\w^t,\lambda x) \leq \lambda^{\overline{\gamma}}\big(U_t^P(\w^t,x)+C_t(\w^t)\big)\label{elas_gammafPt_pj}\label{elas_gammaft_pj}.\\
U_t(\w^t,\lambda x)\leq \lambda^{\underline{\gamma}}\big(U_t(\w^t,x)+C_t(\w^t)\big) \;\; &\mbox{and}&\;\; U_t^P(\w^t,\lambda x) \leq \lambda^{\underline{\gamma}}\big(U_t^P(\w^t,x)+C_t(\w^t)\big)\label{elas_gammafPt2_pj}\label{elas_gammaft2_pj}.
\end{eqnarray}
\end{proposition}
\proof{Proof.}
\textit{Proof of (i) to (iv) under Assumptions \ref{S_borel_pj} and \ref{analytic_graph_pj}.}\\ 
Let $P:=q_1^P \otimes \cdot \cdot \cdot \otimes q_T^P\in \mathcal{Q}^T$. We first show by backward induction that $U_t$ and $U_t^P$ are projective, and that (ii) and (iii) hold.
The initialization step follows from $U_T=U_T^P=U$, and Definition \ref{U_hp_pj}. 
Fix $0\leq t\leq T-1$, and assume that the induction hypothesis holds at time $t+1$. Recall  \eqref{state_val_t_p_pj} and \eqref{state_val_t_rob_pj}. Using that $U_{t+1}$ and $U_{t+1}^P$ are projective, Lemma \ref{lemma_mes_u_pj} shows that $U_t$ and $U_t^P$  are projective. Now, (ii) and (iii) at $t$ follow from (ii) and (iii) at $t+1$. This concludes the induction.\\
Let $0\leq t\leq T$. As $U_t$ is  nondecreasing, using  \eqref{UeqClU}, 
we get that for all $(\w^t,x)\in\Omega^t\times \mathbb{R}$, 
\begin{eqnarray*}
\textup{Cl}(U_t)(\w^t,x) &=& \inf_{n\geq 1} U_t\Big(\w^t,x+\frac{1}{n}\Big) \leq U_t(\w^t,x+1).
\end{eqnarray*}
We have already proved that $U_t$ is projective. So, Proposition \ref{base_hierarchy_pj} (vi) shows that $(\w^t,x) \mapsto U_{t+1}(\w^t,x+1/n)$ is also projective. 
Choosing $D:= \Omega^t\times \mathbb{R} \times \mathbb{N}\setminus \{0\}$ (which is a Borel and, thus a projective set) in Proposition \ref{base_hierarchy_pj} (vii) shows that 
$\textup{Cl}(U_t)$ is projective. The same holds for $\textup{Cl}(U_t^P)$. 

 \textit{Proof of \eqref{elas_gammaft_pj} and  \eqref{elas_gammafPt2_pj} under Assumptions  \ref{S_borel_pj}, \ref{analytic_graph_pj} and \ref{AE_pj}.}\\ 
 We proceed again by backward induction. We only show the left-hand side of \eqref{elas_gammaft_pj} as the proofs of its right-hand side, and of  \eqref{elas_gammaft2_pj} are very similar, and thus omitted. Assumption \ref{AE_pj} ensures that the left-hand side of \eqref{elas_gammaft_pj} holds at time $T$, as $C_T=C$. Fix $0\leq t\leq T-1$. Assume that the left-hand side of \eqref{elas_gammaft_pj} holds at $t+1$. Let $\w^t\in \Omega^t$ such that $C_t(\w^t)<+\infty$, $(x,h)  \in\mathbb{R} \times \mathbb{R}^d$ and $\lambda\in\mathbb{R}$. 
Take any $p\in\mathcal{Q}_{t+1}(\w^t)$. Then, $\{\w_{t+1}\in\Omega_{t+1},\;\; C_{t+1}(\w^t,\w_{t+1})<+\infty\}$ is a $p$-full measure set. Otherwise, we get a contradiction with \eqref{finite_CJ_pj}. So, the left-hand side of (\ref{elas_gammaft_pj}) at time $t+1$ implies
\begin{eqnarray*}
\mathbb{E}_p U_{t+1}\big(\w^t,\cdot,\lambda x+\lambda h\Delta S_{t+1}(\w^t,\cdot)\big) \leq \lambda^{\overline{\gamma}} \Big(\mathbb{E}_p U_{t+1}\big(\w^t,\cdot,x+h\Delta S_{t+1}(\w^t,\cdot)\big)+\mathbb{E}_p C_{t+1}(\w^t,\cdot)\Big).
\end{eqnarray*}
Thus, taking the infimum over all $p\in\mathcal{Q}_{t+1}(\w^t)$, and using \eqref{C_t_eq_pj}
\begin{eqnarray*}
\inf_{p\in\mathcal{Q}_{t+1}(\w^t)} \mathbb{E}_p U_{t+1}\big(\w^t,\cdot,\lambda x+\lambda h\Delta S_{t+1}(\w^t,\cdot)\big) 
&\leq &\lambda^{\overline{\gamma}} \inf_{p\in\mathcal{Q}_{t+1}(\w^t)} \mathbb{E}_p U_{t+1}\big(\w^t,\cdot,x+h\Delta S_{t+1}(\w^t,\cdot)\big) + \lambda^{\overline{\gamma}} C_t(\w^t).
\end{eqnarray*}
Now, the left-hand side of (\ref{elas_gammaft_pj}) at $t$ follows taking the supremum over every $h\in\mathbb{R}^d$.
\Halmos \\ \endproof

\begin{remark}
Proposition \ref{U_t_well_pj} is the pendent of \citep[Proposition 6]{refnotre_pj}. The main difference is that $U_t$ (resp. $U_t^P$) is now only projective and no longer lower-semianalytic (resp. universally measurable). Recall that $U_t$ in \citep[(8)]{refnotre_pj} is defined using countable supremum and closure, under the assumption that $U$ is concave, so that it remains lower-semianalytic through dynamic programming. Moreover, \citep[Proposition 8]{refnotre_pj} shows that it coincides with the definition that we take in this paper on a full-measure set. Such a result seems possible only for a concave random utility function. Otherwise, there is no guarantee that a countable supremum would coincide with an uncountable one. 
\label{rem_link_value_art_pj}\\
\end{remark}

Assume that Assumptions \ref{S_borel_pj},  \ref{analytic_graph_pj} and \ref{H_nonempty_pj}  hold true.
Then, $\mathcal{H}^T\neq \emptyset$, and we fix for the rest of the paper some $P^*\in\mathcal{H}^T$ with the following given disintegration 
\begin{eqnarray}
P^*:=q^{P^*}_1\otimes \cdot\cdot\cdot \otimes q^{P^*}_T.
\label{P^*_exp_pj}
\end{eqnarray}
Lemma \ref{simi_qt_na_pj} shows the existence of the functions $\alpha_t^{P^*}$ : $\Omega^t\to (0,1]$, and also that $\Omega^{t,P^*}_{qNA}$ defined in \eqref{set_qt_na_H_pj} is a $\mathcal{Q}^t$-full-measure set. The stochastic kernels $(q^{P^*}_t)_{1\leq t\leq T}$ will be of special interest for the statements of the one-period Assumptions \ref{P^*_pj}, \ref{integ_V+_pj}, and  \ref{AE_one_pj} in the multiple-period contexts. On the other hand, $(\alpha_t^{P^*})_{0\leq t\leq T-1}$ will serve for the one of Assumption \ref{pb_inequality_pj}.\\ 

We now formally present the two different contexts, already introduced in \citep[Definition 7]{refnotre_pj} (with a different choice of $c_t^P$), where we will apply the one-period results. The robust context will be used to prove Theorem \ref{one_step_strategy_pj}, while the $P$-prior one will be used to prove Theorem \ref{optimality_M_typeA_pj}.
\begin{definition}
\label{def_ctxt_pj}
\label{def_one_prior_pj}
\label{def_ctxt_gen_pj}
Let $0\leq t\leq T-1$ and $\w^t\in\Omega^t.$ \\
We call context $(t+1)$, the one-period market where $\overline{\Omega}:=\Omega_{t+1}$, $Y(\cdot):=\Delta S_{t+1}(\w^t,\cdot)$, $C(\cdot):= C_{t+1}(\w^t,\cdot)$, and $\underline{\gamma}$ and $\overline{\gamma}$ are introduced in Assumption \ref{AE_pj}.\\ 
The robust $(t+1)$ context postulates in addition that $\mathcal{Q}:=\mathcal{Q}_{t+1}(\w^t)$, $p^*:=q^{P^*}_{t+1}[\cdot|\w^t]$, $\alpha^*:=\alpha_t^{P^*}(\w^t)$,  and 
$V(\cdot,\cdot):=U_{t+1}(\w^t,\cdot,\cdot)$. Then, $\Psi(x,h)={u}_t(\w^t,x,h),$  
$v(x)= U_t(\w^t,x)$, see \eqref{state_val_t_rob_pj},  \eqref{v_pj} and \eqref{robust_phi_pj}. \\
Let $P:=q_1^P\otimes \cdots \otimes q_T^P\in\mathcal{H}^T$. We are in the $P$-prior $(t+1)$ context if $\mathcal{Q}:=\{q^{P}_{t+1}[\cdot|\w^t]\}$, $p^*:=q^{P}_{t+1}[\cdot|\w^t]$, $\alpha^*:=\alpha_t^{P}(\w^t)$,  and 
$V(\cdot,\cdot):=U_{t+1}^P(\w^t,\cdot,\cdot)$. Then, $\Psi(x,h)={u}^p_t(\w^t,x,h)$, $v(x)= U_t^P(\w^t,x)$, see \eqref{state_val_t_p_pj}, \eqref{v_pj} and \eqref{robust_phi_pj}.\\
\end{definition}  
Let $0\leq t\leq T-1$ and $P:=q_1^P \otimes \cdots \otimes q_T^P\in \mathcal{H}^T$. For all $\w^t\in\Omega^t$, we define
\begin{eqnarray}
c_t^P(\w^t):=\mathbb{E}_{q_{t+1}^P[\cdot|\w^t]}\; C_{t+1}(\w^t,\cdot) & \mbox{ and } & 
i_t^P(\w^t):=1+2\frac{c_t^P(\w^t)}{\alpha_t^P(\w^t)}.
\label{ct^*_pj} \label{ct^P_pj}
\label{It_P_pj}
\end{eqnarray}
Fix $\w^t\in\Omega^t$. The multiple-period counterpart of $c^*$, $l^*$, $n_0^*$ (see Assumption \ref{AE_one_pj}, Proposition \ref{well_def_one_pj}, Assumption \ref{pb_inequality_pj}) in the robust $(t+1)$ context are $c_t^{P^*}(\w^t)$, 
\begin{eqnarray}
l_t^*(\w^t)&:=&\sum_{\theta\in\{-1,1\}^d}\mathbb{E}_{q^{P^*}_{t+1}[\cdot|\w^t]}\; U_{t+1}^+\big(\w^t,\cdot,1+\theta \Delta S_{t+1}(\w^t,\cdot)\big),\nonumber \\ 
N_t^*(\w^t) &:=&\inf \Big\{k\geq 1,\;  q^{P^*}_{t+1}\big[U_{t+1}(\w^t,\cdot, -k)\leq - i_t^{P^*}(\w^t) | \w^t \big]\geq 1-\frac{\alpha_t^{P^*}(\w^t)}{2}\Big\}, \label{eq_Nt_pj}
\end{eqnarray}
with the convention (which will be used until the end of the paper) that $\inf \emptyset = +\infty$.\\ Now, the counterpart of $c^*$, $l^*$, $n_0^*$ in the $P$-prior $(t+1)$ context are $c_t^P(\w^t)$, 
\begin{eqnarray}
l_t^{P}(\w^t)&:=&\sum_{\theta\in\{-1,1\}^d}\mathbb{E}_{q_{t+1}^P[\cdot|\w^t]}\; (U_{t+1}^P)^+\big(\w^t,\cdot,1+\theta \Delta S_{t+1}(\w^t,\cdot)\big), \label{l_tP_pj}\\
N_t^P(\w^t) &:=&\inf \Big\{k\geq 1,\;  q_{t+1}^P\big[U_{t+1}^P(\w^t,\cdot, -k)\leq - i_t^P(\w^t) | \w^t \big]\geq 1-\frac{\alpha_t^P(\w^t)}{2}\Big\}.\label{eq_NtP_pj}
\end{eqnarray}
Note that for all $\w^t\in\Omega^t$, using \eqref{ct^P_pj}, \eqref{C_t_eq_pj} and Proposition \ref{CJ_t_pj}, we get that
\begin{eqnarray}
0\leq c_t^P(\w^t) &\leq& \sup_{p\in\mathcal{Q}_{t+1}(\w^t)}\mathbb{E}_{p}\; C_{t+1}(\w^t,\cdot)= C_t(\w^t). \label{ineq_ctp2_pj}
\end{eqnarray} 

We first show that all the previous functions are projective.
\begin{lemma}
Assume the (PD) axiom. Assume that Assumptions \ref{S_borel_pj}, \ref{analytic_graph_pj}, \ref{H_nonempty_pj} and \ref{AE_pj} hold. Let $P\in\mathcal{H}^T$, and $0\leq t \leq T-1$. Then, $l_t^*$, $N_t^*$, $i_t^P$, $c_t^P$, $l_t^P$, and $N_t^P$ are projective functions.
\label{lemma_mes_N_pj}
\end{lemma}
\proof{Proof.}
Let $P := q_1^P \otimes \cdots \otimes q_T^P\in\mathcal{H}^T$ and $0\leq t \leq T-1$. 
As $q_{t+1}^P$ is a projectively measurable stochastic kernel and $C_{t+1}$ is a projective function (see Proposition \ref{CJ_t_pj}), Proposition \ref{univ_cvt_pj}  (ii) shows that $c_t^P$ is a projective function. Recalling that $\alpha_t^P$ is projective (see Lemma \ref{simi_qt_na_pj}), Proposition \ref{base_hierarchy_pj} (v) shows that $i_t^P$ is projective.   Propositions   \ref{U_t_well_pj} and \ref{base_hierarchy_pj} (v) show that $U_{t+1}^+$ and $(U_{t+1}^P)^+$ are projective.  
Now, using Lemma \ref{lemma_mes_u_pj} for $U_{t+1}^+$ and $(U_{t+1}^P)^+,$ as well as Proposition \ref{base_hierarchy_pj} (v) and (vi), we get that $l_t^*$ and $l_t^P$ are projective.\\ 
We now show that $N_t^P$ is a projective function. The proof for $N_t^*$ is completely similar and thus omitted. Let $n\geq 1$. By definition of $N_t^P$ in \eqref{eq_NtP_pj},
$$\{N_t^P\leq n\} = \bigcup_{k=1}^n \Big\{ \w^t\in\Omega^t,\; \int_{\Omega^t} 1_{A(k)}(\w^t,\w_{t+1})q_{t+1}^{P}[d\w_{t+1}|\w^t] -1 + \frac{\alpha^{P}_t(\w^{t})}{2} \geq 0\Big\},$$ 
where $A(k):= \{(\w^{t},\w_{t+1})\in\Omega^{t}\times \Omega_{t+1},\; U_{t+1}^P(\w^{t},\w_{t+1},-k) +i_t^{P}(\w^t)\leq 0\}$. 
As $i_t^P,$ $\alpha_t^P$  and  $U_{t+1}^P$ are projective, we have that $U_{t+1}^P$ is $\Delta_p^1(\Omega^t\times \Omega_{t+1}\times \mathbb{R})$-measurable, $i_t^P$ is $\Delta_q^1(\Omega^t)$-measurable, and $\alpha_t^P$ is $\Delta_r^1(\Omega^t)$-measurable for some $p, q, r \geq 1$. We may assume that $r\leq q\leq p$. Then, using (i) in Proposition \ref{base_hierarchy_pj}, $i_t^P$ and $\alpha_t^P$ are $\Delta_{p+1}^1(\Omega^t)$-measurable. Fix some $k\geq 1$.  We get that $U_{t+1}^P(\cdot,\cdot,-k)$ is $\Delta_{p+1}^1(\Omega^t\times \Omega_{t+1})$-measurable applying Proposition \ref{base_hierarchy_pj} (vi) to the Borel, and thus $\Delta_{1}^1(\Omega^t\times \Omega_{t+1})$-measurable, function $(\w^{t},\w_{t+1}) \mapsto (\w^{t},\w_{t+1},-k)$ (see \eqref{equation_borel_set}). 
Thus, $A(k) \in \Delta_{p+1}^1(\Omega^t\times\Omega_{t+1})$ (see Proposition \ref{base_hierarchy_pj} (v)). Now, recalling that $q_{t+1}^P$ is a projectively measurable stochastic kernel, there exists some $i \geq 1$ such that $\w^t\mapsto q_{t+1}^P[\cdot|\w^t]$ is $\Delta_i^1(\Omega^t)$-measurable. Then,  Proposition \ref{univ_cvt_pj} (i) shows that $\w^t \mapsto \int_{\Omega^t} 1_{A(k)}(\w^t,\w_{t+1})q_{t+1}^{P}[d\w_{t+1}|\w^t]$ is $\Delta_{p+i+3}^1(\Omega^t)$-measurable. As $\alpha_t^P$ is $\Delta_{p+1}^1(\Omega^t)$-measurable, and also $\Delta_{p+i+3}^1(\Omega^t)$-measurable, we find that 
$$\Big\{\w^t\in\Omega^t,\; \int_{\Omega^t} 1_{A(k)}(\w^t,\w_{t+1})q_{t+1}^{P}[d\w_{t+1}|\w^t] -1 + \frac{\alpha^{P}_t(\w^{t})}{2} \geq 0\Big\} \in \Delta_{p+i+3}^1(\Omega^t).$$ 
Finally, as $\Delta_{p+i+3}^1(\Omega^t)$ is a $\sigma$-algebra (see (i) in Proposition \ref{base_hierarchy_pj}) and $p+i+3$ is independent of $k$, we obtain that $\{N_t^P\leq n\}\in \Delta_{p+i+3}^1(\Omega^t)$ for all $n\geq 1:$  $N_t^P$ is a projective function. \Halmos \endproof

The following sets describe the paths $\w^t\in\Omega^t$ for which the one-period Assumptions \ref{P^*_pj} to \ref{pb_inequality_pj} are satisfied in the robust $(t+1)$ context, and/or, in the $P$-prior $(t+1)$ context for a prior $P := q_1^P \otimes \cdots \otimes q_T^P\in\mathcal{H}^T$. Note that Assumption \ref{integ_V-_bis_pj} will only be used in the proof of Theorem 2.

\begin{definition}
Let $0\leq t\leq T-1$. For $i\in\{\ref{P^*_pj}, \ref{V_pj}, \ref{integ_V+_pj}, \ref{AE_one_pj}\}$, let 
\begin{eqnarray*}
\Omega^t_{i}&:=&\{\w^t\in\Omega^t,\; \mbox{Assumption \textit{i} holds true in the robust } (t+1) \mbox{ context}\}\\
\Omega^{t,P}_{i}&:=&\{\w^t\in\Omega^t,\; \mbox{Assumption \textit{i} holds true in the } P\mbox{-prior } (t+1) \mbox{ context}\}\\
\Omega^t_{\ref{pb_inequality_pj}}&:=& \Omega^{t,P^*}_{qNA} \cap \{ N_t^*<+\infty \} \mbox{ and }
\Omega^{t,P}_{\ref{pb_inequality_pj}}= \Omega^{t,P}_{qNA} \cap \{N_t^P<+\infty\},
\end{eqnarray*}
recall \eqref{set_qt_na_H_pj} for the definitions of $\Omega^{t,P^*}_{qNA}$ and $\Omega^{t,P}_{qNA}$. Moreover, we set 
\begin{eqnarray}
\widetilde{\Omega}^t := \bigcap_{i=8}^{12} \Omega^t_{i}\quad &\mbox{and}& \quad \widetilde{\Omega}^{t,P}:=\bigcap_{i=8}^{12} \Omega^{t,P}_{i}. \label{tildeOmega_pj}\label{tildeOmegaP_pj}
\end{eqnarray}
\label{def_ctxt_pj}
\label{def_ctxtP_pj}
\end{definition}
The following lemma, whose proof is trivial and thus omitted, shows that if we choose $\w^t$ in $\widetilde{\Omega}^t$ or $\widetilde{\Omega}^{t, P}$, the one-period assumptions are valid in the associated $(t+1)$ context.
\begin{lemma}
Assume the (PD) axiom. Assume that  Assumptions \ref{S_borel_pj}, \ref{analytic_graph_pj}, \ref{H_nonempty_pj} and \ref{AE_pj} hold. Let $P\in\mathcal{H}^T$ and $0\leq t\leq T-1$. If $\w^t\in \widetilde{\Omega}^t$ (resp. $\w^t\in \widetilde{\Omega}^{t,P}$), then Assumptions \ref{P^*_pj}, \ref{V_pj}, \ref{integ_V+_pj}, \ref{AE_one_pj} and \ref{pb_inequality_pj} hold true in the robust $(t+1)$ context (resp. $P$-prior $(t+1)$ context).
\label{lemma_Assumption_true_pj}
\end{lemma}

We now prove that the $\Omega^t_{i}$ are $\mathcal{Q}^t$-full-measure sets, while the $\Omega^{t,P}_{i}$ are $P^t$-full-measure sets. The proof needs the technical Lemmata \ref{lemma1+_pj} and \ref{lemma_N_finite_pj}, which are relegated to Appendix \ref{miss_proof_pj}.
\begin{proposition}
Assume the (PD) axiom. Assume that Assumptions \ref{S_borel_pj}, \ref{analytic_graph_pj}, \ref{H_nonempty_pj} and \ref{AE_pj} hold. Let $P\in\mathcal{H}^T$, and $0\leq t \leq T-1$. Then, for all $i\in \{\ref{P^*_pj},\ref{V_pj},\ref{AE_one_pj}\}$, $\Omega_{i}^t$ is a $\mathcal{Q}^t$-full-measure set, while $\Omega_{i}^{t,P}$ is a $P^t$-full-measure set. Assume furthermore that Assumptions \ref{nncst_pj}, \ref{U0_pj} and \ref{well_def_hp_pj} hold true. Then, $\Omega_{\ref{integ_V+_pj}}^t$ and $\Omega_{\ref{pb_inequality_pj}}^t$ are $\mathcal{Q}^t$-full-measure sets, while $\Omega_{\ref{integ_V+_pj}}^{t,P}$ and $\Omega_{\ref{pb_inequality_pj}}^{t,P}$ are $P^t$-full-measure sets. Thus, there exists a $\mathcal{Q}^t$-full-measure set $\widehat{\Omega}^t \in\mathcal{B}(\Omega^t)$ that satisfies $\widehat{\Omega}^t\subset\widetilde{\Omega}^t$.
\label{U_t^-&C_pj}
\end{proposition}
\proof{Proof.}
Fix $P:=q_1^P\otimes \cdots \otimes q_T^P \in\mathcal{H}^T$.

\textit{Let $0\leq t\leq T-1$ and $i\in \{\ref{P^*_pj}, \ref{V_pj}, \ref{AE_one_pj}\}$. The sets $\Omega_{i}^t$ and $\Omega_{i}^{t,P}$ are of $\mathcal{Q}^t$-full-measure and also of $P^t$-full-measure under Assumptions \ref{S_borel_pj}, \ref{analytic_graph_pj},  \ref{H_nonempty_pj} and \ref{AE_pj}.}\\ 
Recall $P^*$ from \eqref{P^*_exp_pj}. By Definitions \ref{def_ctxt_gen_pj} and \ref{def_ctxt_pj}, we see that for $Q\in q_1^Q \otimes \cdots \otimes q_T^Q \in \mathcal{H}^T$,
\begin{eqnarray*}
\Omega^{t,Q}_{\ref{P^*_pj}}&:=& \big\{\w^t\in\Omega^t,\; 0 \in \textup{ri}\big(\textup{conv}(D_{Q}^{t+1})\big)(\w^t)\big\}\\
\Omega^{t}_{\ref{P^*_pj}}&:=& \big\{\w^t\in\Omega^t,\; 0 \in \textup{ri}\big(\textup{conv}(D_{P^*}^{t+1})\big)(\w^t), \textup{Aff}(D_{P^*}^{t+1})(\w^t)=\textup{Aff}(D^{t+1})(\w^t)\big\},
\end{eqnarray*}
and $\Omega^{t}_{\ref{P^*_pj}} \subset \Omega^{t,P^*}_{\ref{P^*_pj}}.$ As $P^*$ and $Q$ belong to $\mathcal{H}^T$,   $\Omega_{\ref{P^*_pj}}^t$, $\Omega^{t,P^*}_{\ref{P^*_pj}}$, and $\Omega_{\ref{P^*_pj}}^{t,Q}$ are $\mathcal{Q}^t$-full-measure sets, see \eqref{H_def_pj}.\\ 
Proposition \ref{U_t_well_pj} at $t+1$ is now in force. Under the (PD) axiom, Assertion (i) and  Theorem \ref{proj_is_univ_pj} (i) show that $U_{t+1}$ and $U_{t+1}^P$ are $\mathcal{B}_c(\Omega^{t+1} \times \mathbb{R})$-measurable, and \citep[Lemma 7.29, p174]{ref1_pj} gives that $U_{t+1}(\w^t,\cdot,\cdot)$ and $U_{t+1}^P(\w^t,\cdot,\cdot)$ are $\mathcal{B}_c(\Omega_{t+1}\times \mathbb{R})$-measurable for all $\w^t \in \Omega^t$. Thus, assertion (ii) shows that $\Omega^{t}_{\ref{V_pj}}=\Omega_{\ref{V_pj}}^{t,P}=\Omega^t$, which is of course of $\mathcal{Q}^t$-full-measure.\\ 
Now, $C_{t+1}(\w^t,\cdot)$ is $\mathcal{B}_c(\Omega_{t+1})$-measurable, see Proposition \ref{CJ_t_pj} and again  Theorem \ref{proj_is_univ_pj} (i) and \citep[Lemma 7.29, p174]{ref1_pj}. We also have that $C_{t+1}(\w^t,\cdot) \geq 0$. Fix $\w^t\in\Omega^t$ such that $C_t(\w^t)<+\infty$. Then, 
$$c_t^{P^*}(\w^t)= \mathbb{E}_{q_{t+1}^{P^*}[\cdot|\w^t]} C_{t+1}(\w^t,\cdot) \leq C_{t}(\w^t)<+\infty,$$ 
see \eqref{ineq_ctp2_pj}. Moreover, the left-hand sides of \eqref{elas_gammaft_pj} and \eqref{elas_gammaft2_pj} show that the inequalities in \eqref{V_pos_pj}  are indeed satisfied for $\w_{t+1}\in \{C_{t+1}(\w^t,\cdot)<+\infty\}$ in the robust $(t+1)$ context. Thus, $\{\w^t\in\Omega^t,\;  C_t(\w^t)<+\infty\} \subset \Omega^t_{\ref{AE_one_pj}}$. So, \eqref{finite_CJ_pj} in Proposition \ref{CJ_t_pj} shows that $\Omega^t_{\ref{AE_one_pj}}$ is a $\mathcal{Q}^t$-full-measure set. The same arguments apply for $U_{t+1}^P$ (using the right-hand sides of \eqref{elas_gammafPt_pj} and \eqref{elas_gammafPt2_pj} instead of the left-hand sides, and $c_t^P(\w^t)$ instead of $c_t^{P^*}(\w^t)$), and $\Omega^{t,P}_{\ref{AE_one_pj}}$ is also a $\mathcal{Q}^t$-full-measure set.\\ 
Note that, as $\mathcal{H}^T \subset \mathcal{Q}^T$ (see \eqref{H_def_pj}), $\Omega^t_{i}$ and $\Omega^{t,P}_{i}$ are also $P^t$-full-measure sets for all $i\in\{\ref{P^*_pj},\ref{V_pj},\ref{AE_one_pj}\}$.

\textit{Let $0\leq t\leq T-1$ and $i \in \{\ref{integ_V+_pj}, \ref{pb_inequality_pj}\}$. The set $\Omega_{i}^t$ is of $\mathcal{Q}^t$-full-measure, and $\Omega^{t,P}_{i}$ is of $P^t$-full-measure under Assumptions \ref{S_borel_pj}, \ref{analytic_graph_pj},  \ref{H_nonempty_pj}, \ref{AE_pj}, \ref{nncst_pj}, \ref{U0_pj} and \ref{well_def_hp_pj}.}\\
The assertions for $\Omega^t_{\ref{integ_V+_pj}}$ and $\Omega^{t,P}_{\ref{integ_V+_pj}}$ are proved in Lemma \ref{lemma1+_pj} in the Appendix. Recall $\Omega^t_{\ref{pb_inequality_pj}}$ and $\Omega^{t,P}_{\ref{pb_inequality_pj}}$ from Definition \ref{def_ctxt_pj}.
We prove by backward induction that $\Omega^t_{\ref{pb_inequality_pj}}$ is a $\mathcal{Q}^t$-full-measure set, and that $\Omega^{t,P}_{\ref{pb_inequality_pj}}$ is a $P^t$-full-measure set for $P\in\mathcal{H}^T$. 
The initialization step at $T-1$ is a direct consequence of \eqref{NtoPfiniteT_pj} in Lemma \ref{lemma_N_finite_pj} and of the fact that $\Omega^{T-1,P^*}_{qNA}$ and $\Omega^{T-1,P}_{qNA}$ are $\mathcal{Q}^{T-1}$-full-measure sets (see Lemma \ref{simi_qt_na_pj}). 
Assume now that the induction hypothesis holds for some $1\leq t\leq T-1$. 
We have already proved  that the set $\Omega^{t,P}_{i}$, for all $i\in\{\ref{P^*_pj},\ref{V_pj},\ref{integ_V+_pj},\ref{AE_one_pj}\}$, is of $P^t$-full-measure for any $P\in\mathcal{H}^T$. So, the induction hypothesis implies that $\widetilde{\Omega}^{t,P}$ (see \eqref{tildeOmegaP_pj}) is also a $P^t$-full-measure set for all $P\in\mathcal{H}^T$, and Lemma \ref{lemma_N_finite_pj} can be applied for $t$. Thus, \eqref{NtP_finite}, \eqref{NtoPfinite_pj}, and the fact that $\Omega^{t-1,P^*}_{qNA}$ and $\Omega^{t-1,P}_{qNA}$ are $\mathcal{Q}^{t-1}$-full-measure sets, show the heredity step. This concludes the backward induction.\\ 

Finally, under all the assumptions, $\widetilde{\Omega}^t$ is a $\mathcal{Q}^t$-full-measure set, and we choose $\widehat{\Omega}^t\in\mathcal{B}(\Omega^t)$  such that $\widehat{\Omega}^t$ is a $\mathcal{Q}^t$-full-measure set, and $\widehat{\Omega}^t\subset \widetilde{\Omega}^t$.
\Halmos \endproof

\section{Proof of Theorem \ref{one_step_strategy_pj}}
\label{proof_th1_pj}
We now prove Theorem \ref{one_step_strategy_pj}. First, we show in Proposition \ref{opt_strat_exist_pj} that for all $0\leq t\leq T-1$, there exists a projectively measurable optimal investment strategy at time $t$ for problem \eqref{utcl-pb_pj}, when starting with a cash position $x$ (see \eqref{exist_opt_t_eq_pj}).   A measurable selection is  performed on a projective set, and there is no need to show that some mathematical objects are normal integrands (see \citep[Definition 14.27]{ref5_pj}): our proof is simpler than that of \citep[Proposition 8]{refnotre_pj}. 
Proposition \ref{opt_strat_exist_pj} is not provable in the usual ZFC theory, as $U_t$ is only shown to be projective.
\begin{proposition}
\label{opt_strat_exist_pj}
Assume the (PD) axiom. Assume that  Assumptions \ref{S_borel_pj}, \ref{analytic_graph_pj}, \ref{H_nonempty_pj}, \ref{AE_pj}, \ref{nncst_pj}, \ref{U0_pj} and \ref{well_def_hp_pj} hold. Let $0\leq t\leq T-1$. There exists a projective function $H_{t+1}^* : \Omega^t\times \mathbb{R} \to \mathbb{R}^d$ such that $H_{t+1}(\w^t,\cdot)\in\textup{Aff}(D^{t+1})(\w^t)$ for all $\w^t\in\widehat{\Omega}^t$ (where $\widehat{\Omega}^t$ has been defined in Proposition \ref{U_t^-&C_pj}). Moreover, for all $\w^t\in\widehat{\Omega}^t$ and $x\in\mathbb{R}$, 
\begin{eqnarray}
U_t(\w^t,x) \leq \textup{Cl}(U_t)(\w^t,x)\leq u_t^{cl}(\w^t,x)= \sup_{h\in\mathbb{R}^d}\textup{Cl}(u_t)(\w^t,x, h) = \textup{Cl}(u_t)\big(\w^t,x, H_{t+1}^*(\w^t,x)\big),\label{exist_opt_t_eq_pj}
\end{eqnarray}
see \eqref{u_t_rob_pj} for the definition $u_t$ and $U_t$, and \eqref{utcl-pb_pj} for the one of $u_t^{cl}$.
\label{exist_opt_t_pp_pj}
\end{proposition}
\proof{Proof.}
We first remark that $u_t$ is projective (see Lemma \ref{lemma_mes_u_pj} and \eqref{u_t_rob_pj}). We prove now that $\textup{Cl}(u_t)$ is also projective. 
For all $(\w^t,x,h)\in \Omega^t \times \mathbb{R}\times \mathbb{R}^d$, remark that 
\begin{eqnarray*}
\textup{Cl}(u_t)(\w^t,x,h) 
&=& \inf_{\substack{p\geq 1, q \geq 1}} \sup_{\substack{x'\in \mathbb{R},h'\in \mathbb{R}^d }}\upsilon_t(\w^t,x,h,p,q,x',h') \\
\upsilon_t(\w^t,x,h,p,q,x',h') &:=& u_t(\w^t,x+x',h+h')1_{|x'|<\frac{1}{p}}1_{|h'|<\frac{1}{q}}+ (-\infty)1_{|x'|\geq\frac{1}{p}}1_{|h'|\geq \frac{1}{q}}.
\end{eqnarray*}
Proposition \ref{base_hierarchy_pj} (v) and (vi) shows that $\upsilon_t$ is projective. Now, using Proposition  \ref{base_hierarchy_pj} (vii) with $D:= \Omega^t\times\mathbb{R}\times \mathbb{R}^d \times (\mathbb{N}\setminus \{0\})^2 \times \mathbb{R}\times \mathbb{R}^d$  (which is a projective set), we get that
$(\w^t,x,h,p,q) \mapsto \sup_{\substack{x'\in \mathbb{R},h'\in \mathbb{R}^d }}\upsilon_t(\w^t,x,h,p,q,x',h')$ is projective. Then, $\textup{Cl}(u_t)$ is also projective using again Proposition \ref{base_hierarchy_pj} (vii) with $D:= \Omega^t\times\mathbb{R}\times \mathbb{R}^d \times (\mathbb{N}\setminus \{0\})^2$, and so is $\sup_{l\in\mathbb{R}^d} \textup{Cl}(u_t)(\cdot,\cdot,l)$
choosing $D:=\Omega^t\times \mathbb{R} \times \mathbb{R}^d$. 
So, Proposition \ref{base_hierarchy_pj} (v) implies  that $(\w^t,x,h)\mapsto \sup_{l\in\mathbb{R}^d} \textup{Cl}(u_t)(\w^t,x,l) - \textup{Cl}(u_t)(\w^t,x,h)$ is projective, and 
\begin{eqnarray*}
\overline{A} &:=& \Big\{(\w^t,x,h)\in\Omega^t\times\mathbb{R}\times \mathbb{R}^d,\; \sup_{l\in\mathbb{R}^d}\textup{Cl}(u_t)(\w^t,x,l) = \textup{Cl}(u_t)(\w^t,x,h)\Big\}\in \textbf{P}(\Omega^t\times\mathbb{R}\times \mathbb{R}^d).
\end{eqnarray*}
Using \citep[Proposition 11]{refprojnotre_pj}, we see that $\textup{Graph}(\textup{Aff}(D^{t+1})\in \textbf{P}(\Omega^t\times \mathbb{R}^d)$. 
Let $T(\w^t,x,h):=(\w^t,h)$, and $A := T^{-1}(\textup{Graph}(\textup{Aff}(D^{t+1})) \cap \overline{A}$. As $T$ is Borel,  Proposition \ref{base_hierarchy_pj} (ii) shows that $A\in \textbf{P}(\Omega^t\times\mathbb{R}\times \mathbb{R}^d)$. Theorem \ref{proj_is_univ_pj} (ii) applied to $A$ shows the existence of a projective function $H_{t+1} : \textup{proj}_{\Omega^t\times \mathbb{R}}(A) \to \mathbb{R}^d$ such that for all $(\w^t,x)\in\textup{proj}_{\Omega^t\times \mathbb{R}}(A)$, $H_{t+1}(\w^t,x)\in \textup{Aff}(D^{t+1})(\w^t)$ and
\begin{eqnarray*}
\sup_{h\in\mathbb{R}^d}\textup{Cl}(u_t)(\w^t,x, h) = \textup{Cl}(u_t)\big(\w^t,x, H_{t+1}(\w^t,x)\big).
\end{eqnarray*} 
Let $(\w^t,x)\in \widehat{\Omega}^t\times \mathbb{R}$. Note that the set $\widehat{\Omega}^t \subset \widetilde{\Omega}^t,$ introduced in Proposition \ref{U_t^-&C_pj},  does not depend from $x$. As $\w^t \in \widehat{\Omega}^t$, Lemma \ref{lemma_Assumption_true_pj} shows that  Assumptions \ref{P^*_pj} to 
\ref{pb_inequality_pj} hold true in the robust $(t+1)$ context, and \eqref{maximizer_pj} in Proposition \ref{existence_uni_pj} shows immediately that $(\w^t,x)\in \textup{proj}_{\Omega^t\times \mathbb{R}}(A)$. Note that \eqref{maximizer_pj} in Proposition \ref{existence_uni_pj} also shows the inequalities in \eqref{exist_opt_t_eq_pj} for such an $\w^t$. As $\widehat{\Omega}^t\times \mathbb{R}\subset \textup{proj}_{\Omega^t\times \mathbb{R}}(A)$, $H_{t+1}$ is defined on $\widehat{\Omega}^t\times \mathbb{R}$, and can extended to $\Omega^t\times \mathbb{R}$ as follows. Let $H^*_{t+1}$ be defined by $H^*_{t+1}(\w^t,x) := H_{t+1}(\w^t,x)1_{\widehat{\Omega}^t}(\w^t)$ for all $(\w^t,x)\in \Omega^t \times \mathbb{R}$. As $\widehat{\Omega}^t\in\mathcal{B}(\Omega^t)$, $H^*_{t+1}$ remains projective, see \eqref{equation_borel_set} and Proposition  \ref{base_hierarchy_pj} (v). Moreover, for all $(\w^t,x)\in\widehat{\Omega}^t \times \mathbb{R}$, $H_{t+1}^*(\w^t,x)=H_{t+1}(\w^t,x)$, and (\ref{exist_opt_t_eq_pj}) remains true.
\Halmos\\ \endproof

We are now able to prove Theorem \ref{one_step_strategy_pj}. 
\noindent \proof{Proof of Theorem \ref{one_step_strategy_pj}}
Using Assumption \ref{U0_pj}, we show first that an admissible strategy in a $P$-prior context leads to a finite value function. Then, we show that $U_0(x)\geq u(x)$ and that the admissibility of a given subset of strategies is stable in time. After that, we prove that there exists an appropriate strategy satisfying \eqref{one_step_strategy_temp2_pj}. Finally, we show that when this strategy is admissible, \eqref{eg_fdmt_pj} and \eqref{bound_for_opt_pj} hold.\\ 
We first introduce another set of admissible strategies of interest for the proof. 
For all $0\leq t\leq T$, and $x\in \mathbb{R}$, let
$$\widehat{\Phi}_{|t}(x,U,\mathcal{Q}^T):=\Big\{\phi\in\Phi,\; \inf_{P\in\mathcal{Q}^t} \mathbb{E}_P U_t(\cdot,V_t^{x,\phi}(\cdot))>-\infty\Big\}.$$ 
We have that $\widehat{\Phi}_{|t}(x,U_t,\mathcal{Q}^t) \subset \Phi_{|t}(x,U_t,\mathcal{Q}^t)$, where $\Phi_{|t}(x,U_t,\mathcal{Q}^t)$ is the set of admissible strategies for the random utility $U_t$ with time horizon $t$, see Definition \ref{admissibility_def_pj}. 
Indeed, let $\phi\in \widehat{\Phi}_{|t}(x,U_t,\mathcal{Q}^t)$, and assume that $\phi \notin \Phi_{|t}(x,U_t,\mathcal{Q}^t)$. Then, there exists some $P\in\mathcal{Q}^t$ such that $\mathbb{E}_P  U_t^-(\cdot,V_t^{x,\phi}(\cdot)) = +\infty$. Recalling convention \eqref{cvt_inf_pj}, $\mathbb{E}_P  U_t(\cdot,V_t^{x,\phi}(\cdot)) = -\infty$, 
and also $\inf_{P\in\mathcal{Q}^t}\mathbb{E}_P  U_t(\cdot,V_t^{x,\phi}(\cdot)) = -\infty$, a contradiction. Moreover, we still trivially have that 
$$u(x) = \sup_{\phi\in \widehat{\Phi}_{|T}(x,U,\mathcal{Q}^T)}\inf_{P\in \mathcal{Q}^T}\mathbb{E}_{P} U\big(\cdot,V_T^{x,\phi}(\cdot)\big).$$

\textit{Finiteness of $\mathbb{E}_{P^t} U_t(\cdot,V_t^{x,\phi}(\cdot))$ for $0\leq t\leq T,$ $P\in\mathcal{H}^T$, $x\in\mathbb{R}$ and $\phi\in\Phi(x,U,\mathcal{Q}^T)$.}\\
For all $0\leq t\leq T,$ $P\in\mathcal{H}^T$, $x\in\mathbb{R}$, $\phi\in \Phi(x,U,\mathcal{Q}^T)$ and $\w^t\in\Omega^t$, we have using \eqref{state_val_t_p_pj} that,
\begin{eqnarray}
U_t^P\big(\w^t,V_t^{x,\phi}(\w^t)\big)&\geq &\mathbb{E}_{q_{t+1}^P[\cdot|\w^t]}\; U_{t+1}^P(\w^t,\cdot, V_{t+1}^{x,\phi}\big(\w^t,\cdot)\big)\geq - \mathbb{E}_{q_{t+1}^P[\cdot|\w^t]}\; (U_{t+1}^P)^-\big(\w^t,\cdot, V_{t+1}^{x,\phi}(\w^t,\cdot)\big).
\label{temp_adm_P_pj}
\end{eqnarray}
We show that for all $0\leq t\leq T$, $x\in\mathbb{R}$, $P\in \mathcal{H}^T$ and $\phi\in \Phi(x,U,\mathcal{Q}^T)$, we have that, 
\begin{eqnarray}
\mathbb{E}_{P^t} (U^P_t)^- \big(\cdot,V_t^{x,\phi}(\cdot)\big)<+\infty \;\; \mbox{and} \;\; \mathbb{E}_{P^t} (U^P_t)^+ \big(\cdot,V_t^{x,\phi}(\cdot)\big)<+\infty. \label{adm_P_statement_pj}
\end{eqnarray}
First, we show by backward induction that $\mathbb{E}_{P^t}(U^P_t)^-(\cdot,V_t^{x,\phi}(\cdot))<+\infty$. 
The initialization step follows from the definition of $\Phi(x, U,\mathcal{Q}^T)$, and  $U_T=U$. Let $0\leq t\leq T-1$, and assume that the induction hypothesis holds at time $t+1$.
Then, \eqref{temp_adm_P_pj}, 
Fubini's theorem, and the induction hypothesis, show that 
$$\mathbb{E}_{P^t} (U^P_t)^-\big(\cdot,V_t^{x,\phi}(\cdot)\big)\leq \mathbb{E}_{P^{t+1}} (U^P_{t+1})^-\big(\cdot,V_{t+1}^{x,\phi}(\cdot)\big)<+\infty.$$ 
This concludes the proof of the left-hand side of \eqref{adm_P_statement_pj}. We now prove the right-hand side by forward induction. 
For the initialization step, we show that $U_0^P(x)<+\infty$ for all $x\in\mathbb{R}$. As $U_0^P(1)<+\infty$ (by Assumption \ref{U0_pj}), and $U_0^P$ is nondecreasing (see Proposition \ref{U_t_well_pj}) for all $x\leq 1,$  $U_0^P(x)<+\infty$. 
Now, using the right-hand side of \eqref{elas_gammafPt_pj}, we find that for all $x>1$, $U_0^P(x) \leq x^{\overline{\gamma}}(U_0^P(1) + C_0)<+\infty$, as $C_0<+\infty$ by Proposition \ref{CJ_t_pj}. As $(U_0^P)^+ = U_0^P + (U_0^P)^-$, and $(U_0^P)^-<+\infty$ from the first induction, the initialization step is proved. Let $0\leq t\leq T-1$, and assume that the induction hypothesis holds at time $t$. Thus, using \eqref{temp_adm_P_pj}, we get that
\begin{eqnarray*}
\mathbb{E}_{P^{t+1}} U^P_{t+1}\big(\cdot,V_{t+1}^{x,\phi}(\cdot)\big)   =  \mathbb{E}_{P^t} \big[\mathbb{E}_{q_{t+1}^P[\cdot|\w^t]} U^P_{t+1}\big(\w^t,\cdot,V_{t+1}^{x,\phi}(\w^t,\cdot)\big) \big] & \leq  & \mathbb{E}_{P^t} U^P_t \big(\cdot,V_t^{x,\phi}(\cdot)\big)\\
& \leq  & \mathbb{E}_{P^t} (U^P_t)^+ \big(\cdot,V_t^{x,\phi}(\cdot)\big)<+\infty,
\end{eqnarray*}
where we have used Fubini's theorem for the first equality (which can be applied because of the left-hand side of \eqref{adm_P_statement_pj} at $t+1$), and also the induction hypothesis.   
This concludes the second induction for the right-hand side of \eqref{adm_P_statement_pj}.

\textit{Upper bound for $u(x)$, and stability of $\widehat{\Phi}_{|T}$.}\\
We show that for any $x\in\mathbb{R}$, \begin{eqnarray}
U_0(x)\geq \sup_{\phi\in \widehat{\Phi}_{|T}(x,U,\mathcal{Q}^T)}\inf_{P\in \mathcal{Q}^T}\mathbb{E}_P U\big(\cdot,V_T^{x,\phi}(\cdot)\big)=u(x),
\label{ineg_fdmt_pj}
\end{eqnarray}
and that if $\phi\in \widehat{\Phi}_{|T}(x,U,\mathcal{Q}^T)$, then $\phi\in \widehat{\Phi}_{|t}(x,U_t,\mathcal{Q}^t)$ for all $0\leq t\leq T$. 
Fix $x\in\mathbb{R}$, and $\phi\in \widehat{\Phi}_{|T}(x,U,\mathcal{Q}^T)$. We proceed by backward induction with the following induction hypothesis : 
\begin{eqnarray}
\phi\in \widehat{\Phi}_{|t}(x,U_t,\mathcal{Q}^t) \;\;\mbox{and}\;\; \inf_{P\in\mathcal{Q}^T} \mathbb{E}_P U\big(\cdot,V_T^{x,\phi}(\cdot)\big) \leq \inf_{P\in\mathcal{Q}^t} \mathbb{E}_P U_t\big(\cdot,V_t^{x,\phi}(\cdot)\big).
\label{temp_ind_eq_pj}
\end{eqnarray}
The initialization step is trivial as $U_T=U$. Let $0\leq t\leq T-1$, and assume that the induction hypothesis holds at time $t+1$. Proposition \ref{U_t_well_pj} shows that $U_{t+1}$ is projective, and Lemma \ref{lemma_mes_u_pj} that 
\begin{eqnarray*}
(\w^t,x,h,p)\mapsto \overline{u}_t(\w^t,x,h,p):= \mathbb{E}_p U_{t+1}\big(\w^t,\cdot,x+h\Delta S_{t+1}(\w^t,\cdot)\big)
\end{eqnarray*}
is also projective. Note that $\overline{u}_t(\w^t,\cdot,\cdot,p)$ is equal to $\Psi_p$ (see \eqref{Phi_pj}) in the robust $(t+1)$ context. Let $\epsilon>0$. Proposition \ref{base_hierarchy_pj} (vii) applied to the projective set 
$D:= \{(\w^t,x,h,p),\; (\w^t,p)\in \textup{Graph}(\mathcal{Q}_{t+1})\}$ 
(see Assumption \ref{analytic_graph_pj} and  Proposition \ref{base_hierarchy_pj} (ii)) show that there exists a projective function $q^\epsilon_{t+1} : \Omega^t\times \mathbb{R} \times \mathbb{R}^d \to \mathfrak{P}(\Omega_{t+1})$ such that  for all $ (\w^t,x,h)\in\Omega^t \times \mathbb{R} \times \mathbb{R}^d$ (recall that $\mathcal{Q}_{t+1}\neq \emptyset$
), $q^\epsilon_{t+1}[\cdot|\w^t,x,h]\in \mathcal{Q}_{t+1}(\w^t)$, and 
\begin{eqnarray}
\overline{u}_t(\w^t,x,h,q^{\epsilon}_{t+1}[\cdot|\w^t,x,h)]\leq -\frac{1}{\epsilon}1_{\{u_t(\w^t,x,h)=-\infty\}} +(u_t(\w^t,x,h)+\epsilon)1_{\{u_t(\w^t,x,h)>-\infty\}},
\label{temp_ind_eq2_u_t_pj}
\end{eqnarray}
as $u_t(\w^t,x,h) = \inf_{p\in\mathcal{Q}_{t+1}(\w^t)} \overline{u}_t(\w^t,x,h,p)$, see \eqref{u_t_rob_pj}. Fix $\w^t\in \Omega^t$. As $U_t(\w^t,x)\geq u_t(\w^t,x,h)$ for all $(x,h)\in\mathbb{R}\times \mathbb{R}^d$, using \eqref{temp_ind_eq2_u_t_pj} with $x=V_t^{x,\phi}(\w^t)$, $h=\phi_{t+1}(\w^t)$, and setting $\overline{q}^{\epsilon}_{t+1}[\cdot|\w^t]:=q^{\epsilon}_{t+1}[\cdot |\w^t,V_t^{x,\phi}(\w^t),\phi_{t+1}(\w^t)]$, we have that for all $\w^t\in \Omega^t$, 
\begin{equation}
\begin{split}
\overline{u}_t\big(\w^t,V_t^{x,\phi}(\w^t),\phi_{t+1}(\w^t),\overline{q}^{\epsilon}_{t+1}[\cdot|\w^t]\big)  \leq  -& \frac{1}{\epsilon} 1_{\{u_t(\w^t,V_t^{x,\phi}(\w^t),\phi_{t+1}(\w^t))=-\infty\}} \\ &+ \big(U_{t}\big(\w^t,V_t^{x,\phi}(\w^t)\big)+\epsilon\big)1_{\{u_t(\w^t,V_t^{x,\phi}(\w^t),\phi_{t+1}(\w^t))>-\infty\}}. \label{temp_base_fub_adm_pj}
\end{split}
\end{equation}

We prove now that $\overline{q}_{t+1}^{\epsilon}$ is a projectively measurable stochastic kernel, and that $\overline{q}_{t+1}^{\epsilon}[\cdot|\w^t] \in \mathcal{Q}_{t+1}(\w^t)$ for all $\w^t\in\Omega^t$. First, $V_t^{x,\phi} = x+ \sum_{t=1}^t \phi_t \Delta S_t$ is projective. Indeed, by Assumption \ref{S_borel_pj}, and Proposition \ref{base_hierarchy_pj} (iv) and (vi), $\Delta S_t$ is projective. 
As $\phi\in\Phi$, $\phi_t$ is projective, and so is $\phi_t \Delta S_t$, and $V_t^{x,\phi}$ using Proposition \ref{base_hierarchy_pj} (iv) again. As $(\w^t,x,h)\mapsto q_{t+1}^{\epsilon}[\cdot|\w^t,x,h]$ is projective, we obtain using Proposition \ref{base_hierarchy_pj} (vi) that $\w^t\mapsto\overline{q}_{t+1}^{\epsilon}[\cdot|\w^t]$ is projective, and $\overline{q}_{t+1}^{\epsilon}$ is a projectively measurable stochastic kernel. 
We also get that for $\w^t\in\Omega^t$, $\overline{q}_{t+1}^{\epsilon}[\cdot|\w^t] \in \mathcal{Q}_{t+1}(\w^t)$ as $q_{t+1}^{\epsilon}[\cdot|\w^t,x,h]\in\mathcal{Q}_{t+1}(\w^t)$ for all $(x,h)\in \mathbb{R}\times \mathbb{R}^d$. 

Let $P := q_1^P \otimes \cdots \otimes q_T^P\in\mathcal{Q}^T$, and $P^* := q_1^{P^*} \otimes \cdots \otimes q_T^{P^*}\in\mathcal{H}^T$ defined in \eqref{P^*_exp_pj}. We set
$$\overline{P} : = \frac{q_1^{P^*} + q_1^{P}}{2} \otimes \cdots \otimes \frac{q_t^{P^*} + q_t^P}{2}\otimes q_{t+1}^{P^*} \otimes \cdots \otimes q_T^{P^*}.$$ 
Then, $\overline{P}\in\mathcal{H}^T$ (see \eqref{setPT_pj}), and $\overline{P}^t\otimes \overline{q}_{t+1}^{\epsilon}\in\mathcal{Q}^{t+1}$. 
Taking the expectation in \eqref{temp_base_fub_adm_pj} under $\overline{P}^t$, and using Fubini's theorem as $\phi\in\widehat{\phi}_{|t+1}(x,U_{t+1},\mathcal{Q}^{t+1}) \subset \phi_{|t+1}(x,U_{t+1},\mathcal{Q}^{t+1})$, we get that
\begin{eqnarray}
 \nonumber
 \inf_{P\in\mathcal{Q}^{t+1}}\mathbb{E}_{P} U_{t+1}\big(\cdot,V_{t+1}^{x,\phi}(\cdot)\big) & \leq & \mathbb{E}_{\overline{P}^t\otimes \overline{q}^{\epsilon}_{t+1}} U_{t+1}\big(\cdot,V_{t+1}^{x,\phi}(\cdot)\big)\\
 & \leq &  -\frac{1}{\epsilon} \overline{P}^t\big[u_t(\cdot,V_t^{x,\phi}\big(\cdot),\phi_{t+1}(\cdot)\big)=-\infty\big]   
  + \mathbb{E}_{\overline{P}^t}\; U_t^+\big(\cdot,V_t^{x,\phi}(\cdot)\big) + \epsilon.\label{temp_CJt_gen_u_t_pj}
 \label{ineq_temp_J_t_2_u_t_pj}
\end{eqnarray}
As $U_t\leq U_t^P$ (see Proposition \ref{U_t_well_pj}),  \eqref{adm_P_statement_pj} implies  that $\mathbb{E}_{\overline{P}^t}\; U_t^+(\cdot,V_t^{x,\phi}(\cdot)) <+\infty$. So, if $\overline{P}^t[u_t(\cdot,V_t^{x,\phi}(\cdot),\phi_{t+1}(\cdot))=-\infty]>0$, taking the limit, when $\epsilon$ goes to $0$ in \eqref{ineq_temp_J_t_2_u_t_pj}, gives that $\inf_{P\in\mathcal{Q}^{t+1}}\mathbb{E}_{P} U_{t+1}(\cdot,V_{t+1}^{x,\phi}(\cdot))=-\infty$, a contradiction with $\phi\in\widehat{\Phi}_{|t+1}(x,U_{t+1},\mathcal{Q}^{t+1})$ of the induction hypothesis. Thus, $\overline{P}^t[u_t(\cdot,V_t^{x,\phi}(\cdot),\phi_{t+1}(\cdot))=-\infty]=0$. Using now \citep[Proposition 12]{refnotre_pj}, we have that $P^t\ll\overline{P}^t$, and  $P^t[u_t(\cdot,V_t^{x,\phi}(\cdot),\phi_{t+1}(\cdot))=-\infty]=0$. 
As $P$ is arbitrary, we have that 
$$u_t\big(\cdot,V_t^{x,\phi}(\cdot),\phi_{t+1}(\cdot)\big)>-\infty\quad \mathcal{Q}^t\mbox{-q.s}.$$
Thus, taking again the expectation in \eqref{temp_base_fub_adm_pj}, but under $P^t$, and using Fubini's theorem, we get that for all $P^t\in\mathcal{Q}^t$
\begin{eqnarray*}
\inf_{P\in\mathcal{Q}^{t+1}}\mathbb{E}_{P}   U_{t+1}\big(\cdot,V_{t+1}^{x,\phi}(\cdot)\big) \leq  \mathbb{E}_{{P}^t\otimes \overline{q}^{\epsilon}_{t+1}} U_{t+1}\big(\cdot,V_{t+1}^{x,\phi}(\cdot)\big)
\leq \mathbb{E}_{P^t}\; U_t\big(\cdot,V_t^{x,\phi}(\cdot)\big) + \epsilon.
\end{eqnarray*}
Letting $\epsilon$ go to $0$, and using the second part of the induction hypothesis  \eqref{temp_ind_eq_pj}, we get that
$$\inf_{P\in\mathcal{Q}^T} \mathbb{E}_P U\big(\cdot,V_T^{x,\phi}(\cdot)\big) \leq \inf_{P\in\mathcal{Q}^{t+1}}\mathbb{E}_{P}U_{t+1}\big(\cdot,V_{t+1}^{x,\phi}(\cdot)\big) \leq \inf_{P\in\mathcal{Q}^t}\mathbb{E}_{P}U_{t}\big(\cdot,V_{t}^{x,\phi}(\cdot)\big).$$
In particular, as $\phi\in \widehat{\Phi}_{|{t+1}}(x,U_{t+1},\mathcal{Q}^{t+1})$, 
$$-\infty <\inf_{P\in\mathcal{Q}^{t+1}}\mathbb{E}_{P}U_{t+1}(\cdot,V_{t+1}^{x,\phi}(\cdot)) \leq \inf_{P\in\mathcal{Q}^t}\mathbb{E}_{P}U_{t}(\cdot,V_{t}^{x,\phi}(\cdot)).$$ 
So, $\phi\in\widehat{\phi}_{|t}(x,U_{t},\mathcal{Q}^{t})$, and this concludes the induction.\\
Now,  \eqref{temp_ind_eq_pj} at $t=0$ shows that $\inf_{P\in\mathcal{Q}^T} \mathbb{E}_P U(\cdot,V_T^{x,\phi}(\cdot))  \leq U_0(x)$ for $\phi\in\widehat{\Phi}_{|T}(x,U,\mathcal{Q}^T)$. Thus, \eqref{ineg_fdmt_pj} follows by taking the supremum over all such $\phi$.

\textit{Existence of a one-step optimal strategy.}\\
Let $x\in\mathbb{R}$. We define the strategy $\phi^{*,x}$ recursively as follows: 
$$\phi^{*,x}_1 := x \quad \phi^{*,x}_{t+1}(\w^t):=H_{t+1}^*\Big(\w^t,x+ \sum_{s=1}^t \phi_s^{*,x}(\w^t) \Delta S_{s}(\w^t)\Big), \; 
\forall \w^t\in\Omega^t, \, \forall 1\leq t\leq T-1,$$ 
where $H_{t+1}^* : \Omega^t\times \mathbb{R} \to \mathbb{R}^d$ is defined in Proposition \ref{exist_opt_t_pp_pj}. Let $1\leq t\leq T-1$. Proposition \ref{exist_opt_t_pp_pj} shows that for all $(\w^t,x)\in \widehat{\Omega}^t\times \mathbb{R}$, $\phi^{*,x}_{t+1}(\w^t)\in \textup{Aff}(D^{t+1})(\w^t)$, and that \eqref{exist_opt_t_eq_pj} holds true for $x=V_t^{x,\phi^{*,x}}(\w^t)$ (recall that $\widehat{\Omega}^t$ does not depend from $x$). Thus, (\ref{one_step_strategy_temp2_pj}) also holds.\\ 
Now, we show by induction that $\phi^{*,x}_{t+1}$ is projective for all $0\leq t \leq T-1$. At $t=0$, this is trivial (recall that $\Omega^0$ is a singleton). Suppose that this holds for all $0\leq k \leq t-1$. Then, recalling Assumption \ref{S_borel_pj} and Proposition \ref{base_hierarchy_pj} (iv) and (vi), $V_t^{x,\phi^{*,x}} = x+ \sum_{k=1}^t \phi^{*,x}_{k} \Delta S_k$ is projective. Thus, as $H_{t+1}^*$ is projective (see Proposition \ref{exist_opt_t_pp_pj}), Proposition \ref{base_hierarchy_pj} (vi) show that $\phi^{*,x}_{t+1} = H_{t+1}^*(\cdot,V_t^{x,\phi^{*,x}}(\cdot))$ is also projective. This concludes the induction, and $\phi^{*,x}\in\Phi$.

\textit{Optimality of an admissible one-step optimal strategy.}\\
Assume now that $\phi^{*,x} \in \Phi(x,U,\mathcal{Q}^T)$. Let $0\leq t \leq T-1$, and $P^*:=q_1^{P^*}\otimes \cdot\cdot\cdot \otimes q_T^{P^*}\in\mathcal{H}^T$ as in  \eqref{P^*_exp_pj}. Let $U_t^{cl} : \Omega^t \times \mathbb{R} \times \mathbb{R}^d \to \mathbb{R}\cup\{-\infty,+\infty\}$ be defined by 
$$U_t^{cl}(\w^t,x,h):= \mathbb{E}_{q_{t+1}^{P^*}[\cdot|\w^t]} \textup{Cl}\big(U_{t+1})(\w^t,\cdot, x+h\Delta S_{t+1}(\w^t,\cdot)\big).$$
Remark that $U_t^{cl}(\w^t,\cdot,\cdot) = \Psi_{p^*}^{cl}(\cdot,\cdot)$  in the robust $(t+1)$ context, where $\Psi_{p^*}^{cl}$ is defined in \eqref{psi_p*}. 
Recalling \eqref{u_t_rob_pj}, and the definition of the closure, $u_t \leq  U_t^{cl}\leq \textup{Cl}(U^{cl}_{t})$, and so $ \textup{Cl}(u_t) \leq \textup{Cl}(U^{cl}_{t})$. As Assumptions \ref{P^*_pj}, \ref{V_pj}, \ref{integ_V+_pj} and \ref{AE_one_pj} are satisfied in the robust $(t+1)$ context for $\w^t\in\widehat{\Omega}^t$ (see Lemma \ref{lemma_Assumption_true_pj}), Proposition \ref{well_def_one_pj} shows that 
$U_t^{cl}(\w^t,\cdot,\cdot)$ is usc. 
So, we conclude that for all $\w^t\in\widehat{\Omega}^t$, 
$$ \textup{Cl}(u_t)(\w^t,\cdot,\cdot) \leq  \textup{Cl}(U^{cl}_{t})(\w^t,\cdot,\cdot) =U_t^{cl}(\w^t,\cdot,\cdot).$$ 
We have proved in the preceding step that (\ref{one_step_strategy_temp2_pj}) holds. Thus, we have for all $\w^t\in\widehat{\Omega}^t$ that
\begin{eqnarray}
\nonumber
\textup{Cl}(U_t)\big(\w^t,V_t^{x,\phi^{*,x}}(\w^t)\big) & \leq  & u^{cl}_t\big(\w^t,V_t^{x,\phi^{*,x}}(\w^t)\big)=\textup{Cl}(u_t)\big(\w^t,V_t^{x,\phi^{*,x}}(\w^t), \phi_{t+1}^{*,x}(\w^t)\big) \\
 &\leq & 
U_t^{cl}\big(\w^t,V_t^{x,\phi^{*,x}}(\w^t), \phi_{t+1}^{*,x}(\w^t)\big)= \mathbb{E}_{q_{t+1}^{P^*}[\cdot|\w^t]} \textup{Cl}(U_{t+1})\big(\w^t,\cdot,V_{t+1}^{x,\phi^{*,x}}(\w^t,\cdot)\big).
\label{ineq_temp_max_uti_pp_pj}
\end{eqnarray}
Applying recursively \eqref{ineq_temp_max_uti_pp_pj} from $t=0$ to $t=T-1$, we obtain that 
\begin{eqnarray}
\textup{Cl}(U_0)(x)\leq u^{cl}_0(x)=\textup{Cl}(u_0)(x)\leq \int_{\Omega_1}\cdots \int_{\Omega_T} \textup{Cl}(U)\big(\w^T,V_T^{x,\phi^{*,x}}(\w^T)\big) q_T^{P^*}[d\w^T|\w_{T-1}]\cdots q_1^{P^*}[d\w_1].\label{ineq_temp_max_uti_pp2_pj}
\end{eqnarray}
Now, remark that $(\textup{Cl}(U))^- \leq U^-$. So, using $U_T^{P}=U,$ and $\phi^{*,x}\in\Phi(x,U,\mathcal{Q}^T)$, we get that for all $Q\in\mathcal{Q}^T$, 
\begin{eqnarray}
\mathbb{E}_{Q} (\textup{Cl}(U))^-\big(\cdot,V_T^{x,\phi^{*,x}}(\cdot)\big) \leq \mathbb{E}_Q U^-\big(\cdot,V_T^{x,\phi^{*,x}}(\cdot)\big) <+\infty. \label{neg_part_integCl_pj}
\end{eqnarray}
Thus, Fubini's theorem applies in \eqref{ineq_temp_max_uti_pp2_pj}, and  
$\textup{Cl}(U_0)(x)\leq u^{cl}_0(x) =\textup{Cl}(u_0)(x)\leq \mathbb{E}_{P^*} \textup{Cl}(U)(\cdot,V_T^{x,\phi^{*,x}}(\cdot)).$ As $P^*$ is arbitrary in $\mathcal{H}^T$, we have that
\begin{eqnarray}
\textup{Cl}(U_0)(x)\leq u^{cl}_0(x) =\textup{Cl}(u_0)(x)\leq \inf_{P\in\mathcal{H}^T}\mathbb{E}_{P} \textup{Cl}(U)\big(\cdot,V_T^{x,\phi^{*,x}}(\cdot)\big).
\label{ineq_temp_max_uti_pp3_pj}
\end{eqnarray}
Now, using Proposition \ref{U_t_well_pj} (iv), we have that for all $x\in\mathbb{R}$, and for all $P\in\mathcal{H}^T$, 
\begin{eqnarray}
\mathbb{E}_P (\textup{Cl}(U))^+\big(\cdot,V_T^{x,\phi^{*,x}}(\cdot)\big) \leq \mathbb{E}_P U^+\big(\cdot,V_T^{x+1,\phi^{*,x}}(\cdot)\big)<+\infty, \label{ineq_temp_max_uti_pp3_pj00}
\end{eqnarray}
see again \eqref{adm_P_statement_pj}. By definition of the closure of $U_0$ and recalling \eqref{ineg_fdmt_pj} and \eqref{ineq_temp_max_uti_pp3_pj}, we get that
$$u(x)\leq U_0(x)\leq \textup{Cl}(U_0)(x)\leq u^{cl}_0(x)=\textup{Cl}(u_0)(x) \leq \inf_{P\in\mathcal{H}^T}\mathbb{E}_{P} \textup{Cl}(U)\big(\cdot,V_T^{x,\phi^{*,x}}(\cdot)\big)= \inf_{Q\in\mathcal{Q}^T}\mathbb{E}_{Q} \textup{Cl}(U)\big(\cdot,V_T^{x,\phi^{*,x}}(\cdot)\big),$$
where the last equality follows from Lemma \ref{lemma_equ_QH_pj} applied to $X = \textup{Cl}(U)(\cdot,V_T^{x,\phi^{*,x}}(\cdot)).$ Indeed, \eqref{neg_part_integCl_pj} and \eqref{ineq_temp_max_uti_pp3_pj00} hold. Moreover,  $\textup{Cl}(U_{t+1})$ is projective (see Proposition \ref{U_t_well_pj} (i)), and we have already proved that  $V_T^{x,\phi^{*,x}}$ is projective. So, Proposition \ref{base_hierarchy_pj} (vi) shows that $X$ is also projective. This conludes the proof of \eqref{eg_fdmt_pj}.  
Finally, for $x\in \mathbb{R}$, 
\begin{eqnarray*}
\inf_{P\in \mathcal{Q}^T}\mathbb{E}_P U\big(\cdot,V_T^{x,\phi^{*,x}}(\cdot)\big) \leq u(x) \leq \inf_{P\in \mathcal{Q}^T}\mathbb{E}_P U\big(\cdot,V_T^{x,\phi^{*,x}}(\cdot)\big) + \sup_{P\in \mathcal{Q}^T}\mathbb{E}_P \Delta_+ U\big(\cdot,V_T^{x,\phi^{*,x}}(\cdot)\big),
\end{eqnarray*}
where the first inequality follows from the definition of $u(x),$ as $\phi^{*,x}\in\Phi(x,U,\mathcal{Q}^T),$ and the second one from  \eqref{UeqClU}.  Thus, \eqref{bound_for_opt_pj} holds, which concludes the proof. 
\Halmos \endproof

\section{Proof of Theorem \ref{optimality_M_typeA_pj}}
\label{proof_optimality_M_pj}
In this part, we prove Theorem \ref{optimality_M_typeA_pj} by applying Theorem \ref{one_step_strategy_pj}. Recall that Assumptions \ref{AE_pj} and \ref{nncst_pj} hold by definition of a random utility of type $(A)$. Moreover, we prove that if Assumptions \ref{S_borel_pj}, \ref{analytic_graph_pj} and \ref{H_nonempty_pj} hold, and if $U$ is a random utility of type $(A)$, Assumptions \ref{U0_pj} and \ref{well_def_hp_pj} also hold true. For that, we use the results of Section \ref{muti_per_pj} for $\Omega^{t,P}$ and $U_t^P$.            
The proof follows the same path as the proof of \citep[Theorem 2]{refnotre_pj}. There are, nevertheless, three main differences. First, the expression of $c_t^P$ differs from the one of \citep[(61)]{refnotre_pj}. Second, we have, for the moment, introduced no control from below on $U_t$. Third, Assumption \ref{integ_V-_bis_pj} is needed to get a bound on the optimal strategy in Proposition \ref{inequ_borne_pj}. So, we introduce for all $0\leq t\leq T-1$, and $P:= q_1^P\otimes \cdots \otimes q_T^P\in\mathcal{H}^T$, the set of paths $\w^t$ for which Assumption \ref{integ_V-_bis_pj} holds (recall \eqref{u_t_p_pj},  \eqref{u_t_rob_pj} and Definition \ref{def_ctxt_pj}) :
\begin{equation}
\begin{aligned}
\Omega^{t}_{\ref{integ_V-_bis_pj}}&:=&\big\{\w^t\in\Omega^t,\; u_t(\w^t,-k, 0) = \inf_{p\in\mathcal{Q}_{t+1}(\w^t)} \mathbb{E}_p U_{t+1}(\w^t,\cdot,-k)>-\infty,\; \forall k \in \mathbb{N}\setminus \{0\}\big\}\\
\Omega^{t,P}_{\ref{integ_V-_bis_pj}}&:=&\big\{\w^t\in\Omega^t,\; u_t^P(\w^t,-k, 0) =  \mathbb{E}_{q_{t+1}^P[\cdot|\w^t]} U_{t+1}^P(\w^t,\cdot,-k)>-\infty,\; \forall k \in \mathbb{N}\setminus \{0\}\big\}.\label{hp_integV-tP_pj}
\end{aligned}
\end{equation}
\noindent \proof{Proof of Theorem \ref{optimality_M_typeA_pj}.}

\textit{\\Assumption \ref{U0_pj} holds.}\\
Let $P:= q_1^P\otimes \cdots \otimes q_T^P\in\mathcal{H}^T$. Recall that $u_t^P$, $u_t$, $c_t^P$, $i_t^P$, $l_t^P$, $N_t^P$, and $\widetilde{\Omega}^{t,P}$ are defined  in \eqref{u_t_p_pj}, \eqref{u_t_rob_pj}, \eqref{ct^P_pj}, 
\eqref{l_tP_pj}, \eqref{eq_NtP_pj} and \eqref{tildeOmegaP_pj} for all $0\leq t\leq T-1$. Recall also that the sets $\mathcal{M}^t(P)$, $\mathcal{M}^t$ are defined in Definition \ref{M_tP_pj} for all $0\leq t\leq T$. For the need of the induction, we define some of them also for $t=-1$ or $t=T$. We set $u_T^P :=0$, $u_T : = 0$,  $N_{-1}^P := 0$, $l_T^P :=0$, $\widetilde{\Omega}^{T,P}:= \Omega^T$, and $\mathcal{M}^{-1}(P) :=\{0\}$. We also set $\Omega^{T}_{\ref{integ_V-_bis_pj}}:= \Omega^T$,  $\Omega^{T,P}_{\ref{integ_V-_bis_pj}}:= \Omega^T$.\\ 
For all $0\leq t\leq T$, we prove by backward induction the following induction hypothesis  : $(U_{t}^{P})^+(\cdot,1)$ and $l_{t}^P$ belong to $\mathcal{M}^{t}(P)$ , $C_t$, $u_t^-(\cdot,x,0)$, $(u_t^P)^-(\cdot,x,0)$ belong to $\mathcal{M}^{t}$ for all $x\in\mathbb{R}$, $N_{t-1}^P\in\mathcal{M}^{t-1}(P)$, $\widetilde{\Omega}^{t,P} \cap \Omega^{t,P}_{\ref{integ_V-_bis_pk}}$ is a $P^t$-full-measure set and there exists  $D_{1,t}\in \mathcal{M}^t$ such that $D_{1,t}\geq 0$, and 
\begin{eqnarray}
U_t(\w^t,x)\geq -D_{1,t}(\w^t)(1+|x|^p), \; \forall (\w^T,x) \in\Omega^T \times \mathbb{R}.
\label{ineq_b_inf_det_A_t_pj}
\end{eqnarray}
Then, we will obtain from the induction hypothesis at $t=0$ that Assumption \ref{U0_pj} holds. Indeed, $\mathcal{M}^0(P) = \mathbb{R}$, and $U_0^P(1) \leq (U_0^P)^+(1)<+\infty$.

\textit{Initialization step.}\\ 
We trivially have that $l_T^P=0 \in \mathcal{M}^T(P)$, $u_T^- = (u_T^P)^-=0 \in \mathcal{M}^T$, and that $\widetilde{\Omega}^{T,P} \cap \Omega^{T,P}_{\ref{integ_V-_bis_pj}}=\Omega^T$ is a $P$-full-measure set. 
Using Definition \ref{typeA_pj}, $C_T=C\in\mathcal{M}^T$ (see \eqref{C_t_eq_pj}), $(U_T^P)^+(\cdot,1)=U^+(\cdot,1)\in\mathcal{M}^T \subset \mathcal{M}^T(P)$, and \eqref{ineq_b_inf_det_A_t_pj} holds true with $D_{1,T}:= D_1,$ see \eqref{ineq_b_inf_det_A_pj}. 
By assumption of Theorem \ref{optimality_M_typeA_pj}, we have that $1/\alpha_{T-1}^P\in\mathcal{M}^{T-1}\subset\mathcal{M}^{T-1}(P)$. Again, as $U$ is of type (A), we know that $\underline{X} \in \mathcal{M}^T \subset \mathcal{M}^T(P)$, and  \eqref{ineq_b_inf_det_A_pj} implies that  
$U^-(\w^T,0)\leq D_1(\w^T)$, and so $U^-(\w^T,0) \in \mathcal{M}^T\subset \mathcal{M}^T(P),$ see Lemma \ref{lemme_fdmt_W_pj}. 
Thus, we can use assertion (A1) in Lemma \ref{lemma_N_finite_pj}, and we get that $N_{T-1}^P \in\mathcal{M}^{T-1}(P)$. \\

\noindent Assume that the induction hypothesis holds at time $t+1$ for some $0\leq t\leq T-1$. The heredity proof is divided in four steps. 

\textit{Heredity step 1 : $C_t\in\mathcal{M}^t$, there exists $D_{1,t}\in \mathcal{M}^t$ such that $D_{1,t}\geq 0$ and \eqref{ineq_b_inf_det_A_t_pj} holds true at $t$, and $u_t^-(\cdot,x,0), (u_t^P)^-(\cdot,x,0) \in \mathcal{M}^t$ for all $x\in\mathbb{R}$.} \\
Using \eqref{finite_CJ_pj},  we have that $C_t<+\infty$ $\mathcal{Q}^t$-q.s. Thus, \eqref{temp_J_qeps_pj} with $\epsilon=1$ provides some $
q_{t+1}\in SK_{t+1}$ such that for all $\w^t\in \Omega^t$, $q_{t+1}[\cdot|\w^t]\in\mathcal{Q}_{t+1}(\w^t)$, and for all $\w^t$ in the $\mathcal{Q}^t$-full-measure set where $C_t<+\infty$,
\begin{eqnarray}
\mathbb{E}_{q_{t+1}[\cdot|\w^t]} C_{t+1}(\w^t,\cdot) \geq C_t(\w^t)-1.
\label{jt_temp_w_pj}
\end{eqnarray}
By the induction hypothesis $C_{t+1} \in \mathcal{M}^{t+1}$, and Lemma \ref{M_hered_pj}  shows that $\w^t\mapsto \mathbb{E}_{q_{t+1}[\cdot|\w^t]} C_{t+1}(\w^t,\cdot)$ belongs to $\mathcal{M}^t$. Proposition \ref{CJ_t_pj} shows that $C_{t}(\cdot)$ is a non-negative, projective function. Thus, \eqref{jt_temp_w_pj} and Lemma \ref{lemme_fdmt_W_pj} ensure that $C_t\in\mathcal{M}^t$.\\ 
Now, let $\w^t\in\Omega^t$, and $x\in\mathbb{R}$. Using \eqref{state_val_t_rob_pj} and \eqref{ineq_b_inf_det_A_t_pj} at $t+1$, we get that
\begin{eqnarray}
U_t(\w^t,x) \geq u_t(\w^t,x,0) = \inf_{p\in\mathcal{Q}_{t+1}(\w^t)} \mathbb{E}_p U_{t+1}(\w^t,\cdot,x) \geq - D_{1,t}(\w^t) (1+|x|^p), \label{temp_neg_integration_finite_pj}
\end{eqnarray}
where $D_{1,t}(\w^t):=\sup_{p\in\mathcal{Q}_{t+1}(\w^t)} \mathbb{E}_p D_{1,t+1}(\w^t,\cdot)$. As in the proof of Proposition \ref{CJ_t_pj}, we find that $D_{1,t}$ is a non-negative, projective function. As $D_{1,t+1}\in\mathcal{M}^{t+1}$, a very similar reasoning to the one that shows that $C_t\in\mathcal{M}^t$ proves that $D_{1,t}\in\mathcal{M}^t$. So, \eqref{ineq_b_inf_det_A_t_pj} holds true at $t$.\\
Let $x\in\mathbb{R}$. Using \eqref{temp_neg_integration_finite_pj}, we have that $u_t^-(\cdot,x,0) \leq D_{1,t}(\cdot) (1+|x|^p)$. Thus, as $u_t^-(\cdot,x,0)$ is projective (see Lemma \ref{lemma_mes_u_pj}, Proposition \ref{base_hierarchy_pj} (v) and (vi)) and $D_{1,t} \in \mathcal{M}^t$, Lemma \ref{lemme_fdmt_W_pj} shows that $u_t^-(\cdot,x,0)\in\mathcal{M}^t$. Now, as $U_{t+1}\leq U_{t+1}^P$ (see Proposition \ref{U_t_well_pj}), we have that $(u_t^P)^-(\cdot,x,0)\leq u_t^-(\cdot,x,0)$. As $(u_t^P)^-(\cdot,x,0)$ is projective (see again Lemma \ref{lemma_mes_u_pj}, Proposition \ref{base_hierarchy_pj} (v) and (vi)), Lemma \ref{lemme_fdmt_W_pj} shows that $(u_t^P)^-(\cdot,x,0)\in\mathcal{M}^t$.

\textit{Heredity step 2 : $l_t^P\in\mathcal{M}^t(P)$ and $\widetilde{\Omega}^{t,P} \cap \Omega^{t,P}_{\ref{integ_V-_bis_pj}}$ is a $P^t$-full-measure set.}\\
We first show that $l_t^P\in\mathcal{M}^t(P)$. Let $\theta\in\{-1,1\}^d$.  Cauchy-Schwarz inequality, (ii), and the right-hand side of \eqref{elas_gammafPt_pj} at time $t+1$ in Proposition \ref{U_t_well_pj} show that $\mathcal{Q}^{t+1}$-q.s.,
\begin{eqnarray}
(U_{t+1}^P)^+\big(\cdot,1+\theta \Delta S_{t+1}(\cdot)\big)&\leq&  (U_{t+1}^P)^+\big(\cdot,1+|\theta ||\Delta S_{t+1}(\cdot)|\big)\nonumber\\
&\leq & \big(1+\sqrt{d}|\Delta S_{t+1}(\cdot)|\big)^{\overline{\gamma}}\big((U_{t+1}^P)^+(\cdot,1)+C_{t+1}(\cdot)\big).\label{temp_iCP_pj}
\end{eqnarray}
as $C_{t+1}<+\infty$ $\mathcal{Q}^{t+1}$-q.s. , see \eqref{finite_CJ_pj}. 
As $U_{t+1}^P$ is projective (see Proposition \ref{U_t_well_pj}), and Assumption \ref{S_borel_pj} holds, $(U_{t+1}^P)^+(\cdot,1+\theta \Delta S_{t+1}(\cdot))$ is also projective using Proposition \ref{base_hierarchy_pj} (v) and (vi). Now, recalling that $(U_{t+1}^P)^+(\cdot,1)\in\mathcal{M}^{t+1}(P)$, $C_{t+1}\in\mathcal{M}^{t+1}$ by the induction hypothesis, and that $|\Delta S_{t+1}|\in\mathcal{M}^{t+1}$ by assumption of Theorem \ref{optimality_M_typeA_pj}, 
we deduce from \eqref{temp_iCP_pj} and Lemma \ref{lemme_fdmt_W_pj} that 
\begin{eqnarray}
(U_{t+1}^P)^+\big(\cdot,1+\theta \Delta S_{t+1}(\cdot)\big)\in\mathcal{M}^{t+1}(P). \label{pour_well_def_integ_pj}
\end{eqnarray}
Thus, $\w^t\mapsto \mathbb{E}_{q_{t+1}^P[\cdot|\w^t]} (U_{t+1}^P)^+(\w^t,\cdot,1+\theta \Delta S_{t+1}(\w^t,\cdot))$ belongs to $\mathcal{M}^t(P)$ for all $\theta\in\{-1,1\}^d$ thanks to Lemma \ref{M_hered_pj}. So, \eqref{l_tP_pj} and Lemma \ref{lemme_fdmt_W_pj} shows that $l_t^P\in\mathcal{M}^t(P)$.\\
We now prove that $\widetilde{\Omega}^{t,P}\cap \Omega_{\ref{integ_V-_bis_pj}}^{t,P}$ is a $P^t$-full-measure set. 
For that, we first show that $\Omega_{\ref{pb_inequality_pj}}^{t,P}$ is a $P^t$-full-measure set. Indeed, by the induction hypothesis, $N_t^P\in\mathcal{M}^t(P)$, which implies that $N_t^P<+\infty\; P^t-\mbox{a.s.}$ 
Lemma \ref{simi_qt_na_pj} shows that $\Omega^{t,P}_{qNA}$ is a $\mathcal{Q}^{t}$-full-measure set, and also a $P^t$-full-measure set, as $\mathcal{H}^T \subset \mathcal{Q}^T$. Thus, $\Omega^{t,P}_{\ref{pb_inequality_pj}}$ is a $P^t$-full-measure set. 
Moreover, using the first part of Proposition \ref{U_t^-&C_pj}, which do not required Assumptions \ref{U0_pj} and \ref{well_def_hp_pj}, we also have that $\Omega_{i}^{t,P}$ is a $P^t$-full-measure set for all $i\in\{\ref{P^*_pj},\ref{V_pj},\ref{AE_one_pj}\}$. 
As $l_t^P \in\mathcal{M}^t(P)$, we have that $l_t^P<+\infty\; P^t-\mbox{a.s.}:$  $\Omega^{t,P}_{\ref{integ_V+_pj}}$, and also $\widetilde{\Omega}^{t,P}$ are $P^t$-full-measure sets. As $(u_t^P)^-(\cdot,-k,0)\in\mathcal{M}^t$ (see Heredity step 1), $(u_t^P)^-(\cdot,-k,0)<+\infty$ $P^t$-a.s. So, $\Omega^{t,P}_{\ref{integ_V-_bis_pj}}$ is a $P^t$-full-measure set (see \eqref{hp_integV-tP_pj}). Thus, we can find a $P^t$-full-measure set $\widehat{\Omega}^{t,P}\in \mathcal{B}(\Omega^t)$ such that 
$\widehat{\Omega}^{t,P}\subset \widetilde{\Omega}^{t,P} \cap \Omega_{\ref{integ_V-_bis_pj}}^{t,P}$.

\textit{Heredity step 3 : $(U_t^P)^+(\cdot,1)\in\mathcal{M}^t(P)$.}\\ 
We first define the bound on the optimal strategies $K_t^P(\w^t)$. If $\w^t\notin\widehat{\Omega}^{t,P}$, we set $K_t^P(\w^t):=1$. If $\w^t\in \widehat{\Omega}^{t,P}$, Assumptions  \ref{P^*_pj} to \ref{pb_inequality_pj} are satisfied in the $P$-prior $(t+1)$ context (see Definition \ref{def_one_prior_pj}, and Lemma \ref{lemma_Assumption_true_pj}), Proposition \ref{sub_optimal_pj} applies in this context, and we set $K_t^P(\w^t):= K_1(1)$, i.e.,
\begin{small}
$$K_t^P(\w^t)= \max\Bigg(1,\frac{1+N_t^P(\w^t)}{\alpha_{t}^P(\w^t)},\Big(\frac{1+N_t^P(\w^t)}{\alpha_{t}^P(\w^t)}\Big)^{\frac{1}{1-\eta}}, \Big(\frac{6 l_t^P(\w^t)}{\alpha_{t}^P(\w^t)}\Big)^{\frac{1}{\eta \overline{\gamma}-\underline{\gamma}}},\Big(\frac{6c_t^P(\w^t)}{\alpha_{t}^P(\w^t)}\Big)^{\frac{1}{\eta \overline{\gamma}-\underline{\gamma}}},\Big(\frac{6(u_t^P)^-(\w^t,x,0)}{\alpha_{t}^P(\w^t)} \Big)^{\frac{1}{\eta\overline{\gamma}}}\Bigg).$$
\end{small}
We prove that $K_t^P\in\mathcal{M}^t(P)$. As $C_{t}\in\mathcal{M}^{t}(P)$ (see Heredity step 1), we have using \eqref{ineq_ctp2_pj}, Lemmata \ref{lemma_mes_N_pj} and \ref{lemme_fdmt_W_pj} that $c_t^P \in \mathcal{M}^t(P)$. Moreover, $N_t^P\in\mathcal{M}^t(P)$ by the induction hypothesis, and $1/\alpha_t^P\in\mathcal{M}^t(P)$ by assumption of Theorem \ref{optimality_M_typeA_pj}. We have proved in Heredity step 1 that $(u_t^P)^-(\cdot,x,0) \in \mathcal{M}^t$ for all $x\in\mathbb{R}$, and in Heredity step 2 that $l_t^P\in\mathcal{M}^t(P)$. Thus, we deduce from Lemma \ref{lemme_fdmt_W_pj} that $K_t^P$ restricted to $\widehat{\Omega}^{t,P}$ belongs to $\mathcal{M}^t(P)$. We also get that $K_t^P \in \mathcal{M}^t(P)$ as $\widehat{\Omega}^{t,P}\in\mathcal{B}(\Omega^t)$, see \eqref{equation_borel_set} and Lemma \ref{M_hered_pj}.\\
 We now prove that $(U_t^P)^+(\cdot,1)\in \mathcal{M}^t(P)$. For all $\w^t\in\widehat{\Omega}^{t,P}$,  Assumptions  \ref{P^*_pj} to \ref{integ_V-_bis_pj} are satisfied in the $P$-prior $(t+1)$ context, and \eqref{eq_value_valboun1_pj} in Proposition \ref{sub_optimal_pj} applies 
\begin{eqnarray}
U_t^{P}(\w^t,1)&\leq & \sup_{\substack{|h|\leq K_t^P(\w^t) \\ h\in \textup{Aff}(D_{t+1}^P)(\w^t)}}\textup{Cl}(u_t^P)(\w^t,1,h) \leq \sup_{|h|\leq K_t^P(\w^t)}\textup{Cl}(u_t^P)(\w^t,1,h). \label{temp_utp_bound1_pj}
\end{eqnarray}
Now, let $U_t^{P,cl} : \Omega^t \times \mathbb{R} \times \mathbb{R}^d \to \mathbb{R}\cup\{-\infty,+\infty\}$ be defined by 
$$U_t^{P,cl}(\w^t,x,h):= \mathbb{E}_{q_{t+1}^{P}[\cdot|\w^t]} \textup{Cl}(U_{t+1}^P)\big(\w^t,\cdot,x+h\Delta S_{t+1}(\w^t,\cdot)\big).$$
Remark that $U_t^{P,cl}(\w^t,\cdot,\cdot) = \Psi_{p}^{cl}(\cdot,\cdot)$ in the $P$-prior $(t+1)$ context, where $\Psi_{p}^{cl}$ is defined in \eqref{psi_p*}. As Assumptions \ref{P^*_pj}, \ref{V_pj}, \ref{integ_V+_pj} and \ref{AE_one_pj} are satisfied in the $P$-prior $(t+1)$ context for $\w^t\in\widehat{\Omega}^{t,P}$, Proposition \ref{well_def_one_pj} shows that 
$U_t^{P,cl}(\w^t,\cdot,\cdot)$ is usc. 
Then, by definition of the closure $u_t^P \leq U_t^{P,cl}$  (see \eqref{u_t_p_pj}), and also $\textup{Cl}(u_t^P) \leq \textup{Cl}(U_t^{P,cl}).$ So, we obtain for $\w^t\in\widehat{\Omega}^{t,P}$ that 
\begin{eqnarray}
U_t^{P}(\w^t,1)&\leq &  \sup_{|h|\leq K_t^P(\w^t)}\textup{Cl}(u_t^P)(\w^t,1,h) \leq \sup_{|h|\leq K_t^P(\w^t)} \textup{Cl}(U_t^{P,cl})(\w^t,1,h) = \sup_{|h|\leq K_t^P(\w^t)} U_t^{P,cl}(\w^t,1,h)\nonumber \\ 
&=& 
\sup_{|h|\leq K_t^P(\w^t)}  \mathbb{E}_{q_{t+1}^{P}[\cdot|\w^t]} \textup{Cl}\big(U_{t+1}^P)(\w^t,\cdot,1+h\Delta S_{t+1}(\w^t,\cdot)\big)\nonumber\\ 
& \leq &  
\sup_{|h|\leq K_t^P(\w^t)} \mathbb{E}_{q_{t+1}^P[\cdot|\w^t]} U_{t+1}^P \big(\w^t,\cdot,2+ h \Delta S_{t+1}(\w^t,\cdot)\big)
\label{fautbien}\\&\leq & 
\mathbb{E}_{q_{t+1}^P[\cdot|\w^t]}(U_{t+1}^{P})^+(\w^t,\cdot,2+K_t^P(\w^t) |\Delta S_{t+1}(\w^t,\cdot)|),\label{temp_Ut+1Ut_pj} \label{temp_utp_bound2_pj}
\end{eqnarray}
where \eqref{fautbien} follows from Proposition \ref{U_t_well_pj} (iv), and \eqref{temp_utp_bound2_pj} from $U_{t+1}^{P}(\w^t,\cdot)$ being nondecreasing. The last step is to use Lemma \ref{lemme_fdmt_W_pj} in \eqref{temp_Ut+1Ut_pj} (as $\widehat{\Omega}^{t,P}$ is a $P^t$-full-measure set), to prove that  $(U_t^{P})^+(\cdot,1)\in\mathcal{M}^t(P)$. First, Propositions \ref{U_t_well_pj} (i) and  \ref{base_hierarchy_pj} (vi) show that $(U_t^P)^+(\cdot,1)$ is projective. So, it remains to prove that 
$\w^t \mapsto \mathbb{E}_{q_{t+1}^P[\cdot|\w^t]} (U_{t+1}^P)^+(\w^t,\cdot, 2+K_t^P(\w^t) |\Delta S_{t+1}(\w^t,\cdot)|)$ belongs to $\mathcal{M}^t(P)$. This will follows from Lemma \ref{M_hered_pj}, if $(U_{t+1}^{P})^+(\cdot,2+K_t^P(\cdot) |\Delta S_{t+1}(\cdot)|)\in\mathcal{M}^{t+1}(P)$. 
The right-hand side of (\ref{elas_gammafPt_pj}) implies that
\begin{eqnarray}
(U_{t+1}^{P})^+(\cdot,2+K_t^P(\cdot) |\Delta S_{t+1}(\cdot)|) \leq \big(2+K_t^P(\cdot) |\Delta S_{t+1}(\cdot)|\big)^{\overline{\gamma}}\big((U_{t+1}^{P})^+(\cdot,1)+C_{t+1}(\cdot)\big).
\label{marre}
\end{eqnarray}
As $U_{t+1}^P$, $|\Delta S_{t+1}|$, and $K_t^P$ are projective, Proposition \ref{base_hierarchy_pj} (v) and (vi) show that $(U_{t+1}^{P})^+(\cdot,2+K_t^P(\cdot) |\Delta S_{t+1}(\cdot)|)$ is projective. So, Lemma \ref{lemme_fdmt_W_pj} proves the claim if the right-hand side of \eqref{marre} is in $\mathcal{M}^{t+1}(P)$. This 
holds as we have proved that $K_t^P\in\mathcal{M}^t(P)$, that $|\Delta S_{t+1}(\cdot)|\in\mathcal{M}^{t+1}$ by assumption of Theorem \ref{optimality_M_typeA_pj}, and that $C_{t+1}\in\mathcal{M}^{t+1}$ and $(U_{t+1}^{P})^+(\cdot,1)\in\mathcal{M}^{t+1}(P)$ from the induction hypothesis. 

\textit{Heredity step 4 : $N_{t-1}^P\in\mathcal{M}^{t-1}(P)$}.\\ 
If $t=0$, we trivially have that $N_{-1}^P=0\in\mathcal{M}^{-1}(P) = \{0\}$. So, assume that $t\geq 1$. Recall from Heredity steps 1 and 2 that $l_t^P\in\mathcal{M}^t(P)$, $C_t\in\mathcal{M}^t\subset \mathcal{M}^t(P)$, from the induction hypothesis that $N_t^P\in\mathcal{M}^t(P)$, and from the assumptions of Theorem \ref{optimality_M_typeA_pj} that $1/\alpha_{t-1}^P\in \mathcal{M}^{t-1}\subset \mathcal{M}^{t-1}(P)$ and $1/\alpha_{t}^P\in \mathcal{M}^{t}\subset \mathcal{M}^{t}(P)$. Thus, assertion (A2) in Lemma \ref{lemma_N_finite_pj} for $1\leq t\leq T-1$ (recall that $\widetilde{\Omega}^{t,P}$ is a $P^t$-full-measure set for all $P\in\mathcal{H}^T$ from Heredity step 2) shows that $N_{t-1}^P\in\mathcal{M}^{t-1}(P).$ This concludes the heredity step.

\textit{Assumption \ref{well_def_hp_pj} holds true.}\\
Assumption \ref{well_def_hp_pj} follows from \eqref{pour_well_def_integ_pj} at $t$ if $\theta\in\{-1,1\}^d$, and from $(U_t^P)^+(\cdot,1)\in\mathcal{M}^t(P)$ for all $P\in\mathcal{H}^T$ if $\theta = 0$.

\textit{Application of Theorem \ref{one_step_strategy_pj}.}\\
Let $x\in\mathbb{R}$. As Assumptions \ref{S_borel_pj}, \ref{analytic_graph_pj}, \ref{H_nonempty_pj}, \ref{AE_pj}, \ref{nncst_pj}, \ref{U0_pj} and \ref{well_def_hp_pj} hold true, Theorem \ref{one_step_strategy_pj} applies, and there exists $\phi^{*,x}\in \Phi$ satisfying \eqref{one_step_strategy_temp2_pj}. As $U$ is usc ($U$ is of type (A)), if we prove that $\phi^{*,x}\in \Phi(x,U,\mathcal{Q}^T),$ \eqref{eg_fdmt_ok_pj} shows that $\phi^{*,x}$ is an optimal strategy for \eqref{RUMP_pj}, which will conclude the proof.

\textit{For all $P\in\mathcal{Q}^T$, $\mathbb{E}_P\; U^-(\cdot, V_T^{x,\phi^{*,x}}(\cdot))<+\infty:$ $\phi^{*,x}\in \Phi(x,U,\mathcal{Q}^T)$.}\\
As \eqref{ineq_b_inf_det_A_pj} holds with $D_1\in\mathcal{M}^T$, if $U^-(\cdot, V_T^{x,\phi^{*,x}}(\cdot))$ is projective, and $V_T^{x,\phi^{*,x}}\in\mathcal{M}^{T}(P)$ for all $P\in\mathcal{Q}^T$,  Lemma \ref{lemme_fdmt_W_pj} will show that $U^-(\cdot, V_T^{x,\phi^{*,x}}(\cdot))\in\mathcal{M}^{T}(P)$ for all $P\in\mathcal{Q}^T$, and thus the claim. \\
The first assertion follows from Proposition \ref{base_hierarchy_pj} (vi) as $U^-$  is projective (see Proposition \ref{U_t_well_pj}), and we have already proved that $V_T^{x,\phi^{*,x}}$ is projective. \\
Fix $0\leq t \leq T-1$, and recall $\widehat{\Omega}^t$ from Proposition \ref{U_t^-&C_pj}. We know from the induction that for all $k\geq 1$, $u_t^-(\cdot,-k,0) \in\mathcal{M}^t$, and so that $u_t^-(\cdot,-k,0)<+\infty$ $\mathcal{Q}^t$-q.s. Thus, $\Omega^t_{\ref{integ_V-_bis_pj}}$ (see \eqref{hp_integV-tP_pj}) is a $\mathcal{Q}^t$-full-measure set, and so, is $\widehat{\Omega}^t \cap \Omega^t_{\ref{integ_V-_bis_pj}}$. We can find a $\mathcal{Q}^t$-full-measure set $\overline{\Omega}^{t}\in \mathcal{B}(\Omega^t)$ such that 
$\overline{\Omega}^{t}\subset \widehat{\Omega}^{t} \cap \Omega_{\ref{integ_V-_bis_pj}}^{t}$.  For all $\w^t\in\overline{\Omega}^{t}$, Assumptions  \ref{P^*_pj} to \ref{integ_V-_bis_pj} are satisfied in the robust $(t+1)$ context (see  Definition \ref{def_ctxt_pj}), and recalling the definition of $\phi^{*,x}$, Proposition \ref{existence_uni_pj} in the robust $(t+1)$ context shows that for all $\w^t\in\overline{\Omega}^{t}$, 
$|\phi^{*,x}_{t+1}(\w^t)|\leq K_t(\w^t),$ where for $\w^t\in\overline{\Omega}^t$, $K_t(\w^t):= K_1(V_t^{x,\phi^{*,x}})$ with $K_1$ defined in Proposition \ref{sub_optimal_pj} in the robust $(t+1)$ context, and $K_t(\w^t):=1$ when $\w^t\notin\overline{\Omega}^{t}$. 
We set $K_{-1}:=0$ for the  next induction.\\ 
Let $P\in\mathcal{Q}^T$. We show by induction on $0\leq t\leq T$ that $V_{t}^{x,\phi^{*,x}} \in \mathcal{M}^t(P)$, and $K_{t-1}\in\mathcal{M}^{t-1}(P)$. The initialization step is trivial as $V_{0}^{x,\phi^{*,x}} = x$ and $K_{-1} = 0$. Assume now that the induction hypothesis holds at $t$. As $V_{t+1}^{x,\phi^{*,x}} = V_{t}^{x,\phi^{*,x}} + \phi_{t+1}^{*,x} \Delta S_{t+1},$ $V_{t}^{x,\phi^{*,x}}$, $\phi^{*,x}_{t+1}$ and $\Delta S_{t+1}$ are projective, we get that $V_{t+1}^{x,\phi^{*,x}}$ is also projective, see Proposition \ref{base_hierarchy_pj} (iv) and (vi). As $\overline{\Omega}^t$ is a $\mathcal{Q}^t$-full-measure set, 
\begin{eqnarray*}
|V_{t+1}^{x,\phi^{*,x}}| \leq |V_{t}^{x,\phi^{*,x}}| + K_{t} |\Delta S_{t+1}| \; \mathcal{Q}^t\mbox{-q.s.}
\end{eqnarray*}
Thus, as $V_{t}^{x,\phi^{*,x}} \in \mathcal{M}^t(P)$ by the induction hypothesis, and $|\Delta S_{t+1}|\in \mathcal{M}^t$ by assumption of Theorem \ref{optimality_M_typeA_pj}, if we prove that $K_{t}\in \mathcal{M}^{t}(P)$, Lemma \ref{lemme_fdmt_W_pj} shows that $V_{t+1}^{x,\phi^{*,x}} \in \mathcal{M}^{t+1}(P)$,  
and the induction hypothesis at $t+1$ will follow. To prove that $K_t\in\mathcal{M}^t(P)$, as $P$ belongs to $\mathcal{Q}^T$, but not necessarily to $\mathcal{H}^T$, we use $\widehat{P}_{t+1} \in\mathcal{H}^T$ defined in \eqref{hatP_lemmaN_pj} in the Appendix. \\
First, $\mathcal{M}^t(\widehat{P}_{t+1})\subset \mathcal{M}^t(P)$ (see Lemma \ref{lemma_N_finite_pj}). 
Recall also that in the robust $(t+1)$ context, $\alpha^* = \alpha_{t}^{P^*}(\w^{t})$, $n_0^* = N_{t}^*(\w^{t})$, $c^* = c_{t}^{P^*}(\w^{t})$, $l^* = l_{t}^{*}(\w^{t})$ and $\Psi(x,0) = u_t(\w^t,x,0)$. By assumption of Theorem \ref{optimality_M_typeA_pj}, $1/\alpha_t^{P^*}\in\mathcal{M}^t$. So, $1/\alpha_t^{P^*}$ belong both to $\mathcal{M}^t(P)$ and $\mathcal{M}^t(\widehat{P}_{t+1})$.  
We first prove that $l_{t}^*\in \mathcal{M}^{t}(\widehat{P}_{t+1})\subset \mathcal{M}^{t}(P)$. Using that $U_{t}\leq U_{t}^{\widehat{P}_{t+1}}$ (see Proposition \ref{U_t_well_pj}), we have that $l_{t}^*\leq l_{t}^{\widehat{P}_{t+1}}$. 
So, as $l_{t}^{\widehat{P}_{t+1}}\in\mathcal{M}^{t}(\widehat{P}_{t+1})$ (see Heredity step 2), and $l_{t}^*$ is projective (see Lemma \ref{lemma_mes_N_pj}), we get that $l_{t}^*\in \mathcal{M}^{t}(\widehat{P}_{t+1})$ by Lemma \ref{lemme_fdmt_W_pj}. 
We now prove that $N_{t}^*\in\mathcal{M}^{t}(P)$. 
Recall assertions (B1) and (B2) from Lemma \ref{lemma_N_finite_pj}. 
Assertion (B1) applies, and shows that $N_{T-1}^*\in\mathcal{M}^{T-1}(P)$ as $1/\alpha_{T-1}^{P^*} \in \mathcal{M}^{T-1}\subset \mathcal{M}^{T-1}(\widehat{P}_{T})$, and $\underline{X}$, $U^-(\cdot,0)$, $C$ belong to $\mathcal{M}^T\subset \mathcal{M}^T(\widehat{P}_{T})$. \\
Now, Assertion (B2) also applies, and shows that $N_t^* \in \mathcal{M}^t(P),$ as $\widetilde{\Omega}^{t+1,P}$ is a $P^{t+1}$-full-measure set for all $P\in\mathcal{H}^T$, $1/\alpha_{t}^{P^*}\in\mathcal{M}^{t}\subset \mathcal{M}^{t}(\widehat{P}_{t+1})$, $1/\alpha_{t+1}^{\widehat{P}_{t+1}}$, $C_{t+1} \in\mathcal{M}^{t+1} \subset \mathcal{M}^{t+1}(\widehat{P}_{t+1})$, and $l_{t+1}^{\widehat{P}_{t+1}}$, $N_{t+1}^{\widehat{P}_{t+1}}\in \mathcal{M}^{t+1}(\widehat{P}_{t+1})$, see Heredity steps $1$, $2$ and $4$ for $P=\widehat{P}_{t+1} \in\mathcal{H}^T$. 
Now, recalling \eqref{temp_neg_integration_finite_pj}, $D_{1,t} \in \mathcal{M}^t$, and $V_{t}^{x,\phi^{*,x}}\in\mathcal{M}^t(P)$, we have that $u_t^-(\cdot,V_t^{x,\phi^{*,x}}(\cdot),0) \in\mathcal{M}^t(P)$ (see Lemma \ref{lemme_fdmt_W_pj} as $u_t^-(\cdot,V_t^{x,\phi^{*,x}}(\cdot),0)$ is projective, see Heredity step 1, and Proposition \ref{base_hierarchy_pj} (vi)). Finally, as $C_{t}\in \mathcal{M}^{t}$, $c_{t}^{P^*}\in\mathcal{M}^t$ using \eqref{ineq_ctp2_pj}, and Lemmata \ref{lemma_mes_N_pj} and \ref{lemme_fdmt_W_pj}. Thus, recalling again that $V_t^{x,\phi^{*,x}} \in \mathcal{M}^t(P)$, and $1/\alpha_{t}^{P^*} \in \mathcal{M}^{t}$, we deduce from Lemma \ref{lemme_fdmt_W_pj}  that $K_{t}\in\mathcal{M}^{t}(P)$. This concludes the induction, and the proof.
\Halmos \endproof

\section{Counterexamples}
\label{contreex}
This section brings together several important counterexamples. 
\subsection{$U$ need to be usc}
\label{CE_no_cl_pj}
The existence of an optimal strategy may fail when the utility function is not usc and not concave. 
Let $T=1$, $d=1$, $\Omega^1 = \mathbb{R}.$ Let $\Delta S_1$ and $P_0\in \mathfrak{P}(\Omega^1)$ be such that $\Delta S_1$ is a $P_0$-integrable Borel function, and $D_{P_0}^1 = \mathbb{R}$. Note that, in particular, $P_0[\Delta S_1<0]>0$, and $P_0[\Delta S_1>0]>0$. 
Set $\mathcal{Q}^1:= \{P_0\}$. Finally, let
$$U(x):=x \mbox{ if } x\leq 0, \mbox{ and  }U(x):= 1 \mbox{  otherwise.}$$ 
Then, $U$ is nonconcave, lower-semicontinuous (lsc), but not usc at $x=0$, and satisfies Definition \ref{U_hp_pj}. Moreover, Assumptions \ref{S_borel_pj} and \ref{analytic_graph_pj}  hold (see \eqref{equation_borel_set}). 
Here, $D^1 = D^1_{P_0} = \mathbb{R}$, and so, 
$$\textup{Aff}(D^1) = \textup{Aff}(D^1_{P_0}) = \mathbb{R}, \; \; 0 \in \textup{ri}(\textup{Conv}\big(D_{P_0}^1)\big),$$ 
and Assumption \ref{H_nonempty_pj} holds. 
Assumption \ref{AE_pj} holds for $\overline{\gamma}:=1$, $\underline{\gamma}:=0.5$, and $C:=1$. 
Assumption \ref{nncst_pj} holds for $\underline{X}=-2$. Now, as $U$ is bounded from above (by 1), so is $U_0$, and Assumptions \ref{U0_pj} and \ref{well_def_hp_pj} hold. 
We claim that an optimal strategy fails to exist when $x=0$. For all $h\in\mathbb{R},$ let  
$$ \phi(h):=\mathbb{E}_{P_0} U(h\Delta S_1)=\big(P_0[\Delta S_1>0] - h \mathbb{E}_{P_0} (\Delta S_1)^-\big)1_{\{h> 0\}}+
\big(P_0[\Delta S_1<0] + h \mathbb{E}_{P_0} (\Delta S_1)^+\big)1_{\{h> 0\}}.$$
Thus, $\phi$ is nondecreasing on $(-\infty,0)$, nonincreasing $(0,+\infty)$, and the only possible maximizer for $\phi$ is $h=0$. But, as $\phi(0)=0$, $\phi(0^+)=P_0[\Delta S_1>0]>0$,  and $\phi(0^-)=P_0[\Delta S_1<0]>0$, we see that $h=0$ cannot be a maximizer: there is no optimal strategy for \eqref{RUMP_pj}.\\ 
We now claim that there is a strict inequality at $x=0$ in \eqref{one_step_strategy_temp2_pj} or  \eqref{eg_fdmt_pj}. 
From the previous computations, 
$$u(0)= \max\big(P_0[\Delta S_1>0),P_0[\Delta S_1<0]\big)\in (0,1).$$ 
We have that $\textup{Cl}(U)(x) = x$, if $x< 0$, and $\textup{Cl}(U)(x)= 1$ otherwise. 
We see that $h^*=0$ is a maximizer of $h \mapsto \mathbb{E}_{P_0} \textup{Cl}(U)(h\Delta S_1)$. But, $$\mathbb{E}_{P_0}U(h^*\Delta S_1) = 0 < u(0)= U_0(0) <1 = \mathbb{E}_{P_0} \textup{Cl}(U)(h^*\Delta S_1) = \sup_{h\in\mathbb{R}} \mathbb{E}_{P_0} \textup{Cl}(U)(h\Delta S_1).$$

\subsection{If $V$ is continuous, $\Psi$ may be fail to be usc}
\label{CE_no_cl_2}
Even if $V(\w,\cdot)$ is assumed to be finite and concave (and thus continuous), $\Psi$ may fail to be usc and Problem \eqref{v_pj} may have no solution. \\
Let $N_1,$ $N_2$ and $N_3$ be Borel  functions defined on  $\mathbb{R}$. 
Let $p$ be a probability measure on $\mathbb{R}$ such that $(N_1,N_2,N_3)$ are independent standard Gaussian under $p$. Set $\overline{Y}:= \exp(\exp(N_3))$. 
We choose $\overline{\Omega}=\mathbb{R} \times \{0,1\},$ $d=2$, and define $Y(\cdot) = (Y_1(\cdot),Y_2(\cdot))$ for all $\w=(x,i) \in \overline{\Omega}$ by:
\begin{eqnarray*}
Y_1(\w):=N_1(x) 1_{\{0\}}(i)+ \overline{Y}(x) 1_{\{1\}}(i) \quad Y_2(\w):=N_2(x) 1_{\{0\}}(i)- \overline{Y}(x) 1_{\{1\}}(i). 
\end{eqnarray*}
We also define $p^*$ and $p_0$ on $\overline{\Omega}$ as follows: $p^*:=p \otimes \delta_{\{0\}}$ and $p_0:=p \otimes \delta_{\{1\}}$. Set $\mathcal{Q}:= \textup{Conv}(p_0, p^*).$
We see easily that $(Y_1 ,Y_2)= (N_1,N_2)$ $p^*$-a.s. Thus, $Y_1$ and $Y_2$ are independent, and follow some standard Gaussian laws under $p^*$.  We also have that $(Y_1 ,Y_2)= (\overline{Y},- \overline{Y})$ $p_0$-a.s. \\
Now, Assumption \ref{P^*_pj} holds true as 
$$D_{p^*} = \mathbb{R}^2 = \textup{Aff}(D_{p^*}) \subset \textup{Aff}(D) \subset \mathbb{R}^2 \mbox{ and }0 \in \textup{ri}\big(\textup{Conv}(D_{p^*})\big) = \mathbb{R}^2.$$
We now define $V$. For all $(\w,x) \in\mathbb{R}$, we set 
$$V(\w,x) := \widetilde{V}(x-Z(\w)), \mbox{ with }\widetilde{V}(x):= \big(1-\exp(-x)\big) 1_{\{x\leq 0\}} + x1_{\{x>0\}},\mbox{ and }Z(\w):=\exp\big(N_3(x)\big) 1_{\{1\}}(i).$$
Then, $V(\w,\cdot)$ is finite, and strictly concave for all $\w\in\overline{\Omega}$, $Z = 0$ $p^*$-a.s., and $Z = \exp(N_3)$ $p_0$-a.s. 
Assumption \ref{V_pj} trivially holds, as $Z$ and $V$ are Borel. As $V^+(\w,x)  = (x - Z(\w))^+$ for all $(\w,x)\in \overline{\Omega}\times \mathbb{R},$ 
for all $h=(h_1,h_2) \in \mathbb{R}^2,$ we have that 
\begin{eqnarray}
\label{eqfini}
\mathbb{E}_{p^*} V^+\big(\cdot,x+ h Y(\cdot)\big) \leq |x|+  \mathbb{E}_{p} |h_1 N_1+ h_2N_2|  <+\infty,
\end{eqnarray}
as $h_1 N_1+ h_2N_2$ follows a Gaussian law under $p$, and  Assumption \ref{integ_V+_pj} follows. 
We show now that Assumption \ref{AE_one_pj} holds. Note first that 
$$\mbox{AE}_{-\infty}(V(\w,\cdot)) :=\underset{x\rightarrow -\infty}{\lim \inf} \frac{xV\rq{}(\w,\cdot)}{V(\w,\cdot)} =+\infty.$$ 
Thus, applying \citep[Lemma 8 and Proposition 10]{refnotre_pj} to $V(\w,\cdot)$ shows that for all $\lambda \geq 1$ and $x\in\mathbb{R}$,
$$V(\w, \lambda x) \leq \lambda \big(V(\w,x) + C(\w)\big) \quad \mbox{and} \quad V(\w, \lambda x) \leq \lambda^2 \big(V(\w,x) + C(\w)\big),$$
where $C(\w):= \widetilde{V}^+(-Z(\w)) + \widetilde{V}^-(-Z(\w)) + \widetilde{V}^-(-3)$. As $Z = 0$ $p^*$-a.s., we have that $c^*=\mathbb{E}_{p^*}C<+\infty$, and Assumption \ref{AE_one_pj} follows. 
Assumption \ref{pb_inequality_pj} also holds true. Indeed, as $\widetilde{V}$ is unbounded from below, there exists some $n_0^* \geq 1$ such that 
$$V(\cdot,-n_0^*) =\widetilde{V}(-n_0^*) \leq -\Big(1+ 2 \frac{c^*}{\alpha^*}\Big)  \; p^* \mbox{-a.s.}$$
We claim that $\Psi$ is not usc. We first compute $\Psi_{p_0}(0,h)$. Let $h=(h_1,h_2)\in \mathbb{R}^2,$  $\overline{h} := h_1 - h_2$, and $f(x):= \overline{h} \exp(x) - x$. Then, 
$$h Y - Z = h_1 \overline{Y} - h_2 \overline{Y} - Z   =  \overline{h}\exp(\exp(N_3)) - \exp(N_3)=f\big(\exp(N_3)\big) \; p_0 \mbox{-a.s.}$$
If $\overline{h}\leq 0$, $f(\exp(N_3))<0$, and 
$$\Psi_{p_0}(0,h)= \mathbb{E}_{p_0} \widetilde{V}\big(hY(\cdot) -Z(\cdot)\big)= 1- \mathbb{E}_{p}\exp\big[-\overline{h}\exp\big(\exp(N_3)\big) + \exp(N_3)\Big] \leq 1- \mathbb{E}_{p} \exp\big(\exp(N_3)\big) = -\infty.$$
Assume now that $\overline{h}>0$. It is easy to see that $f$ is decreasing on $(-\infty, -\ln(\overline{h}))$, increasing on $(-\ln(\overline{h}), +\infty)$ and that $\lim_{x\to \pm \infty}f(x) = +\infty.$ So, $\min_{x\in\mathbb{R}} f(x) = f(-\ln(\overline{h})) = 1+ \ln(\overline{h})$. \\
If $\overline{h} \geq \exp(-1)$, $f(\exp(N_3)) \geq 0$, and  
$$\Psi_{p_0}(0,h) = \mathbb{E}_{p_0} \widetilde{V}\big(hY(\cdot) -Z(\cdot)\big)= \mathbb{E}_{p_0}(h Y(\cdot) - Z(\cdot)) = \mathbb{E}_{p_0}(\overline{h}\;\overline{Y}(\cdot) - Z(\cdot))  = + \infty,$$ 
as $\mathbb{E}_{p_0} Z(\cdot)  =\mathbb{E}_{p} \exp(N_3)= \exp(1/2) <+\infty$, and $\mathbb{E}_{p_0}(\overline{h}\;\overline{Y}(\cdot)) = \overline{h}\; \mathbb{E}_{p}  \exp(\exp(N_3)) = +\infty$.\\
Assume now that $\overline{h}\in (0,\exp(-1))$. As $\min_{x\in\mathbb{R}} f(x) = f(-\ln(\overline{h})) <0$, and recalling the variations of $f$,  there exist $x_1,x_2$ such that $f(x)\leq 0$ if and only if $x\in [x_1,x_2]$. Thus,
\begin{eqnarray*}
\mathbb{E}_{p_0} V^-\big(\cdot, hY(\cdot)\big) & = &  \mathbb{E}_{p_0} \widetilde{V}^-\big(hY(\cdot)- Z(\cdot)\big)=
\mathbb{E}_{p} \Big[\Big(\exp\Big(-f\big(\exp(N_3)\big)\Big) - 1\Big)1_{\{f(\exp(N_3)) \leq 0\}}\Big] \\ 
&  \leq & \mathbb{E}_{p} \Big[\Big(\exp\big(\exp(N_3)\big) - 1\Big)1_{\exp(N_3) \in [x_1,x_2]}\Big] \leq \exp(x_2) -1 <+\infty\\
\mathbb{E}_{p_0} V^+\big(\cdot, hY(\cdot)\big)  &=  & \mathbb{E}_{p_0} \big[h{Y}(\cdot) - Z(\cdot)\big]^+ \geq \mathbb{E}_{p} \big[f\big(\exp(N_3)\big)\big] = \overline{h}  \mathbb{E}_{p} \exp\big(\exp(N_3)\big) -  \mathbb{E}_{p} \exp(N_3)= +\infty.
\end{eqnarray*}
As a result, $\Psi_{p_0}(0,h) = - \infty$ if $h_1\leq h_2$, and $\Psi_{p_0}(0,h) = + \infty$ otherwise. Now, we turn to $\Psi_{p^*}$. Recalling that $Z = 0$ $p^*$-a.s,  for all $(x,h)\in \mathbb{R}\times \mathbb{R}^2,$
\begin{eqnarray}
\label{lastar}
\Psi_{p^*}(x,h) =  \mathbb{E}_{p^*} \widetilde{V}\big(x+hY(\cdot) -Z(\cdot)\big)=\mathbb{E}_{p^*} \widetilde{V}\big(x+hY(\cdot) \big)=
\mathbb{E}_{p} \widetilde{V}( x+ h_1 N_1 + h_2 N_2).
\end{eqnarray}
As $\widetilde{V}$ is strictly concave, $\Psi_{p^*}$ is strictly concave. Now, \eqref{eqfini} implies that  $\Psi_{p^*}$ is finite, and also continuous. Using Jensen's inequality in \eqref{lastar}, we get that for all $h\in\mathbb{R}^2$, 
$$\Psi_{p^*}(0,h) =\mathbb{E}_{p^*} \widetilde{V}(h Y(\cdot)) \leq \widetilde{V}\big(h \mathbb{E}_{p^*} Y(\cdot) \big) = \widetilde{V}(0)=0=\Psi_{p^*}(0,(0,0)),$$ 
and $(0,0)$ is a maximizer for $\Psi_{p^*}(0,\cdot)$. It is the only one as $\Psi_{p^*}(0,\cdot)$ is strictly concave. \\
We now compute to $\Psi(0,h)$, for all $h\in\mathbb{R}^2$:
$$\Psi(0,h) = \inf_{p\in \textup{Conv}(p_0,p^*)} \Psi_p(0,h) = \min\big(\Psi_{p_0}(0,h), \Psi_{p^*}(0,h)\big)= -\infty 1_{h_1 \leq h_2} +  \Psi_{p^*}(0,h)1_{h_1 >h_2}$$
as $ \Psi_{p^*}$ is finite, see \eqref{eqfini}. 
Now, $\Psi$ is not usc at $(0,0)$ as 
$$\lim\sup_{n\to +\infty} \Psi\big(0,(1/n,0)\big) = \lim\sup_{n\to +\infty} \Psi_{p^*}\big(0,(1/n,0)\big) = \Psi_{p^*}\big(0,(0,0)\big) = 0 > \Psi\big(0,(0,0)\big)=-\infty.$$ 
We claim now that there is no maximizer for $\Psi(0,\cdot)$.  Indeed, assume that a maximizer $h^*=(h_1^*,h_2^*)$ exists for $\Psi(0,\cdot)$. 
Then, as $\Psi_{p^*}$ is finite, $h_1^*> h_2^*$, and $\Psi(0,h^*) = \Psi_{p^*}(0,h^*).$ This implies that $h^* \neq (0,0)$, a contradiction. 
Thus, there is no maximizer for $\Psi(0,\cdot)$, and Problem \eqref{v_pj} doesn't have a solution for $x=0$.

\subsection{ZFC is not sufficient  for Theorem \ref{one_step_strategy_pj}}
\label{ZFC_proof}

\begin{theorem}
There exists a model in ZFC that is inconsistent with the (PD) axiom,  with a financial setting such that  
(i) the graphs of the random sets of priors are Borel sets, the arbitrage-free price process is Borel adapted,  and Assumptions \ref{AE_pj} and \ref{nncst_pj} hold with constants; (ii) the utility function is bounded from above, increasing, concave and continuous as of function of $x$, and analytically measurable (i.e., measurable with respect to the sigma-algebra generated by the analytic sets)  as a function of the path, but (iii) the unique optimal strategy in \eqref{one_step_strategy_temp2_pj} is projective, but not Lebesgue measurable.
\label{not_provable}
\end{theorem}
\begin{remark}
The counterexample has a simple financial interpretation. An agent wants to invest in a stock, and may adjust her position on day $1$ and day $2$. She believes that the price of the stock will not move between day $0$ and day $1$, while it will go up (or down) by $1\$$ between day $1$ and day $2$, but she has no idea about the probability of these events. In addition, she considers the possibility that she may lose one of her stocks between days 1 and 2 for a reason independent of the market, and completely unknown to her (e.g., a computer error triggering a selling order). \\
We show that if the randomness due to the computer error is ``complex" enough, and if the ``real world" is the constructible universe $L$ (a model of ZFC inconsistent with the (PD) axiom), there exists no universally measurable optimal strategy.  
%
Without the (PD) axiom, as $S$ is Borel adapted, it is also projective, and as the graphs of the random sets of priors are Borel sets, they are also projective sets: 
Assumptions \ref{S_borel_pj} and \ref{analytic_graph_pj} hold. 
As ${U}$ is bounded from above, Assumptions \ref{U0_pj} and \ref{well_def_hp_pj} follow immediately. 
Now, as $U(\cdot,x)$ is analytically measurable, it is also projective. So, all the assumptions of Theorem \ref{one_step_strategy_pj}  are satisfied, excepted for the (PD) axiom. \\
The lack of Lebesgue measurability of the unique optimal strategy in \eqref{one_step_strategy_temp2_pj} shows that any relevant set of admissible strategies must include non-Lebesgue measurable functions. 
However, the constructible universe $L$ is inconsistent with the (PD) axiom, since the axiom of Constructibility holds in $L$. Indeed, the axiom of Constructability provides a non-Lebesgue measurable set in the class $\Delta_2^1(\mathbb{R}^2)$, while the (PD) axiom implies that all sets in $\Delta_2^1(\mathbb{R}^2)$ are Lebesgue measurable, see Theorem \ref{proj_is_univ_pj}. 
Having a set of admissible strategies, which includes non-Lebesgue measurable functions, leads to an ill-posed problem for manipulating integrals of such functions, which are not well-defined without the (PD) axiom. For example, setting in the model below $\mathcal{Q}^1 = \{p\},$ where $p$ is any probability measure with a density with respect to the Lebesgue measure,  still yields the same conclusion in Theorem \ref{not_provable}, except for the function $U_0$, which no longer exists due to the lack of measurability of both $\epsilon^*$ and $\epsilon_*$; the dynamic programming may not be implemented. 
So, Theorem \ref{not_provable} does not contradict Theorem \ref{one_step_strategy_pj}, but justifies using models with ZFC+(PD) for multistep optimization, when the utility function is less than lsa. 
\end{remark}


\proof{Proof of Theorem \ref{not_provable}}
We consider the constructible universe $L$ defined in \citep[Definition 13.1]{refjech_pj}, which is a model of ZFC, where the axiom of Constructibility holds (see \citep[Theorem 13.3, 13.18 and 13.16]{refjech_pj}).  
In that model, \citep[Corollary 25.28, p495]{refjech_pj} provides a set $A \in \Delta_2^1(\mathbb{R}^2) \subset \Sigma_2^1(\mathbb{R}^2)$ (see \eqref{Delta_pj}) that does not belong to the Lebesgue (and so the universal) sigma-algebra. 
 Let $C\in\Pi^{1}_1(\mathbb{R}^2 \times \mathcal{N})$ such that $A = \textup{proj}_{\mathbb{R}^2}(C)$.
 
\textit{Financial market.}\\
Let $T=2,$ $\Omega_1:=\mathbb{R}^2$, $\Omega_2 :=\mathbb{R} \times \mathcal{N}$, and $d=1$. We choose $S_0=0$, and for all  $\w^1\in\Omega^1$,  $\w^2 = (\w_1,\w_2) = (\w_1,(a,u))\in\Omega^2=\Omega_1 \times \Omega_2:$
\begin{eqnarray*}
\Delta S_1(\w^1)& :=& 0 \quad \Delta S_2(\w^2) := 1_{\{a\geq 0\}}  -1_{\{a< 0\}}:=r(a)\\
U(\w^2,x) &:= &\widetilde{U}\big(x- \epsilon(\w^2)\Delta S_2(\w^2)\big) \mbox{ with } \widetilde{U}(x) := x1_{\{x< 0\}} +\big( 1-\exp(-x)\big) 1_{\{x\geq 0\}},  \mbox{ and }
\epsilon(\w^2) := 1_C(\w_1,u)\\
\mathcal{Q}_1  & := &  \mbox{Conv}(\{\delta_{\w_1},\; \w_1\in\Omega_1\}) \quad \mathcal{Q}_2(\w^1):= \mathfrak{P}(\Omega_2). 
\end{eqnarray*}
For all $P = q_1^P \otimes q_2^P \in \mathcal{Q}^2$, and $\w^1\in\Omega^1$, $D^1_P=D^1=\{0\}$, $D^2(\w^1)= \{-1,1\}$, and 
$D^2_P(\w^1) = \{1\}$ if $q_2^P[\mathbb{R}^+ \times  \mathcal{N}|\w^1]=1,$ 
 $D^2_P(\w^1) = \{-1\}$ if $q_2^P[(-\infty,0) \times  \mathcal{N}|\w^1]=1,$ 
 and $D^2_P(\w^1) = \{-1,1\}$ otherwise.\\
\textit{(i) holds}\\
It is clear that $S$ is Borel, and that $\textup{Graph}(\mathcal{Q}_{2})=\Omega^1 \times \mathfrak{P}(\Omega_2)$ is a Borel set. 
For all $\w^1\in\Omega^1,$ let $q_2^P[\cdot|\w^1] := \frac{1}{2} \delta_{\{-1\}} + \frac{1}{2} \delta_{\{1\}}.$ Then, 
 $P = \delta_{\{0\}} \otimes q_2^P \in \mathcal{H}^2$, and $NA(\mathcal{Q}^2)$ holds (see \cite[Theorem 3.29]{ref4_pj}). 
Let $\lambda>1$, and  $x\in\mathbb{R}.$ 
It is easy to see that $U(\lambda x)\leq \lambda^{\gamma}(U(x)+1)$ for $\gamma\in \{\underline{\gamma},\overline{\gamma}\}$ with  $
\underline{\gamma} := 1/2$, and $\overline{\gamma} := 1$. 
Now, let $\w^2\in\Omega^2$,
\begin{eqnarray*}
U(\w^2,\lambda x)  &= &  \widetilde{U}\Big(\lambda \big(x - \frac{1}{\lambda}\epsilon(\w^2) \Delta S_2(\w^2)\big)\Big)
\leq  \lambda^{\gamma}\Big[\widetilde{U}\big(x - \frac{1}{\lambda}\epsilon(\w^2) \Delta S_2(\w^2)\big) +1\Big]\\
&= & \lambda^{\gamma}\Big(U(\w^2,x) + 1 + \widetilde{U}\big(x - \frac{1}{\lambda}\epsilon(\w^2) \Delta S_2(\w^2)\big) - \widetilde{U}\big(x - \epsilon(\w^2) \Delta S_2(\w^2)\big)\Big)\\
&\leq & \lambda^{\gamma}\big(U(\w^2,x) + 2 \big),
\end{eqnarray*}
where the second inequality comes the mean value theorem as $|\widetilde{U}'|\leq 1$, and $|\epsilon \Delta S_2 |\leq 1$. So, Assumption \ref{AE_pj} holds for $U$ with $C:= 2$. Finally, as $U(\w^2,x) \leq \widetilde{U}(x+1)$,  Assumption \ref{nncst_pj} holds for $\underline{X}:=-4$.\\
\textit{(ii) holds}\\
As $\widetilde{U}$ is concave, increasing, and bounded from above, it is also continuous, and so is $U(\w^2,\cdot)$ for all $\omega^2 \in \Omega^2$. Now, as $C \in \Pi^{1}_1(\mathbb{R}^2 \times \mathcal{N})$, for all $x,c\in \mathbb{R}$, $\{U(\cdot,x) \leq c\}$ belongs either to $\Pi^{1}_1(\mathbb{R}^2 \times \mathcal{N})$ or  $\Sigma^{1}_1(\mathbb{R}^2 \times \mathcal{N})$, and $U(\cdot,x)$ is  analytically measurable.  \\
\textit{(iii): The (unique) strategy that is a candidate for optimality is not Lebesgue measurable.}\\
We first solve the dynamic programming procedure. First, note that for all $\w^1\in\Omega^1$ and $x\in\mathbb{R}$,
\begin{eqnarray*}
U_1(\w^1,x) &=& \sup_{h\in\mathbb{R}} \inf_{p\in\mathcal{Q}_{2}(\w^1)} \mathbb{E}_p \widetilde{U}\big(x+\big (h -\epsilon(\w^1,\cdot)\big)\Delta S_2(\w^1,\cdot)\big)
= \sup_{h\in\mathbb{R}} \inf_{(a,u) \in\mathbb{R}\times \mathcal{N}} \widetilde{U}\big(x+ \big(h -1_C(\w_1,u)\big)r(a)\big),\\
&=&\widetilde{U}\Big(x+ \sup_{h\in\mathbb{R}} \inf_{(a,u) \in\mathbb{R}\times \mathcal{N}} \big(h -1_C(\w_1,u)\big)r(a)\Big)= \widetilde{U}\big(x+ \sup_{h\in\mathbb{R}} \min(h -\epsilon^*(\w^1), \epsilon_*(\w^1)-h)\big),\\
&=& \widetilde{U}\Big(x+ \frac{\epsilon_*(\w^1)-\epsilon^*(\w^1)}{2}\Big),
\end{eqnarray*}
where $\epsilon^*(\w^1) := \sup_{u\in\mathcal{N}} 1_C(\w_1,u)$, $\epsilon_*(\w^1) := \inf_{u\in\mathcal{N}} 1_C(\w_1,u)$.  The second equality is proved choosing $p= \delta_{(a,u)}$ for $\leq$, and remarking that the infimum in $(a,u)$ do not depend for $\w_2$ for $\geq$. The third equality comes from $\widetilde{U}$ being increasing. Finally, the supremum in $h$ is reached at the (unique) point $h^{*,x}_{\w^1} := H^*(\w^1)$ with $$H^*(\w^1):= \frac{\epsilon^*(\w^1) + \epsilon_*(\w^1)}{2}.$$ 
Now, as $\widetilde{U}$ is continuous, $U_1(\w^1,\cdot)$ is continuous, and $U_1 =  \textup{Cl}(U_1)$. In a similar way, $u_1(\w^1,\cdot,\cdot)$ is also continuous, $\textup{Cl}(u_1) = u_1$, and $U_1 = u_1^{cl}$. So, as the maximizer $H^*$ is independent of  $x$ (and so of any cash position at time $1$), and as $\mathcal{Q}^1$ includes all Dirac measures, we have only one candidate for optimality of $U_1$: $\phi^{*,x}_{1}(\w^1) = H^*(\w^1)$ for all $\w^1\in\Omega^1$. Now, we compute $U_0$. We have for all $x\in\mathbb{R}$, 
\begin{eqnarray*}
U_0(x) &=&
\sup_{h\in\mathbb{R}} \inf_{p\in\mathcal{Q}_{1}}\; \mathbb{E}_p U_1\big(\w^1,x+ h \Delta S_1(\cdot)\big)= 
 \sup_{h\in\mathbb{R}} \inf_{p\in\mathcal{Q}_{1}}\; \mathbb{E}_p \widetilde{U}\Big(x+ \frac{\epsilon_*(\cdot) - \epsilon^*(\cdot)}{2}\Big)\\
& = &  \inf_{\w^1 \in\mathbb{R}^2} \widetilde{U}\Big(x+ \frac{\epsilon_*(\w^1) - \epsilon^*(\w^1)}{2}\Big)
=\widetilde{U}\Big(x+ \inf_{\w_1 \in\mathbb{R}^2} \frac{\epsilon_*(\w^1) - \epsilon^*(\w^1)}{2}\Big),\\
\end{eqnarray*}
as $\widetilde{U}$ is increasing. The supremum in $h$ is trivially attained at any real. We also see that $u_0=U_0$, that both functions are continuous, and coincide with their closure. 
As one-step strategies must belong in $\textup{Aff}(D_{1}) = \{0\}$, the unique strategy that is a candidate for optimality is 
$$\phi^{*,x}=(\phi^{*,x}_1, \phi^{*,x}_2) =(0, H^*).$$
Now, we show that $H^*$ is not Lebesgue measurable. Note first that for all $\w^1\in\Omega^1$,
\begin{eqnarray*}
\epsilon_*(\w^1) = 0 &  \iff & \exists u\in\mathcal{N},\; (\w^1,u)\in (\mathbb{R}^2\times \mathcal{N})\setminus C  \iff \w^1\in \mbox{proj}_{\mathbb{R}^2}((\mathbb{R}^2\times \mathcal{N})\setminus C) \\ 
\epsilon^*(\w^1) = 1 & \iff &  \exists u\in\mathcal{N},\; (\w^1,u) \in C \iff \w^1\in \mbox{proj}_{\mathbb{R}^2}(C)= A\\
H^*(\w^1) & = &  \Bigg\{
    \begin{array}{lll}
        1 & \mbox{if } \w^1 \in (\mathbb{R}^2\setminus\textup{proj}_{\mathbb{R}^2}\big((\mathbb{R}^2 \times \mathcal{N})\setminus C) \big)\cap A\\
        \frac{1}{2} & \mbox{if }  \w^1 \in \textup{proj}_{\mathbb{R}^2}\big((\mathbb{R}^2 \times \mathcal{N})\setminus C\big) \cap A\\
        0 & \mbox{if } \w^1 \in \textup{proj}_{\mathbb{R}^2}\big((\mathbb{R}^2 \times \mathcal{N})\setminus C\big) \cap (\mathbb{R}^2 \setminus A)
    \end{array}
\end{eqnarray*}
If $H^*$ was Lebesgue measurable, $\{H^* = 1\}\cup\{H^* = 1/2\} = A$ would belong to the Lebesgue sigma-algebra, a contradiction to the choice of $A$.
However, $H^*$ is $\Delta_2^1(\mathbb{R}^2)$-measurable, and thus projective. 
\endproof
\section{Appendix}
\label{appendix_pj}

The first part covers the fundamental properties of projective sets and functions.
The second part collects the missing proofs of Section \ref{sett_main_pj} (Lemmata \ref{lemmaH_nonempty_pj} and \ref{simi_qt_na_pj}). The third part contains the other missing results. Lemma \ref{lemma1+_pj} ensures that Assumption \ref{integ_V+_pj} is preserved through dynamic programming, while Lemma \ref{lemma_N_finite_pj} shows that $N_t^*$ (see \eqref{eq_Nt_pj}), and $N_t^P$ (see \eqref{eq_NtP_pj}) are almost-surely finite, and may be integrable. Lemma \ref{lemma_equ_QH_pj} is used directly in the proof of Theorem \ref{one_step_strategy_pj}, and shows that an infimum on $\mathcal{Q}^T$ can coincide with an infimum on $\mathcal{H}^T$. We also provides some properties of the sets $\mathcal{M}^t$ and $\mathcal{M}^t(P),$ 
which are extensively used in the proof of Theorem \ref{optimality_M_typeA_pj}. 

\subsection{Projective sets and functions}
\label{proj_hiera_pj}
\label{subsec_proj_mes_def_pj}
\label{inf_cvt_sub_pj}
This section states some of the results on projective sets and functions proved in \cite{refprojnotre_pj} that are used in this paper. 
The first proposition justifies the term ``hierarchy", and gives crucial properties for $\Sigma_{n}^1(X)$, $\Pi_n^1(X)$, $\Delta_n^1(X)$, and $\textbf{P}(X)$, that are similar to the ones of the analytic sets. It also provides general properties for projective functions.
\begin{proposition}
Let $n\geq 1$ and let $X$, $Y$ and $Z$ be Polish spaces.
\begin{itemize}[align=left]
\item[(i)] We have that  $\Delta_n^1(X) \subset \Delta_{n+1}^1(X)$, $\Sigma_n^1(X) \subset \Sigma_{n+1}^1(X)$. The class $\Delta_n^1(X)$ is a sigma-algebra. 
\item[(ii)] The class $\textbf{P}(X)$ is closed under complements, finite unions and intersections, while the class $\Sigma_n^1(X)$ is closed under countable unions and intersections. If $A\in \textbf{P}(X\times Y)$ (resp. $\Sigma_n^1(X\times Y)$), and $B \in \textbf{P}(Z)$ (resp. $\Sigma_n^1(Z)$), then $\textup{proj}_X(A) \in \textbf{P}(X)$ 
(resp. $\Sigma_n^1(X)$) 
and $f^{-1}(B) \in \textbf{P}(X)$ (resp. $\Sigma_n^1(X)$) for all Borel  functions $f : X \to Z$.
\item[(iii)] We have that  $\Delta_n^1(X)\times \Delta_n^1(Y) \subset \Delta_n^1(X\times Y)$, $\Sigma_n^1(X)\times \Sigma_n^1(Y) \subset \Sigma_n^1(X\times Y)$, $\textbf{P}(X)\times \textbf{P}(Y) \subset \textbf{P}(X\times Y)$, and 
\begin{eqnarray}
\Sigma_n^1(X) \cup \Pi_n^1(X) \subset \Delta_{n+1}^1(X).
\label{eq_ds}
\end{eqnarray}
\item[(iv)] Let $f,g : X \to \mathbb{R}^n$ for some $n\geq 1$. If $f$ and $g$ are projective functions, then $-f$, $fg$, and $f+g$ are also projective functions.
\item[(v)] For all $n\geq 0$, let $f, f_n, g : X\to \mathbb{R}\cup \{-\infty,+\infty\}$. Let $p\geq 1$. Assume that $f$, $f_n$, and $g$ are $\Delta_p^1(X)$-measurable for all $n\geq 0$. Then, $f+g$, $-f$, $\min(f,g)$, $\max(f,g)$, $f^a$ (for $f>0$ and $a\in \mathbb{R}$), $\inf_{n\geq 0} f_n$, and $\sup_{n\geq 0} f_n$ are $\Delta_p^1(X)$-measurable. Now, if $f$ and $g$ are projective functions, then $f+g$, $-f$, $\min(f,g)$, $\max(f,g)$, and $f^a$ (for $f>0$ and $a\in \mathbb{R}$) are also projective functions.
\item[(vi)] Let $g : D \to Y$ and $f: E \to Z$, where $D\subset X$, and $g(D) \subset E \subset Y$. Assume that $f$ is $\Delta_p^1(Y)$-measurable, and $g$ is $\Delta_q^1(X)$-measurable for some $p,q \geq 1$. Then, $f\circ g$ is $\Delta_{p+q}^1(X)$-measurable. Assume that $f$ and $g$ are projective functions. Then, $f\circ g$ is also a projective function.
\item[(vii)] Let $D \in \textbf{P}(X\times Y)$, and $w : X \times Y \to \mathbb{R}\cup\{-\infty,+\infty\}$ be a projective function. Let $D_x := \{y\in Y,\; (x,y)\in D\}$ for all $x\in X$. Then, $w_*, w^* : \textup{proj}_X(D) \to \mathbb{R}\cup\{-\infty,+\infty\}$ defined by
\begin{eqnarray*}
w_*(x) := \inf_{y\in D_x} w(x,y) \;\; \mbox{and} \;\; w^*(x) := \sup_{y\in D_x} w(x,y)
\end{eqnarray*}
are projective functions.\\ 
Assume the (PD) axiom. Let $\epsilon_*>0$, and $\epsilon^*>0$. Then, there exist projective functions $\phi_*, \phi^* : \textup{proj}_X(D) \to Y$ such that $\textup{Graph}(\phi_*) \subset D$, $\textup{Graph}(\phi^*) \subset D$, and for all $x\in \textup{proj}_X(D)$,
\begin{eqnarray*}
w\big(x,\phi_*(x)\big)  < \bigg\{
   \begin{array}{ll}
  w_*(x) +\epsilon_* \;\;\mbox{if}\;\; w_*(x)>-\infty, \\
  -\frac{1}{\epsilon_*} \;\;\mbox{if}\;\; w_*(x)=-\infty.\\
    \end{array}
 & \; & 
w\big(x,\phi^*(x)\big)  > 
\bigg\{
   \begin{array}{ll}
  w^*(x) -\epsilon^* \;\;\mbox{if}\;\; w^*(x)<+\infty, \\
  \frac{1}{\epsilon^*} \;\;\mbox{if}\;\; w^*(x)=+\infty.
    \end{array}
\end{eqnarray*}
\end{itemize}
\label{base_hierarchy_pj}
\end{proposition}
\proof{Proof.}
Assertions (i) to (iii) follow directly from \citep[Proposition 1]{refprojnotre_pj}. Note that for the projection properties in Assertion (ii), we use the direct image with the Borel function  $f = \textup{proj}_X$. 
Assertions (iv), (v), (vi) and (vii) follow respectively from \citep[Lemma 4, Propositions 4, 3, 5 and 8]{refprojnotre_pj}.  
\Halmos \\ \endproof

The next proposition extends \citep[Proposition 7.48, p180]{ref1_pj}, and is a key result to show that the dynamic programming procedure of Section \ref{muti_per_pj} is well-defined.
\begin{proposition}
Assume the (PD) axiom. Let $X$ and $Y$ be Polish spaces, $f : X \times Y \to \mathbb{R}\cup\{-\infty,+\infty\}$, and $q$ be a stochastic kernel on $Y$ given $X$. Let $\lambda : X  \to \mathbb{R}\cup\{-\infty,+\infty\}$ be defined by $$\lambda(x):=\int_{-} f(x,y)q[dy|x].$$
\noindent (i) Assume that $x\mapsto q[\cdot|x]$ is $\Delta_r^1(X)$-measurable for some $r\geq 1$, and that $f$ is $\Delta_p(X\times Y)$-measurable for some $p\geq 1$. Then, $\lambda$ is $\Delta_{p+r+2}^1(X)$-measurable.\\
\noindent (ii) Assume that $x\mapsto q[\cdot|x]$ and $f$ are projective. Then, $\lambda$ is also projective.
\label{univ_cvt_pj} 
\end{proposition}
\proof{Proof.}
See \citep[Proposition 10]{refprojnotre_pj}. Recall these integrals exist (see Remark \ref{rem_proj_int_pj}). 
\Halmos\\ \endproof
This lemma is an application of Propositions \ref{base_hierarchy_pj} and  \ref{univ_cvt_pj}, that solves measurability issues in Section~\ref{muti_per_pj}.
\begin{lemma}
Assume that the (PD) axiom, and Assumptions \ref{S_borel_pj} and \ref{analytic_graph_pj} hold. Let $0\leq t\leq T-1$. 
Let $f : \Omega^{t+1}\times \mathbb{R} \to \mathbb{R}\cup\{-\infty,+\infty\}$ be a projective function. We define $\lambda : \Omega^t\times \mathbb{R}\times \mathbb{R}^d \times \mathfrak{P}(\Omega_{t+1}) \to \mathbb{R}\cup\{-\infty,+\infty\}$, $\lambda_{\inf} : \Omega^t\times \mathbb{R}\times \mathbb{R}^d \to \mathbb{R}\cup\{-\infty,+\infty\}$, and $\lambda_{\sup} : \Omega^t\times \mathbb{R} \to \mathbb{R}\cup\{-\infty,+\infty\}$ as follows
\begin{eqnarray*}
\lambda(\w^t,x,h,p)&:=& \int_{-} f(\w^t,\w_{t+1},x+h\Delta S_{t+1}(\w^t,\w_{t+1})) p[d\w_{t+1}]\\
\lambda_{\inf}(\w^t,x,h)&:=&\inf_{p\in\mathcal{Q}_{t+1}(\w^t)}\lambda(\w^t,x,h,p) \quad \mbox{and} \quad \lambda_{\sup}(\w^t,x):=\sup_{h\in\mathbb{R}^d}\lambda_{\inf}(\w^t,x,h).
\end{eqnarray*}
Let $q \in  SK_{t+1}$. We also define $\lambda^{q} : \Omega^t\times \mathbb{R}\times \mathbb{R}^d \to \mathbb{R}\cup\{-\infty,+\infty\}$, and $\lambda_{\sup}^q : \Omega^t\times \mathbb{R} \to \mathbb{R}\cup\{-\infty,+\infty\}$ as follows
\begin{eqnarray*}
\lambda^{q}(\w^t,x,h)&:=& \int_{-} f\big(\w^t,\w_{t+1},x+h\Delta S_{t+1}(\w^t,\w_{t+1})\big) q[d\w_{t+1}|\w^t]\;\; \mbox{and} \;\;
\lambda_{\sup}^q(\w^t,x):=\sup_{h\in\mathbb{R}^d}\lambda^q(\w^t,x,h).
\end{eqnarray*}
Then, $\lambda$, $\lambda_{\inf}$, $\lambda_{\sup}$, $\lambda^{q}$, and $\lambda_{\sup}^q$ are projective functions.
\label{lemma_mes_u_pj}
\end{lemma}
\proof{Proof.}
Consider the projectively measurable stochastic kernel $q$ defined by $q[\cdot|\w^t,x,h,p]:=p[\cdot]$. Indeed, $(\w^t,x,h,p) \mapsto p[\cdot]$ is Borel, and thus projective, see \eqref{equation_borel_set}. 
Let $g : (\w^t,x,h,p,\w_{t+1})\mapsto f(\w^t,\w_{t+1},x+h \Delta S_{t+1}(\w^t,\w_{t+1}))$. As Assumption \ref{S_borel_pj} holds, Proposition \ref{base_hierarchy_pj}  (iv) and (vi) shows that $g$ is projective.  Proposition \ref{univ_cvt_pj}  (ii) shows that $\lambda$ is projective. \\
Let $q \in SK_{t+1}$, and consider the projectively measurable stochastic kernel $\widehat{q}$ defined by $\widehat{q}[\cdot|\w^t,x,h,p]:=q[\cdot|\w^t]$ (recall that $q \in SK_{t+1}$, and use Proposition \ref{base_hierarchy_pj} (vi)). 
So, Proposition \ref{univ_cvt_pj}  (ii)  again proves that $\lambda^q$ is projective. Assumption \ref{analytic_graph_pj} shows that $\mbox{proj}_{\Omega^t}(\mbox{Graph}(\mathcal{Q}_{t+1}))=\Omega^t$ as $\mathcal{Q}_{t+1}\neq \emptyset$. Let $D := \{(\w^t,x,h,p)\in \Omega^t\times \mathbb{R}\times \mathbb{R}^d \times \mathfrak{P}(\Omega_{t+1}),\; p\in \mathcal{Q}_{t+1}(\w^t)\}$. Using Assumption \ref{analytic_graph_pj} again and (iii) in Proposition \ref{base_hierarchy_pj}, $D\in \textbf{P}(\Omega^t\times \mathbb{R}\times \mathbb{R}^d \times \mathfrak{P}(\Omega_{t+1}))$. So, Proposition \ref{base_hierarchy_pj} (vii) proves that $\lambda_{\inf}$ is projective. 
Finally, Proposition \ref{base_hierarchy_pj} (vii) now with $D = \Omega^t\times \mathbb{R} \times \mathbb{R}^d$ shows that $\lambda_{\sup}$ and $\lambda_{\sup}^q$ are  projective. \Halmos\\ \endproof

\subsection{Section \ref{sett_main_pj}}
\label{miss_proof0_pj}
We state and prove Lemma \Ref{lemmaH_nonempty_pj},  which is used in the proof of Theorems \ref{main_result_NA} and \ref{optimality_pp_pj}, and Lemmata \ref{lemma_2_pj} and \ref{lemma_equ_QH_pj}. We also provide the proof of Lemma \ref{simi_qt_na_pj}.

\begin{lemma}
\label{lemmaH_nonempty_pj}
Assume the (PD) axiom. Assume that Assumptions \ref{S_borel_pj} and \ref{analytic_graph_pj} hold, and that $\mathcal{H}^T \neq \emptyset$.  
We have that (i) 
$\mathcal{H}^T$ and $\mathcal{Q}^T$ have the same polar sets, (ii) $\mbox{NA}(P)$ holds for all $P\in\mathcal{H}^T$, (iii) $\mbox{NA}(\mathcal{Q}^T)$ holds, and (iv) for all $P^*:= q_1^{P^*} \otimes \cdots \otimes q_T^{P^*}\in\mathcal{H}^T$, 
\begin{eqnarray}
\mathcal{P}^T := \big\{(\lambda_1 q_1^{P^*} + (1-\lambda_1)q_1^Q) \otimes \cdots \otimes (\lambda_T q_T^{P^*} + (1-\lambda_T)q_T^Q),\; 0< \lambda_i \leq 1,\; Q\in\mathcal{Q}^T\big\} \subset \mathcal{H}^T.
\label{setPT_pj}
\end{eqnarray}
\end{lemma}
\proof{Proof.}
\textit{Proof of (iv).}\\
Fix $P^* := q_1^{P^*} \otimes \cdots \otimes q_T^{P^*} \in \mathcal{H}^T$. As $\mathcal{Q}_{t+1}$ is convex-valued for all $0\leq t\leq T-1$ (see Assumption \ref{analytic_graph_pj}), we have that $\mathcal{P}^T \subset \mathcal{Q}^T$.  Let $P\in \mathcal{P}^T$. Fix $0\leq t \leq T-1$ and  $\w^t\in\Omega^t.$ One can see easily that $D^{t+1}_{P^*}(\w^t) \subset D_P^{t+1}(\w^t)$.   
Moreover, \eqref{support_def_pj} and \eqref{support_defP_pj} imply that $D_P^{t+1}(\w^t) \subset D^{t+1}(\w^t)$. As a result, we find 
$$\textup{Aff}(D_{P^*}^{t+1}) \subset \textup{Aff}(D_{P}^{t+1})\subset \textup{Aff}(D^{t+1}) \quad \textup{ri}(\textup{conv}(D_{P^*}^{t+1})) \subset \textup{ri}(\textup{conv}(D_P^{t+1})).$$
As $P^*\in\mathcal{H}^T$, $0 \in \textup{ri}(\textup{conv}(D_{P^*}^{t+1}))(\cdot)$ $\mathcal{Q}^t$-q.s., $\textup{Aff}(D_{P^*}^{t+1})(\cdot) = \textup{Aff}(D^{t+1})(\cdot)$ $\mathcal{Q}^t$-q.s. for all $0\leq t\leq T-1$, and we conclude that $P\in\mathcal{H}^T$.

\textit{Proof of (i).}\\
As $\mathcal{H}^T \subset \mathcal{Q}^T$, any $\mathcal{Q}^T$ polar set is a $\mathcal{H}^T$ polar set. Let $A$ be a $\mathcal{H}^T$-polar set. Then, there exists $N\in\mathcal{B}_c(\Omega^T)$ such that $A\subset N$, and $P[N]=0$ for all $P\in\mathcal{H}^T$. Let 
$$Q:=q_1^Q \otimes \cdots \otimes q_T^Q\in\mathcal{Q}^T,\, P:= q_1^{P} \otimes \cdots \otimes q_T^{P}\in\mathcal{H}^T, \mbox{ and } R:= \frac{q_1^Q + q_1^{P}}{2} \otimes \cdots \otimes \frac{q_T^Q + q_T^{P}}{2}.$$
 Then, (iv) proves that $R\in\mathcal{P}^T\subset \mathcal{H}^T$, and so that $R[N]=0$. Now, \citep[Proposition 12]{refnotre_pj} shows that $Q\ll R$, and $Q[N]=0$ follows. As $Q$ is arbitrary in $\mathcal{Q}^T$, 
 $A$ is thus a $\mathcal{Q}^T$-polar set.

\textit{Proof of (ii).}\\
As $\mathcal{H}^T \neq \emptyset$, let $P:=q_1^P\otimes \cdots \otimes q_T^P\in\mathcal{H}^T$. By definition of $\mathcal{H}^T$, for all $0\leq t\leq T-1$, $\w^t\mapsto q_{t+1}[\cdot|\w^t]$ is a projective function,  and  Theorem \ref{proj_is_univ_pj} (i) shows that it is also universally measurable under the (PD) axiom. 
So, $q_{t+1}$ is a universally measurable stochastic kernel. Similarly, Assumption \ref{S_borel_pj} and  Theorem \ref{proj_is_univ_pj} (i) imply that $S_t$ is universally measurable for all $0\leq t\leq T$. Thus, using \citep[Lemma 7.27, p173]{ref1_pj}, one can find for all $0\leq t\leq T$, a $\mathcal{B}(\Omega^t)$-measurable function $\widehat{S}_t$ such that $\widehat{S}_t = S_t$ $P^t$-a.s. In particular, $\Delta\widehat{S}_{t+1} = \Delta S_{t+1}$ $P^{t+1}$-a.s. As a result, using Fubini's theorem, we have that for all $0\leq t\leq T-1$, 
$$E^t := \{\w^t \in\Omega^t,\; q_{t+1}^P[\Delta \widehat{S}_{t+1}(\w^t,\cdot) = \Delta S_{t+1}(\w^t,\cdot) |\w^t] =1 \}$$ is a $P^t$-full-measure set. For all $0\leq t\leq T-1$, let $\widehat{D}^{t+1}_P$ be the conditional support of $\Delta \widehat{S}_{t+1}$ relatively to $P$, see \eqref{support_defP_pj}. Let $0\leq t\leq T-1$ and $\w^t\in E^t$. Then, for all closed subset $A$ of $\mathbb{R}^d$,  
$q_{t+1}^P[\Delta \widehat{S}_{t+1}(\w^t,\cdot)\in A|\w^t]=q_{t+1}^P[\Delta S_{t+1}(\w^t,\cdot)\in A|\w^t]$. So, $D^{t+1}_P(\w^t) = \widehat{D}^{t+1}_P(\w^t)$.
As $E^t$ is a $P^t$-full measure set, and by definition of $\mathcal{H}^T$, we have that 
$$0\in\textup{ri}(\textup{conv}(D_P^{t+1}))(\cdot) = \textup{ri}(\textup{conv}(\widehat{D}_P^{t+1}))(\cdot) \; P^t-a.s.,$$ 
As $\widehat{S}$ is Borel, \citep[Theorem 3]{refjs_pj} applies, and  $NA(P)$ holds for the price process $\widehat{S}=(\widehat{S}_{t})_{0\leq t\leq T}$. Now, let $\phi\in \Phi.$ Note that $V_T^{0,\phi} = \sum_{s=1}^{T} \phi_s \Delta S_s = \sum_{s=1}^{T} \phi_s \Delta \widehat{S}_s =: \widehat{V}_T^{0,\phi}$ $P$-a.s.  Assume that $V_T^{0,\phi}=\widehat{V}_T^{0,\phi}  \geq 0$  $P$-a.s. Then, as $NA(P)$ holds for $\widehat{S}$, $\widehat{V}_T^{0,\phi} = 0$ $P$-a.s. So, $V_T^{0,\phi} = 0$ $P$-a.s.: $NA(P)$  also holds  for $S$.

\textit{Proof of (iii).}\\
Let $\phi\in \Phi$ such that $V_T^{0,\phi}\geq 0$ $\mathcal{Q}^T$-$\mbox{q.s.}$ As $\mathcal{H}^T\subset \mathcal{Q}^T$, for all $P\in\mathcal{H}^T$, $V_T^{0,\phi}\geq 0$ $P$-a.s.,  and from (ii), $V_T^{0,\phi} = 0$ $P$-a.s. Thus, $V_T^{0,\phi} = 0$ $\mathcal{H}^T$-q.s., and also $\mathcal{Q}^T$-q.s. using (i).   Thus, $NA(\mathcal{Q}^T)$ holds true. \Halmos \\

We provide the proof of Lemma \ref{simi_qt_na_pj}, which extends \citep[Proposition 3.35]{ref4_pj} to the projective setup.\\
\textbf{Proof of Lemma \ref{simi_qt_na_pj}}\\
Let $P:=q_1^P\otimes \cdots \otimes q_T^P\in \mathcal{H}^T$, and $0\leq t\leq T-1$.  We define 
$$E^t := \big\{(\w^t,\alpha)\in \Omega^t\times \mathbb{R},\; \forall h\in\mbox{Aff}(D^{t+1})(\w^t),\; h\neq 0,\; \alpha\in (0,1],\; q_{t+1}^P[h\Delta S_{t+1}(\w^t,\cdot)<-\alpha |h| | \w^t ]\geq \alpha\big\}.$$

\textit{Existence of a projective function $\alpha_t^P(\cdot)$}.\\
Assume for a moment that $E^t\in\textbf{P}(\Omega^t\times \mathbb{R})$. Theorem \ref{proj_is_univ_pj} (ii) shows that there exists some projective $\overline{\alpha}_t^P : \textup{proj}_{\Omega^t}(E^t) \to \mathbb{R}$ such that $\textup{Graph}(\overline{\alpha}_t^P) \subset E^t$. As $E^t$ is a projective set, (ii) in Proposition \ref{base_hierarchy_pj} shows that $\textup{proj}_{\Omega^t}(E^t)$ (and its complement) is also a projective set. 
Let $\alpha^t_P : \Omega^t \to (0,1]$ be defined by $\alpha_t^P(\w^t) := \overline{\alpha}_t^P(\w^t)$ if $\w^t\in\textup{proj}_{\Omega^t}(E^t)$, and $\alpha_t^P(\w^t):=1$ otherwise. Then, $\alpha^t_P$ is projective, see Proposition \ref{base_hierarchy_pj} (v). 
Let $\w^t\in\textup{proj}_{\Omega^t}(E^t)$. As $\textup{Graph}(\overline{\alpha}_t^P) \subset E^t$, for all $h\in\mbox{Aff}(D^{t+1})(\w^t)$ such that $h\neq 0$, we have that $q_{t+1}^P[h\Delta S_{t+1}(\w^t,\cdot)<-\alpha_t^P(\w^t)|h| | \w^t ]\geq \alpha_t^P(\w^t).$ Thus, $\textup{proj}_{\Omega^t}(E^t) \subset \Omega^{t,P}_{qNA}$.

\textit{$\Omega^{t,P}_{qNA}$ is a $\mathcal{Q}^t$-full-measure set.} 
Let $$A^t :=\big \{\w^t\in\Omega^t,\; 0 \in \textup{ri}\big(\textup{Conv}(D_P^{t+1})\big)(\w^t),\; \textup{Aff}(D_P^{t+1})(\w^t) = \textup{Aff}(D^{t+1})(\w^t) \big\}.$$ As $P\in\mathcal{H}^T$, $A^t$ is $\mathcal{Q}^t$-full-measure set. Let $\w^t\in A^t$. Then, using for example  \citep[Proposition 3.3]{ref2_pj}, there exist some constants $\beta$, $\kappa\in(0,1]$ such that for all $h\in\textup{Aff}(D_P^{t+1})(\w^t)$, $h\neq 0$, 
$$q_{t+1}^P[h\Delta S_{t+1}(\w^t,\cdot)<-\beta|h| | \w^t ]\geq \kappa .$$ 
As $\w^t\in A^t$, $\textup{Aff}(D_P^{t+1})(\w^t) = \textup{Aff}(D^{t+1})(\w^t)$, and setting  $\alpha := \min(\beta,\kappa)$, we find that $\w^t\in \textup{proj}_{\Omega^t}(E^t)$. Thus, $A^t \subset \textup{proj}_{\Omega^t}(E^t)$, and as $\textup{proj}_{\Omega^t}(E^t) \subset \Omega^{t,P}_{qNA}$ (see first step), $\Omega^{t,P}_{qNA}$ is a $\mathcal{Q}^t$-full-measure set.

\textit{$E^t \in \textbf{P}(\Omega^t\times \mathbb{R})$.}\\ 
For all $(\w^t,\alpha,h)\in\Omega^t\times\mathbb{R}\times\mathbb{R}^d$, we have that 
$$\lambda(\w^t,\alpha, h):= q_{t+1}^P[h\Delta S_{t+1}(\w^t,\cdot)<-\alpha |h| | \w^t ]- \alpha = \int_{\Omega_{t+1}} f(\w^t,\w_{t+1},\alpha,h) q_{t+1}^P[d\w_{t+1}|\w^t],$$ 
where 
$f(\w^t,\w_{t+1},\alpha,h):= 1_{\{h\Delta S_{t+1}(\w^t,\w_{t+1})+\alpha |h|<0\}} - \alpha.$  As $f$ is projective (see Assumption \ref{S_borel_pj}, and Proposition \ref{base_hierarchy_pj} (iv) and (vi)), Proposition \ref{univ_cvt_pj}  (ii) shows that $\lambda$ is projective. Now, we have that
\begin{eqnarray*}
(\Omega^t\times \mathbb{R})\setminus E^t &=& \big\{(\w^t,\alpha)\in \Omega^t\times \mathbb{R},\; \exists h\in\mbox{Aff}(D^{t+1})(\w^t),\; h\neq 0,\; \alpha\in (0,1],\; \lambda(\w^t,\alpha, h)  <  0 \big\}\\
&=& \textup{proj}_{\Omega^t\times \mathbb{R}} (\mathcal{Z}_1 \cap \mathcal{Z}_2 \cap \mathcal{Z}_3),
\end{eqnarray*}
where $\mathcal{Z}_1 := \sigma^{-1}(\textup{Graph}(\textup{Aff}(D^{t+1})) \times (0,1])$ with $\sigma(\w^t,\alpha,h):= (\w^t,h,\alpha)$, $\mathcal{Z}_2 := \Omega^t\times \mathbb{R} \times (\mathbb{R}^d\setminus\{0\})$ and $\mathcal{Z}_3 := \lambda^{-1}((-\infty,0))$. 
\citep[Proposition 11]{refprojnotre_pj}, together with (ii) and (iii) in Proposition \ref{base_hierarchy_pj}, show that $\mathcal{Z}_1 \in \textbf{P}(\Omega^t\times \mathbb{R}\times \mathbb{R}^d)$. 
We have that $\mathcal{Z}_2 \in \mathcal{B}(\Omega^t\times \mathbb{R}\times \mathbb{R}^d) \subset \textbf{P}(\Omega^t\times \mathbb{R}\times \mathbb{R}^d),$ see \eqref{equation_borel_set}. Finally, $\mathcal{Z}_3 \in \textbf{P}(\Omega^t\times \mathbb{R}\times \mathbb{R}^d)$ as $\lambda$ is projective. Thus,  Proposition \ref{base_hierarchy_pj} (ii) shows that $\mathcal{Z}_1 \cap \mathcal{Z}_2 \cap \mathcal{Z}_3 \in \textbf{P}(\Omega^t\times \mathbb{R}\times \mathbb{R}^d)$, $(\Omega^t\times \mathbb{R})\setminus E^t \in \textbf{P}(\Omega^t\times \mathbb{R})$, and $E^t\in \textbf{P}(\Omega^t\times \mathbb{R})$. \Halmos\\ \endproof

\subsection{Sections \ref{muti_per_pj}, \ref{proof_th1_pj} and \ref{proof_optimality_M_pj}}
\label{miss_proof_pj}

The first lemma shows that $\Omega^t_{\ref{integ_V+_pj}}$ is a $\mathcal{Q}^t$-full-measure set, while $\Omega^{t,P}_{\ref{integ_V+_pj}}$ is a $P^t$-full-measure set. It was used in the proof of Proposition \ref{U_t^-&C_pj}. 
\begin{lemma}
Assume the (PD) axiom. Assume that Assumptions \ref{S_borel_pj},  \ref{analytic_graph_pj} and \ref{H_nonempty_pj} hold. Let $P:=q^P_1\otimes \cdot \cdot \cdot \otimes q^P_{T}\in\mathcal{H}^T$, and $0\leq t\leq T-1$. Then, $\Omega^{t}_{\ref{integ_V+_pj}}$ and $\Omega_{\ref{integ_V+_pj}}^{t,P}$ are projective sets, and also universally measurable sets. 
Assume furthermore that Assumption \ref{well_def_hp_pj} holds, and that $U_0^P(1)<+\infty.$ 
Then, $P^t(\Omega_{\ref{integ_V+_pj}}^{t,P})=1$. 
Finally, if Assumptions \ref{S_borel_pj}, \ref{analytic_graph_pj}, \ref{H_nonempty_pj},  \ref{U0_pj}, and \ref{well_def_hp_pj} hold, $\Omega^{t}_{\ref{integ_V+_pj}}$ is a $\mathcal{Q}^t$-full-measure set.
\label{lemma1+_pj}
\label{lemma_2_pj}
\end{lemma}
\proof{Proof.}
 \textit{Assume that Assumptions \ref{S_borel_pj}, \ref{analytic_graph_pj}, and \ref{H_nonempty_pj} holds.} For all $\theta\in\{-1,1\}^d \cup\{0\}$, let
\begin{eqnarray}
\Omega^{t,\theta}_{\ref{integ_V+_pj}} &:=& \Big \{\w^t\in\Omega^t,\; \mathbb{E}_{q^{P^*}_{t+1}[\cdot|\w^t]} U_{t+1}^+\big(\w^t,\cdot,1+\theta \Delta S_{t+1}(\w^t,\cdot)\big)<+\infty  \Big\}\nonumber\\
\Omega^{t,P,\theta}_{\ref{integ_V+_pj},\pm} &:=&  \Big\{\w^t\in\Omega^t,\; \mathbb{E}_{q^P_{t+1}[\cdot|\w^t]} (U^{P}_{t+1})^{\pm}\big(\w^t,\cdot,1 + \theta \Delta S_{t+1}(\w^t,\cdot)\big)<+\infty \Big\}.\label{neg_integ_temp_pj}
\end{eqnarray}
Then, Definitions \ref{def_one_prior_pj} and \ref{def_ctxt_pj} imply that $\Omega^{t,P}_{\ref{integ_V+_pj}} = \bigcap_{\theta\in \{-1,1\}^d} \Omega^{t,P,\theta}_{\ref{integ_V+_pj},+}$, and $\Omega^{t}_{\ref{integ_V+_pj}} = \bigcap_{\theta\in \{-1,1\}^d} \Omega^{t,\theta}_{\ref{integ_V+_pj}}$. 
Let $\theta\in \{-1,1\}^d \cup\{0\}$. 
As $(U_{t+1}^P)^+$ is projective (see Propositions \ref{U_t_well_pj} (i) and \ref{base_hierarchy_pj} (v)), Lemma \ref{lemma_mes_u_pj} applied to  $f=(U_{t+1}^P)^+$,  together with Proposition \ref{base_hierarchy_pj} (vi) show that $\w^t\mapsto \mathbb{E}_{q_{t+1}^P[\cdot|\w^t]} (U_{t+1}^P)^+(\w^t,\cdot,1+\theta \Delta S_{t+1}(\w^t,\cdot))$ is projective. 
This implies that $\Omega^{t,P,\theta}_{\ref{integ_V+_pj},+}$ is a projective set. Similarly, $\Omega^{t,P,\theta}_{\ref{integ_V+_pj},-}$ and $\Omega^{t,\theta}_{\ref{integ_V+_pj}}$ are also projective sets. 
Thus, using Proposition \ref{base_hierarchy_pj} (ii), $\Omega^{t,P}_{\ref{integ_V+_pj}}$ and $\Omega^{t}_{\ref{integ_V+_pj}}$ are projective sets. Now, Theorem \ref{proj_is_univ_pj} (i) shows that all these sets are universally measurable.

 \textit{Assume that Assumptions \ref{S_borel_pj}, \ref{analytic_graph_pj}, \ref{H_nonempty_pj}, and \ref{well_def_hp_pj} holds, and that $U_0^P(1)<+\infty.$ Then, $ P^t(\Omega^{t,P,\theta}_{\ref{integ_V+_pj},+})=1$ for all $\theta\in \{-1,1\}^d\cup\{0\}$ (and so, $P^t(\Omega^{t,P}_{\ref{integ_V+_pj}})=1$).}\\
\noindent By contraposition, assume that there exists $\theta\in \{-1,1\}^d \cup \{0\}$ such that $P^t(\Omega^{t,P,\theta}_{\ref{integ_V+_pj},+})<1$. Then, for all $\w^t\notin \Omega^{t,P,\theta}_{\ref{integ_V+_pj},+}$,
\begin{eqnarray}
\mathbb{E}_{q^P_{t+1}[\cdot|\w^t]}(U^{P}_{t+1})^+\big(\w^t,\cdot,1+\theta \Delta S_{t+1}(\w^t,\cdot)\big) =+\infty. \label{temp_A35P_pj}
\end{eqnarray}

\noindent For all $0\leq k\leq t$, we show by backward induction the following property : there exists some $\widetilde{\Omega}^k_+\in\textbf{P}(\Omega^k)$ such that $P^k[\widetilde{\Omega}^k_+]>0$, and $U_k^P(\w^k,1)=+\infty$ for all $\w^k\in\widetilde{\Omega}_+^k$. The property at $k=0$  will show that $U_0^P(1)=+\infty$ : a contradiction that proves the claim.\\ 
We start with $k=t$. Using \eqref{temp_A35P_pj}, together with Fubini's theorem, we obtain that $\mathbb{E}_{P^{t+1}}(U^{P}_{t+1})^+(\cdot,1+\theta \Delta S_{t+1}(\cdot)) = +\infty$. So, Assumption \ref{well_def_hp_pj} implies that $\mathbb{E}_{P^{t+1}}(U^{P}_{t+1})^-(\cdot,1+\theta \Delta S_{t+1}(\cdot)) < +\infty$. Using again Fubini's theorem,
$$\int_{\Omega^{t}} \mathbb{E}_{q^P_{t+1}[\cdot|\w^t]} (U^{P}_{t+1})^-\big(\w^t,\cdot,1 + \theta \Delta S_{t+1}(\w^t,\cdot)\big) P^t[d\w^t]<+\infty,$$ 
and we deduce that $P^t[\Omega^{t,P,\theta}_{\ref{integ_V+_pj},-}]=1$. Let $\widetilde{\Omega}^t_+ := (\Omega^t \setminus \Omega^{t,P,\theta}_{\ref{integ_V+_pj},+}) \cap \Omega^{t,P,\theta}_{\ref{integ_V+_pj},-}$. Then, $\widetilde{\Omega}^t_+ \in \textbf{P}(\Omega^t)$ (see Proposition \ref{base_hierarchy_pj} (ii)), and 
$P^t[\widetilde{\Omega}^t_+]=P^t[\Omega^t\setminus \Omega^{t,P,\theta}_{\ref{integ_V+_pj},+}]>0$. 
For all $\w^t\in \Omega^{t}$, \eqref{state_val_t_p_pj} implies that
\begin{small}
\begin{eqnarray}
U_t^P(\w^t,1) \geq  \mathbb{E}_{q^P_{t+1}[\cdot|\w^t]} (U^{P}_{t+1})^+\big(\w^t,\cdot,1 + \theta \Delta S_{t+1}(\w^t,\cdot)\big) -\mathbb{E}_{q^P_{t+1}[\cdot|\w^t]} (U^{P}_{t+1})^-\big(\w^t,\cdot,1 + \theta \Delta S_{t+1}(\w^t,\cdot)\big).\label{temp_A35P2_pj}
\end{eqnarray}
\end{small} 
So, \eqref{neg_integ_temp_pj},  \eqref{temp_A35P_pj}, \eqref{temp_A35P2_pj} imply that $U_t^P(\w^t,1)=+\infty$ for all $\w^t\in \widetilde{\Omega}^t_+$: the property is proved for $k=t$.\\ 
Now, we prove the induction step. Assume that the property holds for some $1\leq k+1\leq t$. Define $\widehat{\Omega}^{k}_+:=\{\w^k\in\Omega^k,\;\; q_{k+1}^P[\widetilde{\Omega}^{k+1}_{+,\w^k}|\w^k]>0 \}$, where 
\begin{eqnarray*}
\widetilde{\Omega}^{k+1}_{+,\w^k} := \big\{\w_{k+1}\in\Omega_{k+1},\; (\w^k,\w_{k+1}) \in \widetilde{\Omega}^{k+1}_{+}\big\} &\subset &
\big\{\w_{k+1}\in\Omega_{k+1},\; (U_{k+1}^P)^+(\w^k,\w_{k+1},1) = +\infty\big\}.
\end{eqnarray*}
As $\widetilde{\Omega}^{k+1}_{+}\in\textbf{P}(\Omega^{k+1})$, Proposition \ref{base_hierarchy_pj} (ii) (with the Borel function $\w_{k+1} \mapsto(\w^k,\w_{k+1})$) shows that 
$\widetilde{\Omega}^{k+1}_{+,\w^k} \in \textbf{P}(\Omega_{k+1})$. Thus, $\widehat{\Omega}^{k}_+ \in \textbf{P}(\Omega^k)$ using Proposition \ref{univ_cvt_pj}  (ii). Moreover, we have that 
\begin{eqnarray*}
P^{k+1}\big[\widetilde{\Omega}^{k+1}_{+}\big]
&= &\int_{\widehat{\Omega}^{k}_{+}}q^P_{k+1}[\widetilde{\Omega}^{k+1}_{+,\w^k}|\w^k] P^k[d\w^k]+ \int_{\Omega^k \setminus\widehat{\Omega}^{k}_{+}}q^P_{k+1}[\widetilde{\Omega}^{k+1}_{+,\w^k}|\w^k] P^k[d\w^k]=\int_{\widehat{\Omega}^{k}_{+}}q^P_{k+1}[\widetilde{\Omega}^{k+1}_{+,\w^k}|\w^k] P^k[d\w^k].
\end{eqnarray*}
As $P^{k+1}[\widetilde{\Omega}^{k+1}_{+}]>0$, we get that $P^{k}[\widehat{\Omega}^{k}_{+}]>0$. Now, for all $\w^k\in \widehat{\Omega}^{k}_{+}$, 
\begin{small}
\begin{eqnarray}
\mathbb{E}_{q_{k+1}^P[\cdot|\w^k]} (U_{k+1}^P)^+(\w^k,\cdot,1) \geq \mathbb{E}_{q_{k+1}^P[\cdot|\w^k]} \big[(U_{k+1}^P)^+(\w^k,\cdot,1)1_{\widetilde{\Omega}^{k+1}_{+,\w^k}}(\cdot)\big] = (+\infty)\; q_{k+1}^P\big[\widetilde{\Omega}^{k+1}_{+,\w^k}|\w^k\big]= +\infty, \label{temp_intg_V+_pj}
\end{eqnarray}
\end{small}
using that $q_{k+1}^P[\widetilde{\Omega}^{k+1}_{+,\w^k}|\w^k]>0$ as $\w^k\in \widehat{\Omega}^{k}_{+}$. Now, Fubini's theorem, and $P^{k}[\widehat{\Omega}^{k}_{+}]>0$ show that $\mathbb{E}_{P^{k+1}}(U^{P}_{k+1})^+(\cdot,1)  = +\infty$. Assumption \ref{well_def_hp_pj} implies then that $\mathbb{E}_{P^{k+1}}(U^{P}_{k+1})^-(\cdot,1) < +\infty$, and Fubini's theorem that,
$$\int_{\Omega^{k}} \mathbb{E}_{q^P_{k+1}[\cdot|\w^k]} (U^{P}_{k+1})^-(\w^k,\cdot,1) P^k[d\w^k]<+\infty$$ and we deduce that $P^k[\Omega^{k,P,0}_{\ref{integ_V+_pj},-}]=1$. Let $\widetilde{\Omega}^{k}_{+} := \widehat{\Omega}^{k}_{+} \cap \Omega^{k,P,0}_{\ref{integ_V+_pj},-}$. Then, $\widetilde{\Omega}^{k}_{+}\in\textbf{P}(\Omega^k)$ (using Proposition \ref{base_hierarchy_pj} (ii)), and $P^k[\widetilde{\Omega}^{k}_{+}]=P^k[\widehat{\Omega}^{k}_{+}]>0$. Let $\w^k\in \widetilde{\Omega}^{k}_{+}$. Using  \eqref{state_val_t_p_pj}, 
we see that 
\begin{eqnarray*}
U_k^P(\w^k,1)\geq   \mathbb{E}_{q_{k+1}^P[\cdot|\w^k]} U_{k+1}^P(\w^k,\cdot,1) 
 &\geq & \mathbb{E}_{q^P_{k+1}[\cdot|\w^k]}\big[(+\infty) 1_{(\w^k,\cdot)\in \widetilde{\Omega}^{k+1}_{+}}-   (U^{P}_{k+1})^-(\w^k,\cdot,1)\big]\\
&\geq & (+\infty) q_{k+1}^P\big[\widetilde{\Omega}^{k+1}_{+,\w^k}|\w^k\big] - \mathbb{E}_{q^P_{k+1}[\cdot|\w^k]} \big[(U^{P}_{k+1})^-(\w^k,\cdot,1)\big],
\end{eqnarray*}
using for the last inequality \citep[Lemma 7.11 (a), p139]{ref1_pj} adapted to convention \eqref{cvt_inf_pj}. Recalling \eqref{neg_integ_temp_pj}, and that $q_{k+1}^P[\widetilde{\Omega}^{k+1}_{+,\w^k}|\w^k]>0$ for all $\w^k\in \widehat{\Omega}^k_+$,
we find that $U_k^P(\w^k,1)=+\infty$ for all $\w^k\in \widetilde{\Omega}^{k}_{+}$.

\textit{If Assumptions \ref{S_borel_pj}, \ref{analytic_graph_pj}, \ref{H_nonempty_pj}, \ref{U0_pj} and \ref{well_def_hp_pj} hold, $\Omega^t_{\ref{integ_V+_pj}}$ is a $\mathcal{Q}^t$-full-measure set.}\\
Let $\widetilde{P} := q_{1}^{\widetilde{P}} \otimes \cdots \otimes q_{T}^{\widetilde{P}} \in\mathcal{Q}^T$, and $\widehat{P} := \frac{q_{1}^{P^*}+q_1^{\widetilde{P}}}{2}\otimes \cdot\cdot\cdot \otimes\frac{q_T^{P^*}+q_T^{\widetilde{P}}}{2},$ where $P^*$ is defined in \eqref{P^*_exp_pj}. As $\widehat{P}\in \mathcal{H}^T$ (see \eqref{setPT_pj}), we have that $U_0^{\widehat{P}}(1)<+\infty$ by Assumption \ref{U0_pj}. So, the preceding step shows that $\Omega^{t,\widehat{P}}_{\ref{integ_V+_pj}}$ is a $\widehat{P}^t$-full-measure set. Using now (iii) in Proposition \ref{U_t_well_pj} at $t+1$, $U_{t+1}\leq U_{t+1}^{\widehat{P}}$, and we get that $\Omega^{t,\widehat{P}}_{\ref{integ_V+_pj}}\subset \Omega^{t}_{\ref{integ_V+_pj}}$. Thus,  
$\Omega_{\ref{integ_V+_pj}}^t$ is also a $\widehat{P}^t$-full-measure set. Now, \citep[Proposition 12]{refnotre_pj} shows that $\widetilde{P} \ll  \widehat{P}$ and $\Omega^t_{\ref{integ_V+_pj}}$ is a $\widetilde{P}^t$-full-measure set. As ${\widetilde{P}}$ is arbitrary, $\Omega_{\ref{integ_V+_pj}}^t$ is a $\mathcal{Q}^t$-full-measure set. \Halmos \\ \endproof
The next lemma was used in the proofs of Proposition \ref{U_t^-&C_pj} and Theorem \ref{optimality_M_typeA_pj}.  It ensures that Assumption \ref{pb_inequality_pj} is preserved through dynamic programming, and provides properties of $N_t^P$ and $N_t^*$. 
\begin{lemma}
Assume the (PD) axiom. Assume that Assumptions  \ref{S_borel_pj}, \ref{analytic_graph_pj}, \ref{H_nonempty_pj},  \ref{AE_pj} and \ref{nncst_pj} hold. For all $P:= q_1^P \otimes \cdots \otimes q_T^P\in\mathcal{Q}^T$, let 
\begin{eqnarray}
\widehat{P}_t:= \frac{q_{1}^{P^*}+q_1^{P}}{2} \otimes \cdots \otimes q_{t}^{P^*} \otimes \cdots \otimes \frac{q_T^{P^*}+q_T^{P}}{2}. \label{hatP_lemmaN_pj}
\end{eqnarray}
Then, $\widehat{P}_t \in\mathcal{H}^T$ and $\mathcal{M}^{t-1}(\widehat{P}_{t})\subset \mathcal{M}^{t-1}(P)$, for all  $P\in\mathcal{Q}^T$ and $1\leq t \leq T$.  
Moreover, 
\begin{eqnarray}
\forall P\in\mathcal{H}^T \;  N_{T-1}^{P}<+\infty \;\;P^{T-1}-\mbox{a.s.} \mbox{ and } N^*_{T-1}<+\infty \;\;\mathcal{Q}^{T-1}-\mbox{q.s.}  \label{NtoPfiniteT_pj} \label{Nt_finiteT_pj} 
\end{eqnarray}
\noindent $i)$ Fix $P\in\mathcal{H}^T$. Then, Assertions (A1) and (B1) below hold :\\ 
\noindent (A1): If $1/\alpha_{T-1}^P \in \mathcal{M}^{T-1}(P)$, and $\underline{X}$,   $C$, $U^-(\cdot,0)\in \mathcal{M}^{T}(P)$, then $N_{T-1}^{P}\in\mathcal{M}^{T-1}(P)$.\\
\noindent (B1): If $1/\alpha_{T-1}^{P^*}\in\mathcal{M}^{T-1}(\widehat{P}_T)$, and $\underline{X}$,  $C$, $U^-(\cdot,0) \in \mathcal{M}^T(\widehat{P}_T)$, then $N^*_{T-1} \in \mathcal{M}^{T-1}(P)$.\\
Assume now that there exists $1\leq t\leq T-1$ such that $\widetilde{\Omega}^{t,P}$ (see \eqref{tildeOmegaP_pj}) is a $P^{t}$-full-measure set. Then, 
\begin{eqnarray}
N_{t-1}^{P}<+\infty \;\;P^{t-1}-\mbox{a.s.}
\label{NtP_finite}
\end{eqnarray} and 
Assertion (A2) holds :\\
\noindent (A2): If $1/\alpha_{t-1}^P \in \mathcal{M}^{t-1}(P)$, and $1/\alpha_{t}^P$, $N_{t}^P$, $l_t^P$, $C_t \in \mathcal{M}^{t}(P)$, then $N_{t-1}^{P}\in \mathcal{M}^{t-1}(P)$.\\
\noindent  $ii)$ Finally, assume that there exists $1\leq t\leq T-1$ such that $\widetilde{\Omega}^{t,P}$ (see \eqref{tildeOmegaP_pj}) is a $P^{t}$-full-measure set for all $P\in\mathcal{H}^T$. Then, 
\begin{eqnarray}
N^*_{t-1}<+\infty \;\;\mathcal{Q}^{t-1}-\mbox{q.s.}   \label{NtoPfinite_pj}
\end{eqnarray}
and Assertion (B2) holds for any $P\in\mathcal{H}^T$:\\
\noindent (B2): If $1/\alpha_{t-1}^{P^*}\in\mathcal{M}^{t-1}(\widehat{P}_t)$, and $1/\alpha_t^{\widehat{P}_t}$, $N_{t}^{\widehat{P}_t}$, $l_t^{\widehat{P}_t}$, $C_t\in\mathcal{M}^t(\widehat{P}_t)$, then $N^*_{t-1}\in\mathcal{M}^{t-1}(P)$.
\label{lemma_N_finite_pj}
\end{lemma}
\proof{Proof.}
The proof is completely similar to that of \citep[Lemma 14]{refnotre_pj}, and thus omitted. One only needs to be careful about the definition of $c_t^P$, which differs here. 
For the proof of $\mathcal{M}^{t-1}(\widehat{P}_{t})\subset \mathcal{M}^{t-1}(P)$, it is completely similar to the one of  \citep[Lemma 10]{refnotre_pj}. 
\Halmos \\ \endproof 
The next lemma was used in the proof of Theorem \ref{one_step_strategy_pj}, see \citep[Proposition 3.12]{ref4_pj} for a related result.
\begin{lemma}
Assume the (PD) axiom. Let $X : \Omega^T \to \mathbb{R}\cup\{-\infty,+\infty\}$ be a projective function. Assume that $\mathcal{H}^T\neq \emptyset$, $\mathbb{E}_P X^+ <+\infty$ for all $P\in\mathcal{H}^T$, and $\mathbb{E}_Q X^- <+\infty$ for all  $Q\in \mathcal{Q}^T$. Then, $\inf_{P\in\mathcal{Q}^T} \mathbb{E}_P X = \inf_{P\in\mathcal{H}^T} \mathbb{E}_P X$.
\label{lemma_equ_QH_pj}
\end{lemma}
\proof{Proof.}
As $\mathcal{H}^T \neq \emptyset$, fix $P^*:=q_1^{P^*} \otimes \cdots \otimes q_T^{P^*} \in\mathcal{H}^T$. Let $P:= q_1^P \otimes \cdots \otimes q_T^P\in\mathcal{Q}^T$. Define for all $n\geq 1$ and $1\leq t\leq T$, 
$$P_n^t = \Big(\frac{1}{n}q_1^{P^*}+(1-\frac{1}{n})q_1^P\Big) \otimes \cdots \otimes \Big(\frac{1}{n}q_t^{P^*} + (1-\frac{1}{n})q_t^P\Big).$$
It is clear that $ P_n^T\in \mathcal{H}^T$.  
Fix $n\geq 1$. Let $R_0^1 := q_1^{P^*}$ and $R_0^{t+1} := R_0^t \otimes q_{t+1}^{P^*}$ for all $1\leq t\leq T-1$. Moreover, for all $1\leq k \leq t-1$,
\begin{eqnarray*}
R_k^{t+1} := \frac{\binom{t}{k-1} R_{k-1}^t \otimes q_{t+1}^P + \binom{t}{k} R_{k}^t \otimes q_{t+1}^{P^*}}{\binom{t+1}{k}} & \quad & 
R_t^{t+1} :=  \frac{P^t\otimes q_{t+1}^{P^*} + t R_{t-1}^t \otimes q_{t+1}^P}{\binom{t+1}{t}}.\
\end{eqnarray*}
Then, $(R_k^t)_{0\leq k\leq t-1}\subset \textup{Conv}(\mathcal{Q}^t)$, and are  independent of $n$. One can show by induction on $t$ that 
\begin{eqnarray*}
P_n^t = \Big(1-\frac{1}{n}\Big)^t P^t +  \frac{1}{n^t}\sum_{k=0}^{t-1} \binom{t}{k} (n-1)^k R_k^t.
\label{Pnt_pj}
\end{eqnarray*}
So, we obtain that
\begin{eqnarray}
\inf_{P\in\mathcal{H}^T}\mathbb{E}_{P} X \leq\mathbb{E}_{P_n^T} X &=&  \Big(1-\frac{1}{n}\Big)^T \mathbb{E}_P X +  \frac{1}{n^T}\sum_{k=0}^{T-1} \binom{T}{k} (n-1)^k \mathbb{E}_{R_k^T} X. \label{temp_inf_HQ_pj}
\end{eqnarray}
We show that $\mathbb{E}_P X$ and $\mathbb{E}_{R_k^T} X$ are finite for all $0\leq k\leq T-1$. First, by assumption of the lemma, $\mathbb{E}_Q X^- <+\infty$ for all $Q\in\mathcal{Q}^T$, and thus $\mathbb{E}_R X^- <+\infty$ for all $R\in \textup{Conv}(\mathcal{Q}^{T})$. So, $\mathbb{E}_P X^-$ and $\mathbb{E}_{R_k^T} X^-$ are finite. Now, if $\mathbb{E}_P X^+ = +\infty$ or $\mathbb{E}_{R_k^T} X^+ = +\infty$ for some $0\leq k \leq T-1$, then $\mathbb{E}_{P_n^T} X^+ = +\infty$ using \eqref{temp_inf_HQ_pj}. But this contradicts the assumption of the lemma as $P_n^T\in\mathcal{H}^T$. So, $\mathbb{E}_{P} X^+ <+\infty$, and $\mathbb{E}_{R_k^T} X^+ <+\infty$ for all $0\leq k\leq T-1$. 
So, taking the limit in $n$ in \eqref{temp_inf_HQ_pj} (recall that the $(R_k^T)_{0\leq k \leq T-1}$ do not depend of $n$), we obtain that $\inf_{P\in\mathcal{H}^T}\mathbb{E}_P X \leq \mathbb{E}_P X$. As $P$ is arbitrary in $\mathcal{Q}^T$, we get that
$\inf_{P\in\mathcal{H}^T}\mathbb{E}_P X \leq \inf_{P\in\mathcal{Q}^T}\mathbb{E}_P X$. The reverse inequality is trivial as $\mathcal{H}^T\subset \mathcal{Q}^T$. \Halmos\\ \endproof
We finish with properties of the sets $\mathcal{M}^t(P)$ and $\mathcal{M}^t$, defined in Definition \ref{def_W_pj}. All the results in this part were already proved in \citep{refnotre_pj} for universally measurable functions. Recall that contrary to \citep{refnotre_pj}, which constructs $\mathcal{Q}^T$ using universally measurable stochastic kernels, we construct $\mathcal{Q}^T$ with projectively measurable stochastic kernels (that are in particular universally measurable stochastic kernels under the (PD) axiom, see  Theorem \ref{proj_is_univ_pj} (i)). 
\begin{lemma}
Assume the (PD) axiom. Fix $0\leq t\leq T$, $P\in\mathcal{Q}^T$ and $a\geq 0$. If $X,Y \in\mathcal{M}^t$ (resp. $\mathcal{M}^t(P)$), then $X+Y$, $XY$, $\min(X,Y)$, $\max(X,Y)$ belong to $\mathcal{M}^t$ (resp. $\mathcal{M}^t(P)$), and $X^a$ also belongs to $\mathcal{M}^t$ (resp. $\mathcal{M}^t(P)$) if $X \geq 0$. Let $Z :\Omega^t\to \mathbb{R}\cup\{-\infty,+\infty\}$ be projective. If $0\leq Z\leq Y$ $\mathcal{Q}^t$-q.s. with $Y\in\mathcal{M}^t$, then $Z\in\mathcal{M}^t$. If $0\leq Z\leq Y$ $P^t$-a.s. with $Y\in\mathcal{M}^t(P)$, then $Z\in\mathcal{M}^t(P)$.
\label{lemme_fdmt_W_pj}
\end{lemma}
\proof{Proof.}
The fact that $X+Y$, $XY$, $\min(X,Y)$, $\max(X,Y)$, and $X^a$ (when $X\geq 0$) are projective functions follows from Proposition \ref{base_hierarchy_pj} (v). The rest of the proof is trivial ,and thus omitted. 
\Halmos \\ \endproof
\begin{lemma}
Assume the (PD) axiom. Let $0\leq t\leq  T-1$. Let $X : \Omega^{t+1} \to \mathbb{R}\cup\{-\infty,+\infty\}$ be a projective function, and choose $q_{t+1}\in SK_{t+1}$ such that $q_{t+1}[\cdot|\w^t]\in\mathcal{Q}_{t+1}(\w^t)$ for all $\w^t\in\Omega^t$. Let $\lambda_X : \Omega^t \to \mathbb{R}\cup\{-\infty,+\infty\}$ be defined by $\lambda_X(\w^t):= \mathbb{E}_{q_{t+1}[\cdot|\w^t]} X(\w^t,\cdot)$. Let $P\in\mathcal{Q}^t$. If $X\in \mathcal{M}^{t+1}(P\otimes q_{t+1})$, then $\lambda_X \in \mathcal{M}^t(P)$. If $X\in\mathcal{M}^{t+1}$, then $\lambda_X \in \mathcal{M}^t$.
\label{M_hered_pj}
\end{lemma}
\proof{Proof.}
As $q_{t+1}$ is a projectively measurable stochastic kernel, Proposition \ref{univ_cvt_pj}  shows that $\lambda_X$ is projective. The rest of the proof is entirely similar to that of \citep[Lemma 12]{refnotre_pj}, and so omitted.
\Halmos \endproof

\end{document}